\documentclass[aps,10pt,tightenlines,twocolumn,letterpaper,superscriptaddress,secnumarabic,nofootinbib]{revtex4-1}

\usepackage{amsmath}
\usepackage{amssymb}
\usepackage{wasysym}
\usepackage{bm}
\usepackage{latexsym}
\usepackage{stmaryrd}
\usepackage{textcomp}
\usepackage{verbatim}
\usepackage{slashed}
\usepackage{array}
\usepackage[pdftex]{graphicx}
\usepackage{epstopdf}
\usepackage{layout}
\usepackage{color}
\usepackage{cancel}
\usepackage[pdftex]{hyperref}

\renewcommand{\arraystretch}{1.5}

\newcommand{\lb}{\left[}
\newcommand{\rb}{\right]}
\newcommand{\lc}{\left\{}
\newcommand{\rc}{\right\}}
\newcommand{\abs}[1]{\left|#1\right|}

\newcommand{\of}[1]{\left(#1\right)}
\newcommand{\eqnref}[1]{eqn.~$\left(\ref{#1}\right)$}
\newcommand{\eqnsref}[2]{eqns.~$\left(\ref{#1}\right)$-$\left(\ref{#2}\right)$}
\newcommand{\figref}[1]{fig.~\ref{#1}}

\newcommand{\tabref}[1]{table~\ref{#1}}
\newcommand{\appref}[1]{appendix~\ref{#1}}

\newcommand{\Figref}[1]{Fig.~\ref{#1}}
\newcommand{\Secref}[1]{Section~\ref{#1}}
\newcommand{\secref}[1]{section~\ref{#1}}

\newcommand{\msbar}{\overline{\text{MS}}}

\newcommand{\loopfactor}{\frac{1}{16\pi^2}}

\newcommand{\themodel}{the $Cx$SM}
\newcommand{\bigS}{\mathbb{S}}

\newcommand{\del}{\delta_2}
\newcommand{\theminimum}{$\left( h,S,A\right) = \left(v,x,0\right)$}

\begin{document}

\hfill NPAC-12-01
\title{\Large Complex Scalar Singlet Dark Matter: Vacuum Stability and Phenomenology }
\vspace{4.0cm}
\author{Matthew~Gonderinger}
\email{gonderinger@wisc.edu}
\author{Hyungjun~Lim}
\email{hlim29@wisc.edu}
\affiliation{
{\small Department of Physics, University of Wisconsin-Madison} \\
{\small Madison, WI 53706, USA}}
\author{Michael~J.~Ramsey-Musolf}
\email{mjrm@physics.wisc.edu}
\affiliation{
{\small Department of Physics, University of Wisconsin-Madison} \\
{\small Madison, WI 53706, USA}}
\affiliation{
{\small Kellogg Radiation Laboratory, California Institute of Technology} \\
{\small Pasadena, CA, 91125, USA}}

\begin{abstract}
We analyze one-loop vacuum stability, perturbativity, and phenomenological constraints on a complex singlet extension of the Standard Model (SM) scalar sector containing a scalar dark matter candidate. We study vacuum stability considerations using a gauge-invariant approach and compare with the conventional gauge-dependent procedure. We show that, if new physics exists at the TeV scale, the vacuum stability analysis and experimental constraints from the dark matter sector, electroweak precision data, and LEP allow both a Higgs-like scalar in the mass range allowed by the latest results from CMS and ATLAS and a lighter singlet-like scalar with weak couplings to SM particles.  If instead no new physics appears until higher energy scales,  there may be significant tension between the vacuum stability analysis and phenomenological constraints (in particular electroweak precision data) to the extent that the complex singlet extension with light Higgs and singlet masses would be ruled out. We comment on the possible implications of a scalar with $\sim 125$~GeV mass and future ATLAS invisible decay searches.

\end{abstract}

\maketitle

\section{Introduction}\label{sec:intro}

The Standard Model (SM) of particle physics is known to be an incomplete theory in part because of its inability to explain phenomena such as dark matter and the baryon asymmetry of the universe.  Among the many models vying to supplant the SM, scalar extensions of the SM are among the simplest.  A gauge singlet real scalar extension has been studied as a potential dark matter candidate (see \cite{Bento:2001yk,Bento:2000ah,McDonald:2001vt,McDonald:1993ex,Burgess:2000yq,He:2008qm,Mambrini:2011ik,Guo:2011mi,He:2011gc,Kadastik:2011aa}), for its impact on the electroweak phase transition (EWPT) \cite{Espinosa:2011ax,Profumo:2007wc,Barger:2011vm}, and for its role in Higgs collider phenomenology \cite{Barger:2007im,O'Connell:2006wi}.  The collider phenomenology and dark matter prospects of a complex scalar gauge singlet \cite{Barger:2010yn,Barger:2008jx} and a real scalar $SU(2)_L$ triplet \cite{FileviezPerez:2008bj,Forshaw:2001xq,Chen:2006pb,Cirelli:2007xd} have been studied as well.  Extensions involving four or more new degrees of freedom, such as the 2 Higgs doublet model, have been widely analyzed over the years.  Indeed, such scalar extensions of the SM have resulted in a prolific field of study.

Important assessments of the theoretical self-consistency of scalar extensions are the vacuum stability of the renormalization group (RG) improved one-loop effective potential and perturbativity of the scalar couplings.  Within the SM, vacuum stability and perturbativity have resulted in theoretical bounds on the Higgs mass (see \cite{Sher:1988mj} for a comprehensive review; also, see \cite{Lindner:1985uk,Lindner:1988ww,Sher:1993mf,Casas:1994us,Casas:1994qy,Casas:1996aq}).  In the real singlet extension of the SM, vacuum stability and perturbativity again place bounds on the Higgs mass but also constrain the singlet in dark matter and inflation scenarios \cite{Gonderinger:2009jp,Clark:2009dc,Lerner:2009xg}.  All of these results are dependent upon the cutoff scale of the theory, $\Lambda$.  This is the scale of new physics, the scale above which new massive degrees of freedom can no longer be integrated out of the theory and become relevant for interactions and the effective potential.  It is widely anticipated that new physics, in particular supersymmetry (SUSY), will appear at the TeV scale.  The requirement of vacuum stability and perturbativity up to $\Lambda=1$~TeV is thus a minimal requirement of the scalar extensions of the SM and result in the weakest theoretical constraints on the models.  However, given the lack of signatures of SUSY or other new physics in early LHC data (see for example \cite{ATLAS_Caron:2011ge,ATLAS_Collaboration:2011rv,CMS_Chatrchyan:2011ek,CMS_Chatrchyan:2011qs}), it is possible the scale at which new physics and new massive degrees of freedom become relevant lies beyond the TeV scale.  As the cutoff scale increases, vacuum stability and perturbativity can impose increasingly significant constraints on the scalar extensions.

In this work, we study this issue of vacuum stability and perturbativity --- particularly for higher cutoff scales --- for the complex scalar singlet extension of the SM, referred to as \lq\lq \themodel".  With an appropriate set of symmetries, this model yields both a viable dark matter candidate ($A$) as well as two real neutral scalars $h'$ and $S'$ that are mixtures of the SM Higgs boson and the real part of the complex singlet.  We discuss in detail the requirement of vacuum stability of the effective potential, {\em i.e.},  that the electroweak minimum of $V_\mathrm{eff}$  be deeper than any other minimum. Generally, however, there exists at least one deeper minimum at large values of the scalar field $\varphi$ due to top quark loop contributions. The conventional stability requirement, then, is to restrict the effective theory to energy scales below the value of $\varphi$ for which $V_\mathrm{eff}(\varphi)$ falls below the electroweak minimum\footnote{Alternately, if the electroweak minimum is metastable with a lifetime longer than the age of the Universe, then the stability radius in field space can be increased.}. One then identifies the maximum stability scale $\Lambda$ with this maximum value of $\varphi$. This criterion, however, is gauge-dependent since only the value of the potential at its extrema is gauge-invariant\cite{Nielsen:1975fs}, whereas the field itself remains gauge-dependent. Consequently, identifying the cutoff $\Lambda$ with a value of $\varphi$ is not physically meaningful. As an alternative, we will use an analysis of the RG evolution of the quartic couplings of \themodel~since the effective potential is dominated by terms quartic in the fields.  By restricting the running of these parameters to energy scales below an appropriately chosen value of $\Lambda$, we guarantee in a gauge-invariant way that the effective potential is bounded below and that the EW minimum is stable.  

In addition to stability and perturbativity considerations, we will also apply various phenomenological constraints in our analysis of \themodel: results for electroweak precision observables (EWPO), dark matter relic density and direct detection measurements, and limits from LEP.  We also study scenarios that may lead to a strong, first order electroweak phase transition (EWPT) as is needed for electroweak baryogenesis that may lead to relic gravity waves. We find that, should new physics exist at the TeV scale, \themodel~has regions of parameter space which satisfy all constraints and favor a relatively light and weakly coupled singlet-like scalar in addition to a Higgs-like scalar in the current mass range allowed by searches at ATLAS \cite{ATLAS-CONF-2011-157} and CMS \cite{CMS-PAS-HIG-11-023}.  Conversely, if new physics does not appear until higher energy scales well above a TeV, the vacuum stability considerations are in significant tension with experimental constraints, particularly EWPO data.  Rather generally,  \themodel~can be ruled out should new physics fail to appear just below the grand unification (GUT) scale, $M_{\mathrm{GUT}}\simeq 10^{16}$~GeV.  These conclusions hold for both a relatively light dark matter mass (as indicated by the CoGeNT \cite{Aalseth:2011wp_cogent2011,Hooper:2011hd_cogentreanalysis,Aalseth:2010vx_cogent2010}, DAMA/LIBRA \cite{Bernabei:2010mq_damanew,Savage:2008er_damadarkmatter,Bernabei:2008yi_damafirst}, and CRESST-II \cite{Angloher:2011uu} collaborations) and a heavier dark matter mass.  Furthermore, it is possible for the scalars $h'$ and $S'$ to decay to dark matter; in fact, in scenarios where both the dark matter and one of the scalar eigenstates are light and the scale of new physics is roughly a TeV, the branching fraction to dark matter is sufficiently large that the ATLAS detector could be sensitive to these invisible decays.

Our discussion of these issues is organized as follows.  We begin with an introduction to \themodel~in \secref{sec:cxsm_model}.  In \secref{sec:vacuum_stability}, we describe the requirements of vacuum stability and perturbativity in detail, and discuss in detail the impact of gauge dependence on the traditional vacuum stability analysis.  We then present our analysis of the RG evolution as a gauge-independent substitute.  \Secref{sec:other_constraints} introduces phenomenological constraints on \themodel~from the EWPT, EWPO, dark matter relic density and direct detection measurements, and collider physics at LEP and the LHC.  We present our results in \secref{sec:results}.  \Secref{sec:concl} contains our conclusions.

\section{Complex Singlet Model}\label{sec:cxsm_model}

\subsection{Tree Level Potential}

In \themodel, the SM is supplemented by the addition of a single complex scalar degree of freedom that transforms trivially under the SM gauge groups.  Thus, the only renormalizable tree-level interactions between the complex singlet, $\bigS$, and the SM occur in the scalar potential of \eqnref{eq:potential_z2u1} --- the singlet couples to the SM fermions and gauge bosons only through the Higgs, $H$ (sometimes referred to as the ``Higgs portal''~\cite{Patt:2006fw}).  
\begin{multline}
\label{eq:potential_z2u1}
V\of{H,\bigS} = \frac{1}{2}m^2H^\dagger H + \frac{\lambda}{4}\of{H^\dagger H}^2\\
+ \frac{\del}{2}H^\dagger H\abs{\bigS}^2 + \frac{b_2}{2}\abs{\bigS}^2 + \frac{d_2}{4}\abs{\bigS}^4\\
+\left(\frac{1}{4}\abs{b_1}e^{i\phi_{b1}}\bigS^2 +\abs{a_1}e^{i\phi_{a1}}\bigS + \mathrm{c.c.}\right)
\end{multline}
In the absence of the $b_1$ and $a_1$ terms, $V\of{H,\bigS}$ obeys a global $U(1)$ symmetry: $\bigS\rightarrow e^{i\alpha}\bigS$. By breaking this symmetry both spontaneously and softly (through the last two terms), we obtain a cold dark matter candidate. When the singlet gets a vacuum expectation value (vev), $\langle\bigS\rangle\equiv x/\sqrt{2}$ (the Higgs has its usual vev, $\langle H\rangle=\of{0,v/\sqrt{2}}^T$ where $v\equiv 246$~GeV), the global $U(1)$ symmetry is spontaneously broken, the real part of the singlet mixes with the SM Higgs, and the imaginary part of the singlet becomes a massless Goldstone boson.  To give mass to the imaginary part of the singlet so that it can potentially fill the role of a stable cold dark matter candidate, we include the explicit $U(1)$-breaking terms proportional to $b_1$ and $a_1$. 

Note that for $a_1=0$ the potential retains a $Z_2$ symmetry associated with the components of $\bigS$.   Since spontaneously broken discrete symmetries create issues with cosmological domain walls \cite{Zeldovich:1974uw,Kibble:1976sj,Kibble:1980mv} we also introduce an explicit $Z_2$-breaking term proportional to $a_1$.  These additional terms are chosen so that the potential retains a $Z_2$ symmetry for $\mathrm{Im} (\bigS)$, thereby ensuring stability of the dark matter particle. Moreover,  these operators close under renormalization.  The phase $\phi_{a1}$ can be absorbed in a redefinition of $\bigS$ and $\phi_{b1}$, and we choose $\phi_{b1}=\pi$ to avoid mixing between the real and complex components of $\bigS$ \cite{Barger:2008jx}.  Then, expanding $\bigS = \of{S+iA}/\sqrt{2}$ and\footnote{We always choose the minimum of the potential so that the neutral real component of the Higgs doublet has a non-zero vev and the other components, the would-be Goldstone bosons, have zero vev.} $H=h/\sqrt{2}$ gives the tree level potential
\begin{multline}
\label{eq:potential_tree}
V_0\of{h,S,A} = \frac{m^2}{4}h^2 + \frac{\lambda}{16}h^4 + \frac{\del}{8}h^2\of{S^2+A^2}\\
+ \frac{1}{4}\of{b_2-b_1}S^2 + \frac{1}{4}\of{b_2+b_1}A^2 - \sqrt{2}a_1 S\\
+ \frac{d_2}{8}S^2A^2 + \frac{d_2}{16}\of{S^4 + A^4}\ \ .
\end{multline}

Requiring that the potential in \eqnref{eq:potential_tree} have a minimum at $\langle H\rangle = h/\sqrt{2} = v/\sqrt{2}$ and $\langle \bigS\rangle = S+iA = x+i\cdot 0$ gives the following set of minimization conditions:
\begin{equation}\label{eq:tree_min_cond}
\frac{\partial V_0}{\partial h} = 0,\ \ \frac{\partial V_0}{\partial S} = 0,\ \ \frac{\partial V_0}{\partial A} = 0
\end{equation}
where all derivatives are evaluated at \theminimum.  (Note that other solutions to the minimization equations may exist; however, our vacuum stability analysis described in \secref{sec:vacuum_stability} verifies none of these other critical points is a global minimum given values for all of the parameters.)   These minimization conditions allow the Higgs vev $v$ and the singlet vev $x$ to replace $m^2$ and $b_2$ as parameters in \themodel~according to \eqnref{eq:msq_b2_replace}.
\begin{equation}\label{eq:msq_b2_replace}
\begin{aligned}
&m^2 \equiv -\frac{1}{2}\lambda v^2 - \frac{1}{2}\del x^2\\
&b_2 \equiv b_1 + 2\sqrt{2}\frac{a_1}{x}-\frac{1}{2}d_2x^2 - \frac{1}{2}\del v^2
\end{aligned}
\end{equation}
At the minimum, the  mass (second derivative) matrix is then given by \eqnref{eq:tree_mass_matrix}.
\begin{multline}\label{eq:tree_mass_matrix}
\lb\begin{array}{ccc} m_h^2 & m_{hS}^2 & m_{hA}^2 \\ m_{hS}^2 & m_S^2 & m_{SA}^2 \\ m_{hA}^2 & m_{SA}^2 & m_A^2\end{array}\rb =\\
\lb\begin{array}{ccc} \frac{1}{2}\lambda v^2 & \frac{1}{2}\del xv & 0 \\ \frac{1}{2}\del xv & \frac{1}{2}d_2x^2 + \sqrt{2}a_1/x & 0 \\ 0 & 0 & b_1 + \sqrt{2}a_1/x\end{array}\rb 
\end{multline}
We choose the $U(1)$ and $\mathbb{Z}_2$ symmetry breaking parameter $a_1$ such that $a_1\ll x$ --- {\em i.e.}, we take $a_1 = 10^{-3}~{\mathrm{GeV}}^3$ and $x\geq 10$~GeV.  This choice serves two purposes: first, it simplifies the model by reducing by one the number of unknown parameters that must be varied; second, it ensures that the minimum at \theminimum~is the global minimum of the potential, as we will discuss in \secref{sec:vacuum_stability}.  With this choice for $a_1$, the dark matter mass is given by $m_A\simeq\sqrt{b_1}$.  Meanwhile, the non-zero entry for $m_{hS}^2$ induces mixing between the SM Higgs and the real component of the singlet.  The resulting mass eigenstates, which we denote $h'$ and $S'$, have masses given by the eigenvalues of $\mathbb{M}$, the $2\times 2$ upper left quadrant of \eqnref{eq:tree_mass_matrix}.  These eigenvalues are
\begin{equation}\label{eq:tree_mass_evals}
m_\pm^2 = \frac{1}{2}\lb \text{Tr}\of{\mathbb{M}} \pm \sqrt{\of{\text{Tr}\of{\mathbb{M}}}^2 - 4\text{Det}\of{\mathbb{M}}}\rb 
\end{equation}
with $m_+ > m_-$.  In order for these masses to be positive real numbers, the condition $\text{Det}\of{\mathbb{M}}>0$ must hold.  In the limit of small $a_1$, this condition simplifies to $\del^2<\lambda d_2$.  The eigenstates $h', S'$ are written in terms of $h$ and $S$ according to \eqnref{eq:tree_mass_estates}.
\begin{equation}\label{eq:tree_mass_estates}
\lb\begin{array}{c}h' \\ S'\end{array}\rb = \lb\begin{array}{cc} \cos\phi & \sin\phi \\ -\sin\phi & \cos\phi \end{array}\rb\lb\begin{array}c h \\ S\end{array}\rb 
\end{equation}
The eigenstates $h', S'$ couple to the fermions and gauge bosons via SM Higgs couplings reduced by a factor of $\cos\phi, -\sin\phi$, respectively.  The mixing angle $\phi$ is given at tree level by
\begin{equation}\label{eq:tree_mixing_angle}
\tan 2\phi = \frac{2m_{hS}^2}{m_h^2 - m_S^2}\ \ .
\end{equation}
We take the mixing angle to be $-\pi/4\leq\phi\leq\pi/4$ so that $h'$ is always the ``Higgs-like'' eigenstate and $S'$ is always ``singlet-like''\footnote{In the literature, the mass eigenstates are often denoted as $h_1$ and $h_2$.  We use a different notation to emphasize that one state is always ``Higgs-like'' and the other is always ``singlet-like''.}.  Which eigenstate is heavier will depend on our choice of parameters.  When choosing $\del<0$ (which allows for a strongly first order EWPT while being consistent with LEP bounds, as discussed later in \secref{sec:ewpt}),  \eqnsref{eq:tree_mass_matrix}{eq:tree_mixing_angle} imply that $h'$ will be the heavier eigenstate with $\phi<0$ for relatively large $\lambda$ and relatively small $d_2, x$ whereas $S'$ will be the heavier eigenstate with $\phi>0$ for relatively small $\lambda$ and relatively large $d_2, x$.

\subsection{One-Loop Potential}

For our vacuum stability analysis, we use the full Coleman-Weinberg one-loop effective potential at zero temperature with one-loop renormalization group (RG) running parameters.  
\begin{equation}\label{eq:effective_potential}
V_\mathrm{eff}\of{h,S,A} = V_0\of{h,S,A}+V_1\of{h,S,A}
\end{equation}
$V_0\of{h,S,A}$ is given in \eqnref{eq:potential_tree}, where all fields, couplings, and masses are replaced by their RG running counterparts.  The one-loop contribution, calculated in the Landau gauge and renormalized in the $\msbar$ scheme, is given by
\begin{multline}\label{eq:oneloop_potential}
V_1\of{h,S,A}=\\
\frac{1}{64\pi^2}\sum_i n_i\text{Tr}\lc M_i^4\of{\log\frac{M_i^2}{\mu^2}-c_i}\rc \ \ .
\end{multline}
The sum $i$ runs over scalars, fermions, and gauge bosons.  The field-dependent mass matrices $M_i^2$, the number of associated degrees of freedom $n_i$, and the numerical constants $c_i$ are given in \appref{app:rges}.  $\mu$ is the 't Hooft renormalization scale.  As discussed in the next paragraph, the effective potential is renormalization scale independent to one-loop order; any residual scale dependence is higher order.  To remove this residual scale dependence, we would like to choose $\mu$ to minimize the $\log$s in \eqnref{eq:oneloop_potential}.  However, no single choice for $\mu$ will simultaneously minimize all of the $\log$s, and so we make the simple choice $\mu^2=h^2+S^2+A^2$.

The RG equations for the fields, couplings, and masses are determined by requiring scale invariance of the effective potential to one-loop order: the scale dependence implicit in the parameters of $V_0$ cancels the explicit scale dependence in $V_1$, {\em i.e.},
\begin{equation}\label{eq:scale_invariance}
\mu\frac{dV_\mathrm{eff}}{d\mu}=\mu\frac{dV_0}{d\mu} + \mu\frac{\partial V_1}{\partial\mu} = 0\ \ .
\end{equation}
Applying this condition to \themodel~effective potential gives a series of equations to be solved for the $\beta$ and $\gamma$ (anomalous dimension) functions that determine the running of the fields, couplings, and masses.  The $\beta$ and $\gamma$ functions are given in \appref{app:rges}.  For convenience, we take $\mu=M_Z$ as the input scale for all our running parameters.

In analogy with the tree level potential, we apply the minimization conditions to the one-loop effective potential: requiring that the minimum of the effective potential occur at \theminimum~fixes the boundary conditions for the running mass parameters $m^2\of{M_Z}$ and $b_2\of{M_Z}$.  Furthermore, we obtain the masses of the scalars by diagonalizing the matrix of second derivatives of the effective potential evaluated at the minimum.  The dark matter $A$ is protected by a $Z_2$ symmetry so that it is stable and does not mix with $h$ and $S$ at the minimum of the potential even upon inclusion of the one-loop corrections.  (The necessity of the $Z_2$ symmetry to ensure the dark matter cannot decay is a generic feature of these simple scalar extensions of the SM.  The real scalar singlet dark matter extension of the SM is referred to as the $Z_2x$SM for this reason.)  The mass eigenstates $h',S'$ are defined in terms of $h$ and $S$ as in \eqnref{eq:tree_mass_estates} using the one-loop value of the mixing angle $\phi$.

\section{Vacuum Stability Analysis}\label{sec:vacuum_stability}

\subsection{The Vacuum Stability Analysis}

The requirement of absolute vacuum stability is equivalent to requiring that the electroweak (EW) zero-temperature minimum of the effective potential be a global minimum over the energy range for which the SM is valid.  The common practice for vacuum stability analyses in the literature begins, as described above, with the RG improved effective potential (generically $V_\mathrm{eff}\of{\varphi_i}$) and choice of the renormalization scale $\mu^2=\vec{\varphi}^2\equiv\varphi_i\varphi_i$ to minimize logarithms in the one-loop potential.  Then, the maximum radius in field space, $\varphi_{max}$, is found according to the requirement of absolute vacuum stability:
\begin{equation}\label{eq:traditional_vs_req}
V_\mathrm{eff}\of{\varphi_i} > V_\mathrm{eff}\of{\vec{\varphi}_{EW}}\quad\forall\ \vec{\varphi}^2 < \varphi_{max}^2
\end{equation}
where $\vec{\varphi}_{EW}$ gives the values of the fields at the electroweak minimum --- \theminimum~in \themodel.  This maximum radius in field space is identified with the cutoff scale of the effective theory, $\Lambda$.  It's presumed that $\Lambda=\varphi_{max}$ is the scale at which new physics is required to alter the shape of the potential so the electroweak minimum remains a global minimum.

The requirement of absolute vacuum stability can be relaxed to the case of metastability for which the EW minimum may not be a global minimum, but the tunneling probability from the EW minimum to the true global minimum is sufficiently small (the lifetime of the electroweak vacuum is greater than the present age of the universe).  In the real scalar singlet extension of the SM, the authors of \cite{Profumo:2010kp} showed that the vacuum metastability requirement is indeed less restrictive of the model parameter space than the absolute vacuum stability analysis in \cite{Gonderinger:2009jp}.  To obtain more conservative bounds, and for simplicity, we focus on the absolute stability scenario.

\subsection{Vacuum Stability for Scalar Extensions of the SM}

There are two primary considerations for vacuum stability in \themodel.  The first is the possibility of a $Z_2$ symmetry breaking minimum at tree level; the second is the set of constraints on the quartic couplings and their RG evolution.

\subsubsection{$Z_2$ Symmetry Breaking Minimum}

In the $Z_2x$SM, the singlet mass depends on its quadratic mass parameter and its coupling to the Higgs.  In \cite{Gonderinger:2009jp}, it was shown that for dark matter masses in the range 10-100~GeV there is a tension between having a sufficiently large (positive) Higgs-singlet coupling to avoid oversaturating the dark matter relic density and maintaining a stable EW minimum of the potential.  Obtaining light scalar singlet dark matter that saturates the relic density can require a negative mass-squared parameter in the potential, leading to a minimum along the singlet axis of the potential for which $\langle S\rangle \neq 0$ and $\langle H\rangle =0$; thus the $Z_2$ symmetry is broken and the dark matter is not a stable particle.  In the present analysis of \themodel, the dark matter mass depends on the linear parameter $a_1$ and the quadratic parameter $b_1$ (plus small loop corrections).  We have chosen $a_1$ to be small, and so to obtain a positive dark matter mass we unambiguously choose $b_1$ to be positive.  Thus there is no dangerous $Z_2$ symmetry-breaking minimum along the $\text{Im}\left[\mathbb{S}\right]$-axis and we do not have a tension between the dark matter mass and vacuum stability as in the real scalar singlet model.

\subsubsection{Limits on the Quartic Couplings}\label{sec:quartic_couplings}

In \themodel, the stability of the tree level potential minimum is guaranteed simply by requiring that\footnote{See \cite{Barger:2008jx} for further discussion.}
\begin{equation}\label{eq:tree_stability}
\begin{aligned}
&\del^2<\lambda d_2\ \ ,\\
&\lambda>0\ \ ,\\
&d_2>0\ \ .
\end{aligned}
\end{equation}
The first condition is necessary for obtaining positive mass-squared eigenvalues for the mixing between the Higgs and real component of the singlet at the minimum of the potential, \theminimum.  This is of course equivalent to the second derivative test to ensure that the critical point \theminimum~is actually a minimum.  The second two conditions are required for the potential to be bounded below in all scenarios; the first condition is also required for the potential to be bounded below when $\del<0$.

Going beyond tree level with the one-loop potential and the one-loop RGEs affects the stability of the potential in two ways.
\begin{enumerate}
\item 
As in the SM and $Z_2x$SM, a global minimum for $h\gg v$ can arise due to the running of the Higgs quartic coupling $\lambda$.  The large top Yukawa coupling, $y_t$, causes $\lambda$ to evolve to negative values for large $\mu$ when $\lambda\of{M_Z}$ is sufficiently small, as seen from the $\beta$-function in \eqnref{eq:beta_lambda_sm}.  
\begin{equation}\label{eq:beta_lambda_sm}
\beta_{\lambda} = \mu\frac{d\lambda}{d\mu} = \loopfactor\of{6\lambda^2 - 36y_t^4 + \cdots} 
\end{equation}
Along the $h$-axis of the potential, the Higgs self-coupling dominates, viz.:
\begin{equation}\label{eq:quartic_potential}
V_\mathrm{eff}^{(\mathrm{SM})}\sim \lambda h^4\ \ ,\quad h\gg v\ \ .
\end{equation}
Setting $\mu=h$ to minimize large logs in the effective potential thus combines these two effects: the potential can develop a very deep global minimum $\of{h\gg v,S=0,A=0}$, high above the EW scale.\footnote{This is a minimum and not simply an unbounded direction because $\lambda$ does become positive again at higher scales.}  

In the $Z_2x$SM, the Higgs-singlet coupling $\del$ has a positive contribution to the $\beta$-function of $\lambda$ irrespective of the sign of $\del$.  As was shown in \cite{Gonderinger:2009jp}, the contribution of $\del$ to the running of $\lambda$ decreases the theoretical lower bounds on the Higgs mass from vacuum stability.  In \themodel, the running of $\lambda$ is again tempered by the Higgs-singlet coupling $\del$, so a larger value of $\del$ may push this deep minimum above the cutoff scale $\Lambda$.
\item
The second effect is one specific to the choice $\del<0$ when one-loop corrections are included.  At tree level, the requirement $\del^2<\lambda d_2$ is sufficient to prevent a runaway direction in the potential  between the $h$-, $S$-, and $A$-axes when $\del<0$.  However, as the $\beta$-function for $\del$ shows (\eqnref{eq:beta_delta2}), a negative $\del\of{M_Z}$ will decrease as the scale $\mu$ increases.  This could in principle lead to a runaway direction in the potential for some region of field space between the axes.
\end{enumerate}

\subsection{Gauge Dependence}

It has been pointed out that the SM one-loop effective potential depends on the choice of gauge-fixing condition; equivalently, in the $R_\xi$ gauges, the potential depends on the gauge parameter $\xi$ (see \cite{Patel:2011th,Loinaz:1997td} and references therein).  Hence, the field expectation values at a minimum, $\vec{\varphi}_{min}$, do not correspond to a physical observable.  It was shown in \cite{Patel:2011th} that the value of the effective potential at its extrema can be calculated in a gauge-invariant way through a consistent expansion in $\hbar$, provided that the extrema have classical (tree level) analogs.  The validity of this procedure is a consequence of the Nielsen identity \cite{Nielsen:1975fs}.

\begin{figure}
	\includegraphics[width=\columnwidth]{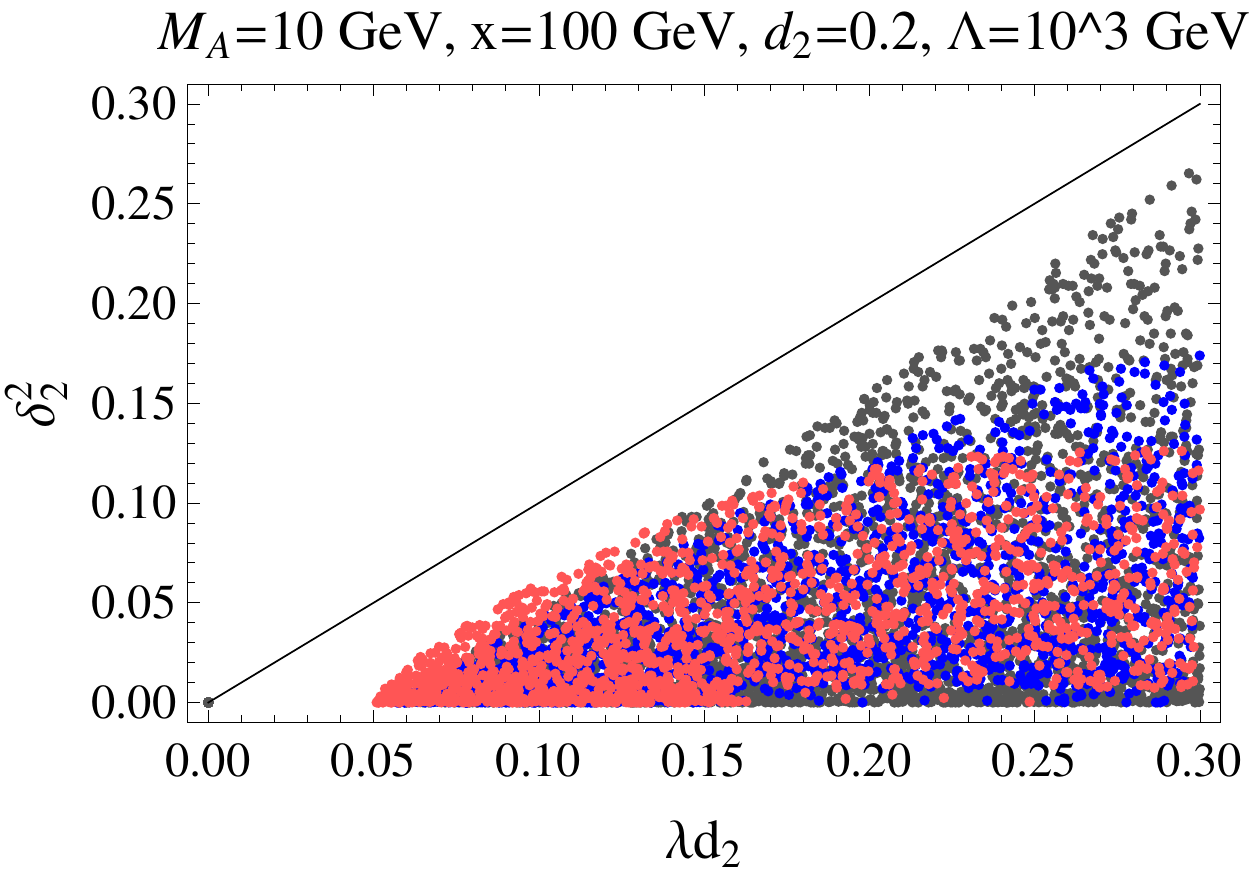}
	\caption{A plot of $\delta_2^2\of{M_Z}$ vs. $\lambda\of{M_Z}d_2\of{M_Z}$.  For all points, $M_{A}=10$~GeV, $x=100$~GeV, and $d_2\of{M_Z}=0.2$.  The tree level vacuum stability requirement, $\del^2<\lambda d_2$, is indicated with the solid line.  All points satisfy the effective potential vacuum stability requirement of \eqnref{eq:traditional_vs_req} with $\Lambda=1$~TeV for some choice of gauge parameter $\xi$.  For gray points, $\xi=0$ (Landau gauge); for blue points, $\xi=1$, and for red points $\xi=50$.  (In this and subsequent figures, the point (0,0) is included for reference only.)}
	\label{fig:del2_vs_lambdad2_gaugedep}
\end{figure}

The gauge dependence of the effective potential presents complications for the vacuum stability analysis despite the existence of the Nielsen identity.  As described above, the vacuum stability analysis generally performed in the literature is interested in a particular radius in field space, $\varphi_{max}$, obtained from \eqnref{eq:traditional_vs_req}.  Because the potential is gauge dependent, it is possible that for one choice of gauge the potential may satisfy the stability requirement below $\varphi_{max}$, but for another choice of gauge the potential may become unstable: 
\begin{equation}\label{eq:eff_pot_gauge_stability}
\begin{aligned}
&V_\mathrm{eff}\of{\varphi_i; \xi_1} > V_\mathrm{eff}\of{\vec{\varphi}_{EW}; \xi_1}\quad \forall\ \vec{\varphi}^2<\varphi_{max}^2\ \ ,\\
&V_\mathrm{eff}\of{\varphi_i; \xi_2} \ngtr V_\mathrm{eff}\of{\vec{\varphi}_{EW}; \xi_2}\quad \forall\ \vec{\varphi}^2<\varphi_{max}^2\ \ .
\end{aligned}
\end{equation}
This ambiguity is dramatically demonstrated in \figref{fig:del2_vs_lambdad2_gaugedep} for one particular choice of \themodel~parameters.  The plotted points are those allowed by the vacuum stability requirement of \eqnref{eq:traditional_vs_req} (with a 1~TeV cutoff) for three different gauge parameters: $\xi=0$ (gray), $\xi=1$ (blue), and $\xi=50$ (red).  Thus, identifying the cutoff scale of the effective theory --- a physical, gauge-independent number --- with $\varphi_{max}$ is problematic.\footnote{Furthermore, in the SM and its extensions such as \themodel, the appearance of a global minimum for $h\gg v$ occurs due to the RG running of the Higgs self-coupling $\lambda$ at one-loop.  There is no classical minimum corresponding to this new global minimum appearing at one-loop, and so the perturbative $\hbar$ expansion described in \cite{Patel:2011th} yields trivial equations when evaluating the potential at this minimum in a gauge-independent way.}

There exist in the literature two methods for performing a gauge-independent analysis of the vacuum stability and corresponding Higgs mass bounds: the ``physical effective potential'' in \cite{Boyanovsky:1997dj}, and the Vilkovisky-DeWitt formalism in \cite{Lin:1998up}.  These methods have been applied to toy models and have derived gauge-independent results that reproduce to within a few percent the results of a traditional vacuum stability analysis done in the Landau gauge \cite{Wainwright:2011qy,Boyanovsky:1997dj}.  To our knowledge, however, no gauge-independent method for analyzing the vacuum stability of the effective potential in the full SM, much less \themodel, has been presented.  

Rather than generalizing either of the above methods, in the current analysis we choose to make vacuum stability arguments based on the running of the quartic couplings that dominate the potential, {\em i.e.},
\begin{multline}
V_\mathrm{eff}\of{h,S,A}\sim \lambda\of{\mu}h^4 \\
+ \del\of{\mu}h^2\of{S^2+A^2} + d_2\of{\mu}\of{S^4+A^4}
\end{multline}
for $\mu^2\gg v^2$.\footnote{A full gauge-invariant vacuum stability analysis of the effective potential in the SM and its scalar extensions is relegated to future work.}  In \themodel, vacuum stability requires that the tree level couplings obey \eqnref{eq:tree_stability}.  We extend these requirements to the one-loop RG running couplings, as in \eqnref{eq:rg_running_stability}.
\begin{equation}\label{eq:rg_running_stability}
\begin{aligned}
&\left.
\begin{aligned}
&\del^2\of{\mu}<\lambda\of{\mu}d_2\of{\mu}\quad\\
&\lambda\of{\mu}>0\\
&d_2\of{\mu}>0
\end{aligned}\right\}
&\quad\forall\ \mu<\Lambda 
\end{aligned}
\end{equation}
After evolving all the parameters of the theory in the effective potential up to the cutoff scale of the theory, if any of these conditions is violated then the potential may become unstable for larger scales\footnote{Minimizing logarithms in the one-loop potential requires the choice $\mu^2=\vec{\varphi}_{min}^2$ when evaluating the potential in a gauge-independent fashion at the minimum $\vec{\varphi}_{min}$, as discussed in \ref{sec:quartic_couplings}.}:
\begin{itemize}
\item If $\lambda\of{\mu}\ngtr 0$ there will be a deep second minimum along the $h$-axis of the potential.
\item If $d_2\of{\mu}\ngtr 0$ there will be a ``runaway direction'' of the potential along the $S$- and $A$-axis, {\em i.e.}, the potential is unbounded from below.
\item If $\del^2\of{\mu}\nless\lambda\of{\mu}d_2\of{\mu}$ and $\del <0$ there will be a runaway direction somewhere between the field axes.
\item If $\del > 0$, the requirement $\del^2\of{\mu}<\lambda\of{\mu}d_2\of{\mu}$ may be overly restrictive.  Though $\del^2<\lambda d_2$ is necessary to ensure that \theminimum~is a minimum, the running of $\del$ at large scales will not affect the shape of the potential at the electroweak minimum.  Thus the EW minimum will remain the global minimum and the potential will be bounded below for large values of the field.
\end{itemize}
The values of the fields where these instabilities occur is immaterial to our analysis; the mere fact that they occur because the conditions of \eqnref{eq:rg_running_stability} are violated implies that the vacuum stability requirement is not satisfied.  Since the RG evolution of all the mass and coupling parameters in \themodel~is gauge-independent, the scale at which any one of the requirements of \eqnref{eq:rg_running_stability} is violated --- which we identify with the cutoff scale of the effective theory, $\Lambda$ --- is also gauge-independent.  We emphasize that the constraints placed on the couplings (and hence the masses of the scalar fields) from \eqnref{eq:rg_running_stability} are motivated by the requirement of vacuum stability.  Since we have not calculated gauge-independent tunneling probabilities for transitions to a non-EW global minimum of the potential, our analysis may give more conservative bounds than those determined by allowing the EW minimum to be metastable.

\subsection{Perturbativity}\label{sec:perturbativity}

We also require that the couplings in the scalar potential remain perturbative for all values of the scale $\mu$.  The definition of ``perturbative'' is somewhat subjective.  At one-loop order in perturbation theory, the quartic scalar couplings all have Landau poles as $\mu$ approaches $\Lambda_{LP}$; minimally, the location of the Landau pole could be taken as the cutoff scale of the theory, $\Lambda = \Lambda_{LP}$.  However, the couplings reach unreasonably large values well before the Landau pole.  Two-loop analysis of the SM RGEs shows that Higgs quartic self-coupling $\lambda$ has a fixed point at large scales where $\beta_\lambda\rightarrow 0$ and $\lambda\of{\mu}\rightarrow \lambda_{FP}$ \cite{Hambye:1996wb}.  Furthermore, it has been shown in \cite{Riesselmann:1996is} that the SM remains perturbative for values of the Higgs quartic self-coupling $\lambda\of{\Lambda}$ in the range $\lambda_{FP}/4$ to $\lambda_{FP}/2$.  A full two-loop analysis of \themodel~ is beyond the scope of our current work, so we impose the an approximate perturbativity constraint on the couplings in \eqnref{eq:coupling_pert_req}.
\begin{equation}\label{eq:coupling_pert_req}
\begin{aligned}
&\left.
\begin{aligned}
&\del\of{\mu}\lesssim\lambda_{FP}/3\quad\\
&\lambda\of{\mu}\lesssim\lambda_{FP}/3\\
&d_2\of{\mu}\lesssim\lambda_{FP}/3
\end{aligned}\right\}
&\quad\forall\ M_Z\leq\mu\leq\Lambda 
\end{aligned}
\end{equation}

\subsection{Analysis Procedure}

In practice, we take as inputs the boundary conditions for the  running Lagrangian parameters (with the boundary conditions for $m^2$ and $b_2$ fixed by the other inputs, the scalar vevs $v = 246$~GeV and $x$).  We then solve the RGEs up to the Planck scale ($\mathcal{O}\of{10^{19}}$~GeV) and determine the scalar masses and mixing angle by diagonalizing the matrix of second derivatives (all evaluated at \theminimum ):
\begin{multline}
\left[
\begin{array}{ccc}
\partial_h^2 V_\mathrm{eff} & \partial_h\partial_S V_\mathrm{eff} & \partial_h\partial_A V_\mathrm{eff} \\
\partial_h\partial_S V_\mathrm{eff} & \partial_S^2 V_\mathrm{eff} & \partial_S\partial_A V_\mathrm{eff} \\
\partial_h\partial_A V_\mathrm{eff} & \partial_S\partial_A V_\mathrm{eff} & \partial_A^2 V_\mathrm{eff}
\end{array}
\right] =
\\
\\
P\cdot \text{Diag}\of{M_{h'}^2, M_{S'}^2, M_A^2}\cdot P^{-1} 
\end{multline}
where the matrix $P$ is the orthogonal matrix containing the mixing parameterized by the angle $\phi$ between the SM Higgs and the real part of the singlet.  We take the cutoff scale $\Lambda$ as an output, obtained by finding the minimum value of $\mu$ for which either \eqnref{eq:rg_running_stability} is violated or one or more of the couplings becomes non-perturbative according to \eqnref{eq:coupling_pert_req}.  We also scan over field space for the minimum radius (if such a point exists) at which the value of the potential is equal to its value at the electroweak minimum --- $V_\mathrm{eff}\of{h,S,A}=V_\mathrm{eff}\of{v,x,0}$ --- to compare the gauge-independent results from the running couplings with the gauge-dependent results of the effective potential in the Landau gauge.  The cutoff scale is required to be at least one TeV.

\section{Implementation of Other Constraints}\label{sec:other_constraints}
\subsection{Electroweak Phase Transition}\label{sec:ewpt}

The electroweak phase transition in a real scalar singlet extension of the SM has been studied extensively \cite{Profumo:2007wc, Espinosa:2011ax, Barger:2011vm}.  If the potential possesses a $Z_2$ symmetry, as in the $Z_2x$SM, which is spontaneously broken by a non-zero singlet vev, the Higgs and singlet mix and the singlet cannot act as a stable dark matter candidate.  Under these circumstances, a negative value of the Higgs-singlet coupling parameter, $\del$, is most suitable for obtaining a strong first-order phase transition and satisfying the LEP constraints on the Higgs mass and mixing angles.  We therefore consider $\del < 0$ for our analysis.  As mentioned in \secref{sec:vacuum_stability}, this choice has the most interesting implications for vacuum stability of the full one-loop potential with running parameters \cite{Chowdhury:2011ga}.

\subsection{Electroweak Precision Observables}\label{sec:ewpo}

In the SM, measurements of electroweak precision observables (EWPO), such as $Z^0$ pole measurements, provide sensitivity to the Higgs mass via loop-level effects.  In \themodel, mixing between the SM Higgs and the real component of the complex singlet alters these loop-level effects.  To determine which values of \themodel~parameters best match EWPO data, we follow the procedure described in \cite{Profumo:2007wc}, which we summarize here.

The EWPO data are parameterized in terms of the oblique parameters $S, T$, and $U$.  Experimental values of the oblique parameters are determined by performing a best fit analysis using all electroweak precision data, as in \cite{Profumo:2007wc}.  Alternatively, the oblique parameters can be calculated analytically as they are defined in terms of the self-energy corrections to the gauge boson propagators: $\Pi_{ZZ}\of{p^2}, \Pi_{WW}\of{p^2}, \Pi_{\gamma\gamma}\of{p^2}, \Pi_{Z\gamma}\of{p^2}$.  Given the direct search limit from LEP on the Higgs mass, a SM reference value, $O^0\equiv O\of{M_h^{\mathrm{SM}}=114.4~\mathrm{GeV}}$, can be computed for each of these oblique parameters ($O=S,T,U$).  The best-fit value determined from electroweak precision data for the difference between $O$ and the SM reference value is defined as
\begin{equation}\label{eq:oblique_deltaO_0}
\Delta O^0\equiv O - O^0\ \ .
\end{equation}

Since the real component of the scalar singlet $\mathbb{S}$ mixes with the Higgs, the propagator corrections $\Pi_{WW}$ and $\Pi_{ZZ}$ in \themodel, and hence the oblique parameters, are different from the SM results (however, $\Pi_{\gamma\gamma}$ and $\Pi_{Z\gamma}$ are unchanged because the scalars are neutral).  The difference can be written as
\begin{multline}\label{eq:oblique_param_diff}
\Delta O \equiv \cos^2\phi\cdot O\of{M_h^{\mathrm{SM}}\rightarrow M_{h'}}\\
+ \sin^2\phi\cdot O\of{M_h^{\mathrm{SM}}\rightarrow M_{S'}} - O^0\ \ .
\end{multline}
Thus, given values for the masses of the scalar eigenstates $h'$ and $S'$ and the mixing angle $\phi$, all extracted from the effective potential, the quantity $\Delta O$ can be computed.  The masses and mixing angle of \themodel~are consistent with EWPO data if the oblique parameter differences $\Delta O$ fall within the 95\% C.L. region of the experimental values $\Delta O^0$.  This is equivalent to $\Delta\chi^2<7.815$, where $\Delta\chi^2$ is defined in \eqnref{eq:deltachisq_def} using the correlation matrix $\rho$ and errors $\sigma$ from \cite{Profumo:2007wc}.
\begin{equation}\label{eq:deltachisq_def}
\Delta\chi^2\equiv \sum_{i,j}\of{\Delta O_i - \Delta O^0_i}\of{\sigma\rho\sigma}^{-1}_{ij}\of{\Delta O_j - \Delta O^0_j}
\end{equation}
The analytic forms of the oblique parameters are given in \appref{app:STU}.

\subsection{Dark Matter Relic Density}\label{sec:relic}

\begin{figure}
\includegraphics[width=\columnwidth]{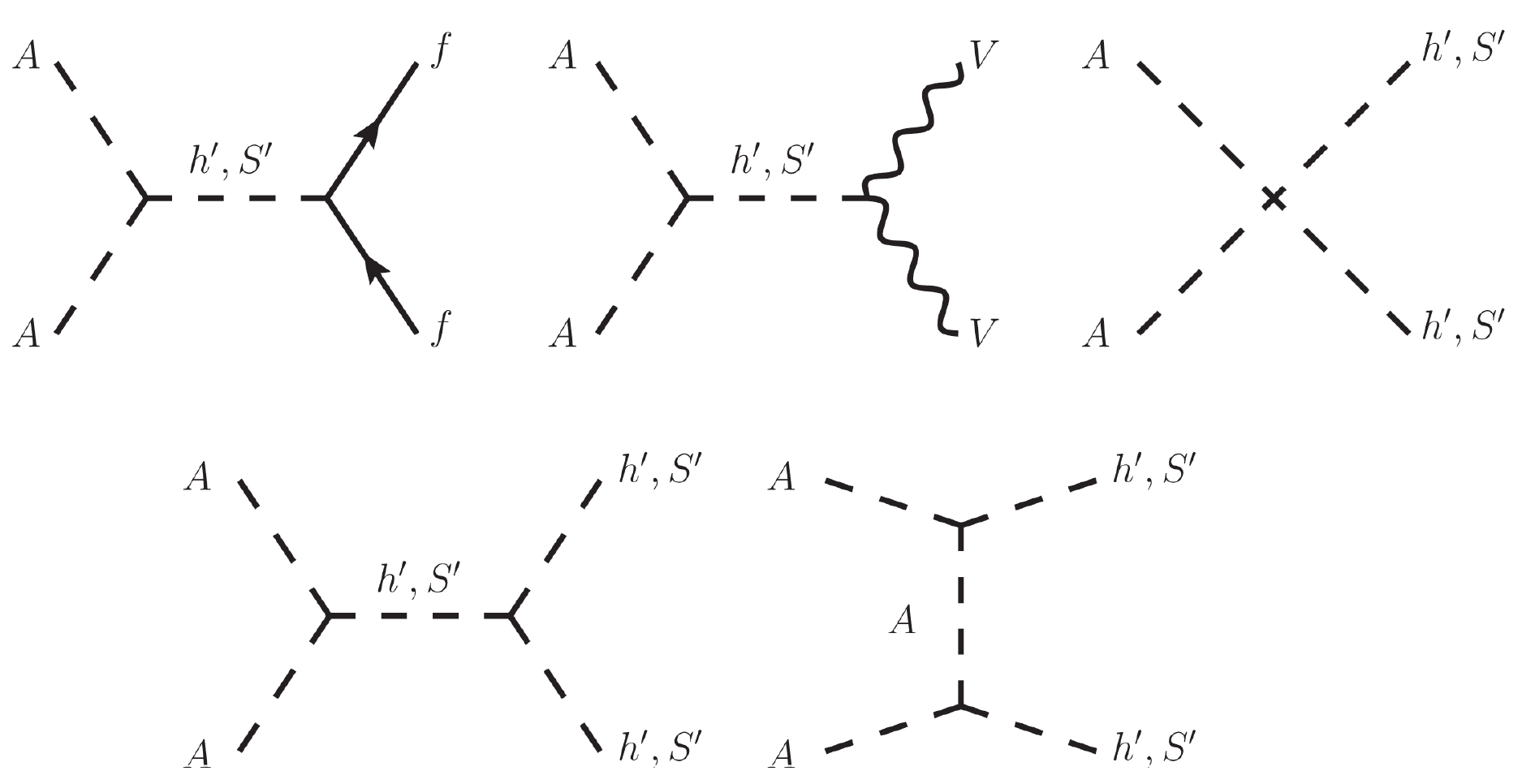}
\label{fig:dm_ann_diagrams}
\caption{Feynman diagrams showing processes contributing to the annihilation cross section of the dark matter particles, $A$.}
\end{figure}

As described in \secref{sec:cxsm_model}, we choose parameters for \themodel~such that the imaginary component of the complex singlet is a stable dark matter candidate.  The thermal relic density of the scalar $A$, $\Omega_A h^2$, is controlled in part by the annihilation cross section of the dark matter particles, $\sigma_{ann}\of{AA\rightarrow XX}$: roughly, $\Omega_A h^2\sim 1/\langle\sigma_{ann} v_{rel}\rangle$ where $\langle\sigma_{ann} v_{rel}\rangle$ is the thermal average of the annihilation cross section times the relative velocity of the dark matter particles in the center-of-mass frame.  Processes that contribute to $\sigma_{ann}$ are shown in \figref{fig:dm_ann_diagrams}.  The kinematical/mass-dependent factors associated with the cross sections for these diagrams can be of particular importance in determining the relic density.  In the limit of non-relativistic dark matter where $\sqrt{s}\simeq 2m_A$ and $v_{rel} \simeq 2\abs{\vec{p}_A}/m_A$,
\begin{multline}\label{eq:dm_ann_cxn}
\of{\frac{d}{d\Omega}\sigma_{ann}\of{AA\rightarrow XX}}v_{rel} \propto \\
\frac{1}{M_A^2}\sqrt{1-\of{\frac{M_X}{M_A}}^2}\abs{\mathcal{M}^2}\ \ .
\end{multline}
In the case of the four-point vertex in \figref{fig:dm_ann_diagrams}, the amplitude $\abs{\mathcal{M}}^2$ is independent of masses and momenta, so the annihilation channels $AA\rightarrow h'h'$ or $AA\rightarrow S'S'$, when kinematically allowed, are largest for $M_{h'}$ or $M_{S'} \ll M_A$.  For the $s$-channel resonances $AA\rightarrow h'\rightarrow XX$ and $AA\rightarrow S'\rightarrow XX$, $\abs{\mathcal{M}}^2\propto \of{4M_A^2-M_{h'}^2}^{-2}$ or $\of{4M_A^2 - M_{S'}^2}^{-2}$; thus the annihilation cross section of course becomes large for $M_{h'}$ or $M_{S'}\simeq 2M_A$.  When the annihilation cross section becomes large, the complex singlet undersaturates the total dark matter relic density, $\Omega_{DM}h^2$, for which we use the WMAP $1\sigma$ measurement $\Omega_{DM}h^2=0.92-0.118$ \cite{Amsler:2008zzb}.  We use the computational tool micrOMEGAs \cite{Belanger:2006is_micromegas_omega} to numerically calculate the relic density.  Though we allow \themodel~to undersaturate the relic density, oversaturation is forbidden.

\subsection{Dark Matter Direct Detection}\label{sec:direct}

A number of experiments have performed searches for dark matter scattering off atomic nuclei and have published limits on the spin-independent scattering cross section as a function of the dark matter mass.  The most restrictive limits at present come from the XENON100 \cite{Aprile:2011hi_xenon2011,Aprile:2010um_xenon2010} and CDMS \cite{Ahmed:2010wy_cdmslowmass,Ahmed:2009zw_cdms2009} experiments.  In apparent conflict with these limits are results from the CoGeNT \cite{Aalseth:2011wp_cogent2011,Hooper:2011hd_cogentreanalysis,Aalseth:2010vx_cogent2010}, DAMA/LIBRA \cite{Bernabei:2010mq_damanew,Savage:2008er_damadarkmatter,Bernabei:2008yi_damafirst}, and CRESST-II \cite{Angloher:2011uu} experiments which have observed signal events corresponding to dark matter particles with $M_{A}\sim 10$~GeV\cite{Hooper:2010uy_damacogent}.  For our analysis, we remain impartial in the debate over these experimental results.  In \themodel, the scattering cross section of the dark matter candidate with a proton is calculated (see \cite{Barger:2010yn,Barger:2008qd}) according to \eqnref{eq:dd_cxn}.
\begin{multline}
\sigma_{dd} = \frac{m_p^4}{2\pi v^2\of{m_p+M_{A}}^2}\\
\times \of{\frac{g_{AAh'}\cos\phi}{M_{h'}^2}-\frac{g_{AAS'}\sin\phi}{M_{S'}^2}}^2 \\
\times \of{f_{pu}+f_{pd}+f_{ps}+\frac{2}{27}\of{3f_G}}^2\label{eq:dd_cxn}
\end{multline}
where
\begin{align}
g_{AAh'}&=\of{\del v\cos\phi + d_2 x\sin\phi}/2\\
g_{AAS'}&=\of{d_2 x\cos\phi -\del v\sin\phi}/2
\end{align}
The proton matrix elements $f$,
\begin{equation}
m_p f_{Tq}^{\of{p}}\equiv \langle p \left| m_q \bar{q}q\right| p\rangle \ \ ,\quad f_{TG}^{\of{p}} = 1-\sum_{q=u,d,s} f_{Tq}^{\of{p}}\ \ ,
\end{equation}
are calculated in \cite{Ellis:2000ds}; we take the central values 
\begin{equation}\label{eq:f_central_vals}
f_{Tu}^{\of{p}} = 0.020\quad f_{Td}^{\of{p}} = 0.026\quad f_{Ts}^{\of{p}} = 0.118
\end{equation}
We consider masses and cross sections that satisfy exactly one of the direct detection experiments: either the upper bound from XENON100 or the signal regions from CoGeNT, or DAMA/LIBRA, or CRESST-II.  We utilize micrOMEGAs \cite{Belanger:2008sj_micromegas_sigma} for numerical calculation of the direct detection cross section.  In comparing this calculated scattering cross section to the limits from the cited experiments, in \eqnref{eq:dd_cxn_scaled} we scale the cross section by the fraction of the total relic density constituted by \themodel~dark matter candidate to account for the reduced flux of dark matter particles in the detectors when the relic density is undersaturated.
\begin{equation}\label{eq:dd_cxn_scaled}
\sigma_{scaled} = \sigma_{dd}\cdot \frac{\Omega_Ah^2}{\Omega_{DM} h^2}
\end{equation}

\subsection{LEP Mixing Angle Constraints}\label{sec:lep}

Application of the LEP limits (this section) and the ATLAS invisibly decaying Higgs search conditions (\secref{sec:invis}) requires calculation of the scalar mass eigenstates' widths.  These are given by \eqnref{eq:scalar_widths}.
\begin{equation}\label{eq:scalar_widths}
\begin{aligned}
&\Gamma_{tot}\of{h'} &&= \cos^2\phi \cdot\Gamma_{\mathrm{SM}}\of{H^*}\ \lb +\ \Gamma\of{h'\rightarrow A A}\rb\\
&&&\quad\lb +\ \Gamma\of{h'\rightarrow S'S'}\rb\ \lb +\ \Gamma\of{h'\rightarrow AAAA}\rb\\
&\Gamma_{tot}\of{S'} &&= \sin^2\phi \cdot\Gamma_{\mathrm{SM}}\of{H^*}\ \lb +\ \Gamma\of{S'\rightarrow A A}\rb\\
&&&\quad\lb +\ \Gamma\of{S'\rightarrow h'h'}\rb\ \lb +\ \Gamma\of{S'\rightarrow AAAA}\rb
\end{aligned}
\end{equation}
In \eqnref{eq:scalar_widths}, $\Gamma_{\mathrm{SM}}\of{H^*}$ is the rate of decays of the SM Higgs to SM final states where the Higgs is assumed to have a mass equivalent to that of the $h'$ or $S'$ eigenstate appropriately.  We calculate the SM Higgs width as a function of the Higgs mass using the program \texttt{HDECAY} \cite{HDECAY_Djouadi:1997yw}.  The decay rates in square brackets in \eqnref{eq:scalar_widths} are only included when the indicated decay is kinematically allowed.  The decay rate of the $h',S'$ eigenstates to pairs of dark matter particles is given in \eqnref{eq:invis_decay_rate}.
\begin{equation}\label{eq:invis_decay_rate}
\Gamma\of{h',S'\rightarrow AA}=\frac{\abs{g_{AAh',AAS'}}^2}{32\pi m_{h',S'}}\sqrt{1-4\frac{m_{DM}^2}{m_{h',S'}^2}}
\end{equation}
The parameters $g_{AAh'}$ and $g_{AAS'}$ are defined in \secref{sec:direct}.  Decays to four dark matter particles have intermediate states of two (possibly off-shell) scalars.

The LEP Working Group for Higgs Boson Searches has made use of the combined data from the four LEP experiments to constrain the mass and $ZZH$ coupling of BSM Higgs-like scalars \cite{Barate:2003sz_lep}.  An upper bound is set on the quantity
\begin{equation}\label{eq:xi_sq}
\xi^2\equiv \of{\frac{g_{ZZH}^{\mathrm{BSM}}}{g_{ZZH}^{\mathrm{SM}}}}^2\times Br\of{H\rightarrow \mathrm{SM}}
\end{equation}
as a function of the Higgs mass.  If the scalar particle $H$ has only SM decays, then $Br\of{H\rightarrow \mathrm{SM}}=1$ and the limits are on the BSM-to-SM ratio of the Higgs-$Z$-$Z$ coupling.  In \themodel, the ratio of the couplings is $\cos^2\phi$ for $h'$ and $\sin^2\phi$ for $S'$.  If additional scalar decays are kinematically allowed, then the widths in \eqnsref{eq:scalar_widths}{eq:invis_decay_rate} are used to calculate $Br\of{H\rightarrow \mathrm{SM}}$.  We apply the LEP limits to both scalar mass eigenstates $h'$ and $S'$.

\subsection{ATLAS Sensitivity to an Invisibly Decaying Higgs}\label{sec:invis}

The mixing between the SM Higgs and the real component of the complex singlet, as well as the potential for one or both eigenstates to decay to an even number of dark matter particles when kinematically allowed, also has implications for Higgs searches at the LHC.  The study in \cite{Aad:2009wy} found that the ATLAS experiment at the LHC would be sensitive, via the vector boson fusion channel, to a Higgs with a mass between 114-200~GeV and an invisible decay mode so long as the condition 
\begin{equation}\label{eq:invis_search_xi}
\xi^2\equiv Br\of{H\rightarrow invis}\times\frac{\sigma_{\mathrm{BSM}}}{\sigma_{\mathrm{SM}}} \gtrsim 60\%
\end{equation}
is satisfied (for masses greater than 200~GeV, the requirement increases to $\xi^2\gtrsim 70\%$).  In \eqnref{eq:invis_search_xi}, $\sigma_{\mathrm{BSM}}$ and $\sigma_{\mathrm{SM}}$ are the Beyond-the-Standard-Model and SM production cross sections, respectively.  In \themodel, $\sigma_{\mathrm{BSM}}\propto\sigma_{\mathrm{SM}}$ where the proportionality factor is either $\cos^2\phi$ for the ``Higgs-like'' eigenstate or $\sin^2\phi$ for the ``singlet-like'' eigenstate.  The invisible decay branching fraction, $Br\of{H\rightarrow invis}$, includes the kinematically allowed decays of the $h'$ or $S'$ to two or four dark matter particles:
\begin{multline}\label{eq:invis_br}
Br\of{h',S'\rightarrow invis} = 
\\\frac{\lb\Gamma\of{h',S'\rightarrow AA}\rb\ \lb +\ \Gamma\of{h',S'\rightarrow AAAA}\rb}{\Gamma_{tot}\of{h',S'}}
\end{multline}
We calculate $\xi^2$ for each choice of \themodel~parameters to determine if the ATLAS experiment is sensitive to decays of the $h'$ or $S'$ eigenstates to dark matter.

\section{Results}\label{sec:results}

In addition to fixing the $Z_2$ breaking parameter $a_1 = 10^{-3}$, we also choose fixed representative values of some of \themodel~parameters for simplicity.  We make the following choices for the parameters:
\begin{itemize}
\item the dark matter mass is 10 or 100~GeV;
\item $d_2\of{M_Z}$ is fixed to 0.2, 0.5, or 0.9;
\item the singlet vev, $x =10, 100,\ \text{or}\ 1000$~GeV.
\end{itemize}
The values for the dark matter mass and the coupling $d_2$ were motivated by the study of the $Z_2x$SM in \cite{Gonderinger:2009jp} which found a dark matter self-coupling of order 0.1-1.0 to be most interesting for satisfying vacuum stability while avoiding problems with non-perturbativity when the dark matter mass is between 10 and 100~GeV.  The chosen order-of-magnitude values for the singlet vev result in masses for the $S'$ state which span a sufficiently large range that allows us to draw conclusions about the parameter space of \themodel.  We summarize in \tabref{table:scans} the values of \themodel~parameters chosen for our analysis.  We will first present our results in detail for a single choice of parameters, and then present our general results for all those values in \tabref{table:scans}.

\begin{table}\label{table:scans}
\begin{tabular}{|c|c|c|}
\hline 
& $M_{A}=10$~GeV & $M_{A}=100$~GeV\\\hline 
$x=10$~GeV & $d_2=0.2, 0.5, 0.9$ & $d_2=0.2, 0.5, 0.9$ \\\hline 
$x=100$~GeV & $d_2=0.2, 0.5, 0.9$ & $d_2=0.2, 0.5, 0.9$ \\\hline 
$x=1000$~GeV & $d_2=0.2, 0.5, 0.9$ & \\
\hline 
\end{tabular}
\caption{A list of parameter scans performed.  $a_1$ is chosen to be $10^{-3}~{\mathrm{GeV}}^3$ so $b_1\simeq M_{A}^2$.  The only free parameters are $\lambda$ and $\del$.  A number between 0 and 1.5 is randomly chosen for $\lambda$; then, a number between 0 and $-\sqrt{\lambda d_2}$ is chosen for $\del$.}
\end{table}

\subsection{A Light Dark Matter Example}\label{sec:light_dark_matter}

\begin{figure*}
	\includegraphics[width=.45\textwidth]{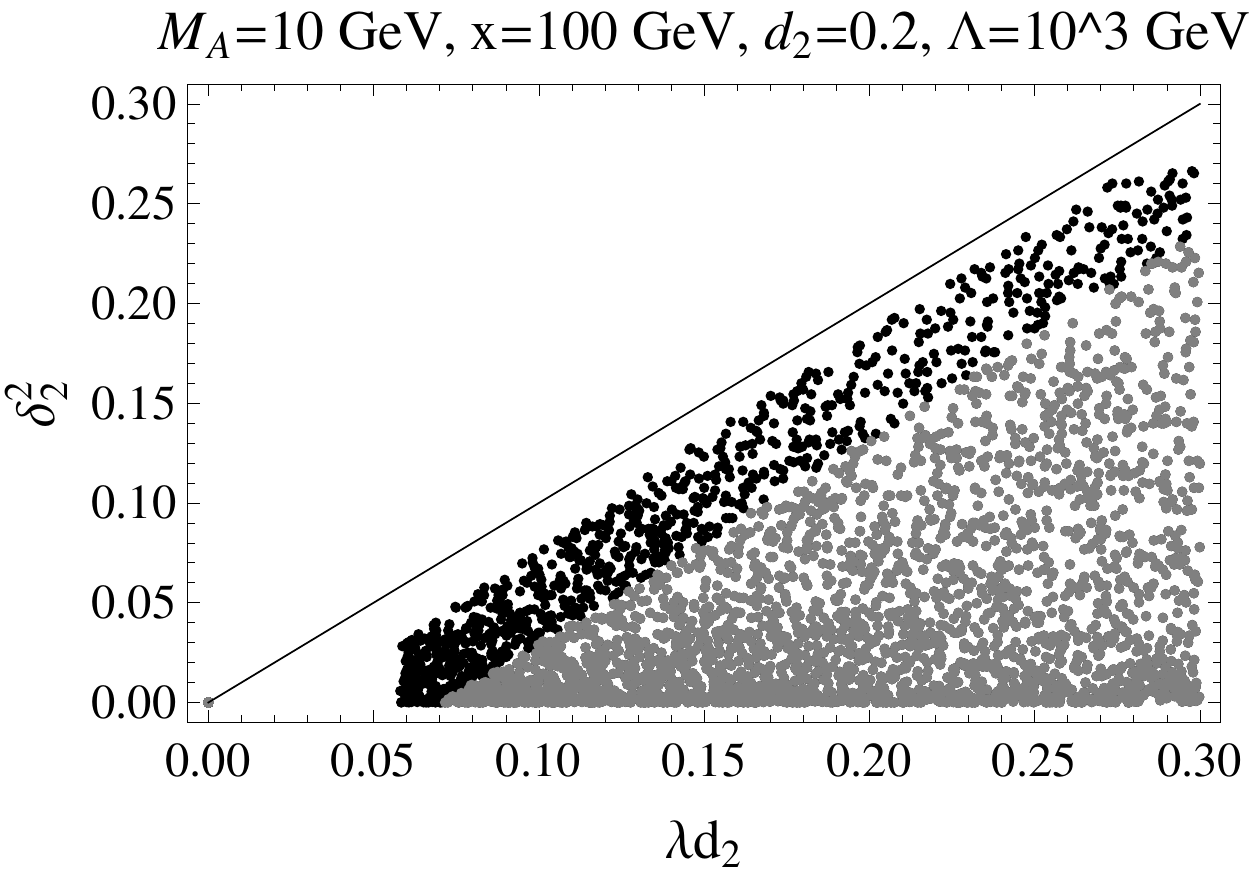}
	\includegraphics[width=.45\textwidth]{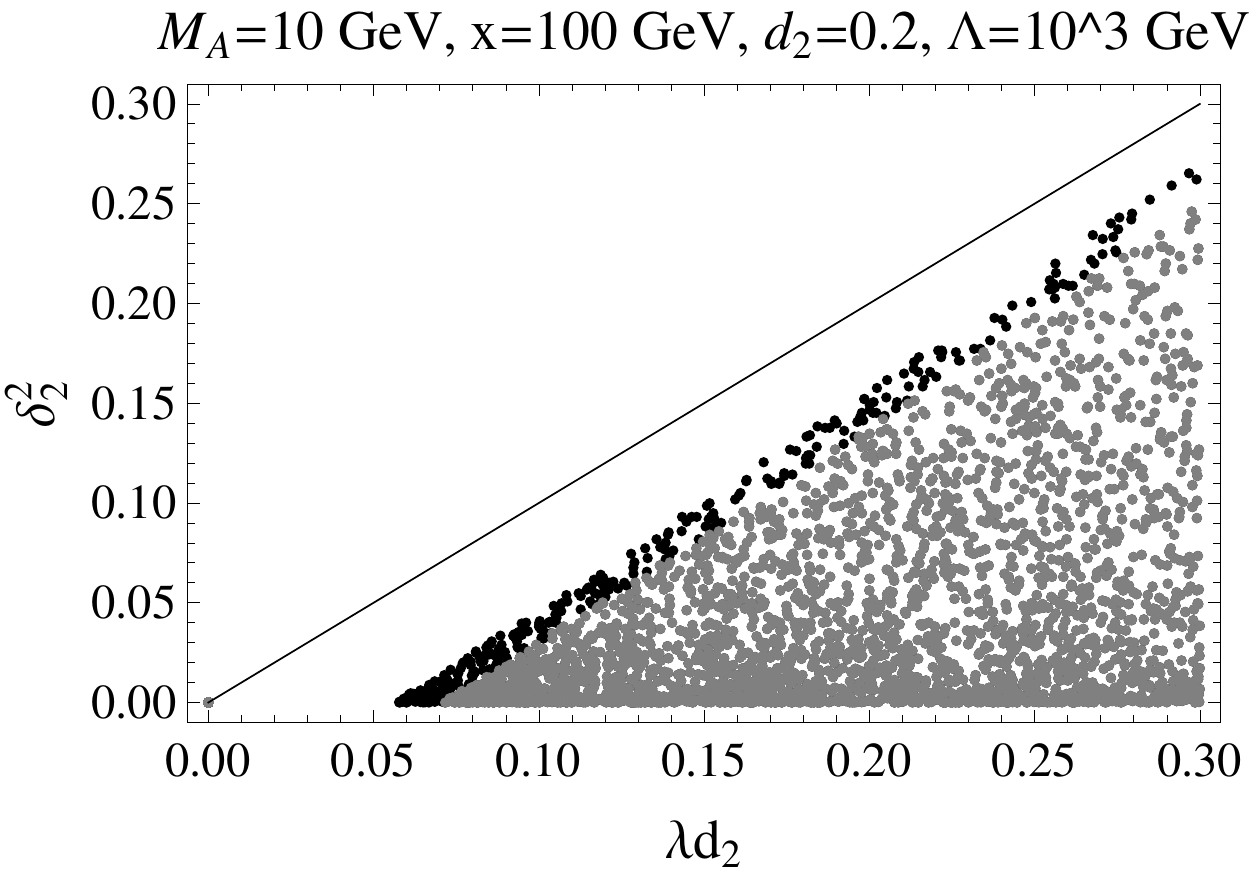}\\
	\includegraphics[width=.45\textwidth]{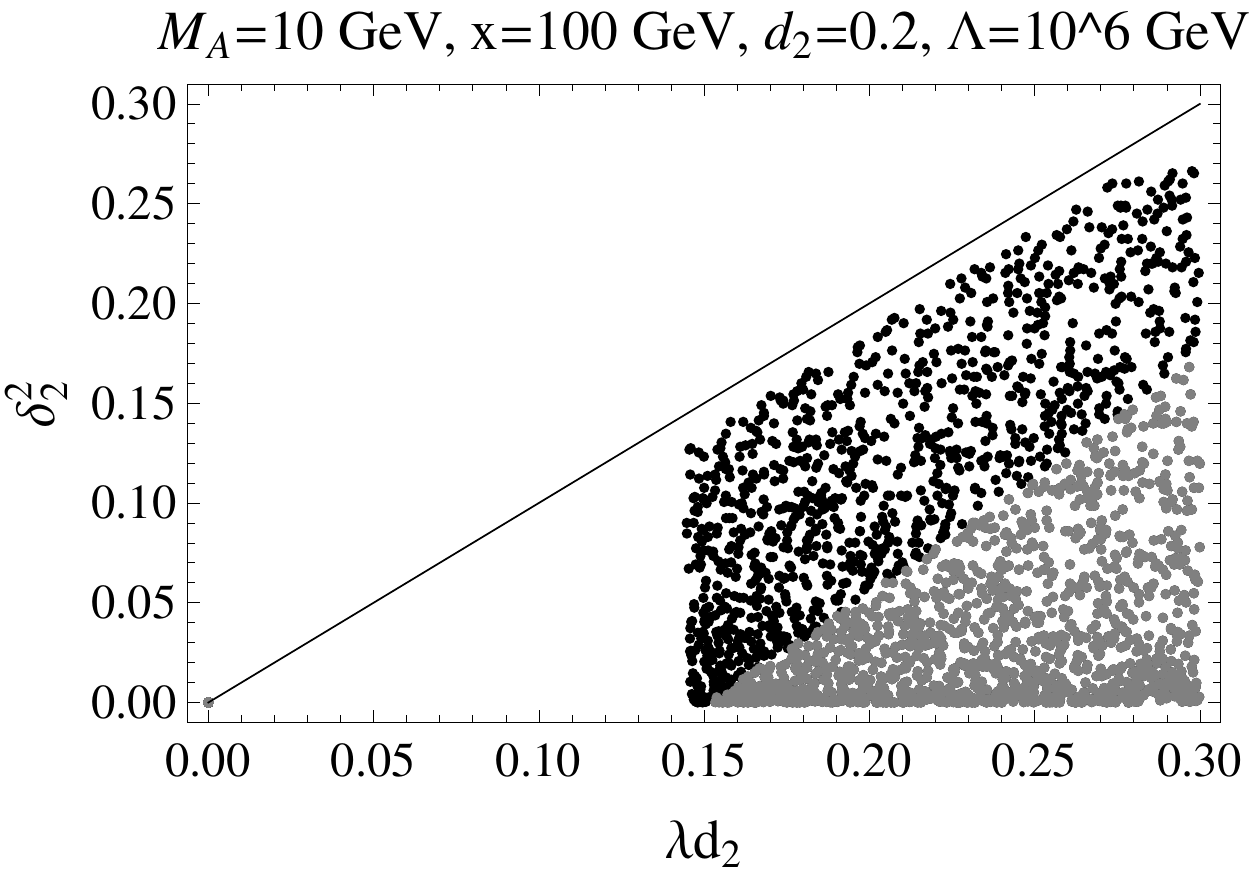}
	\includegraphics[width=.45\textwidth]{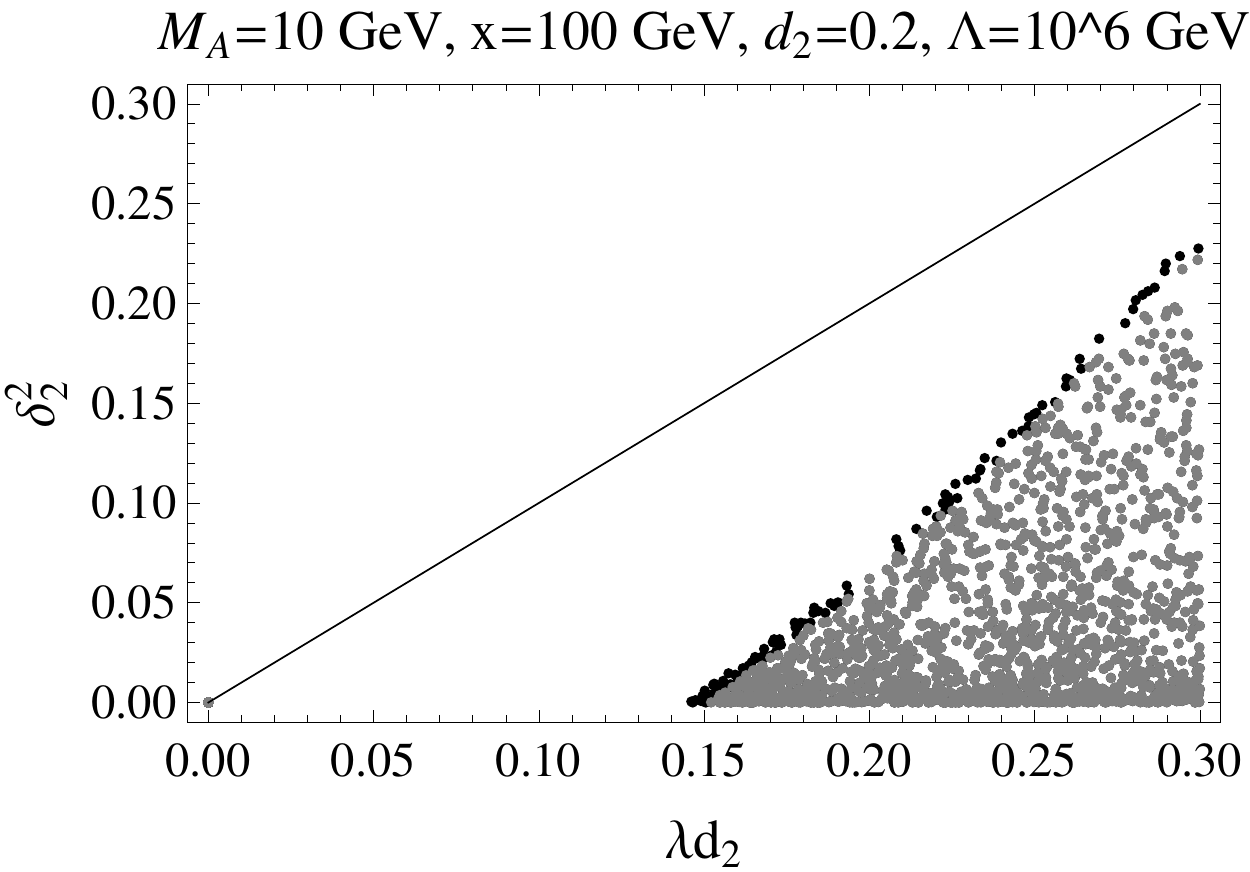}\\
	\includegraphics[width=.45\textwidth]{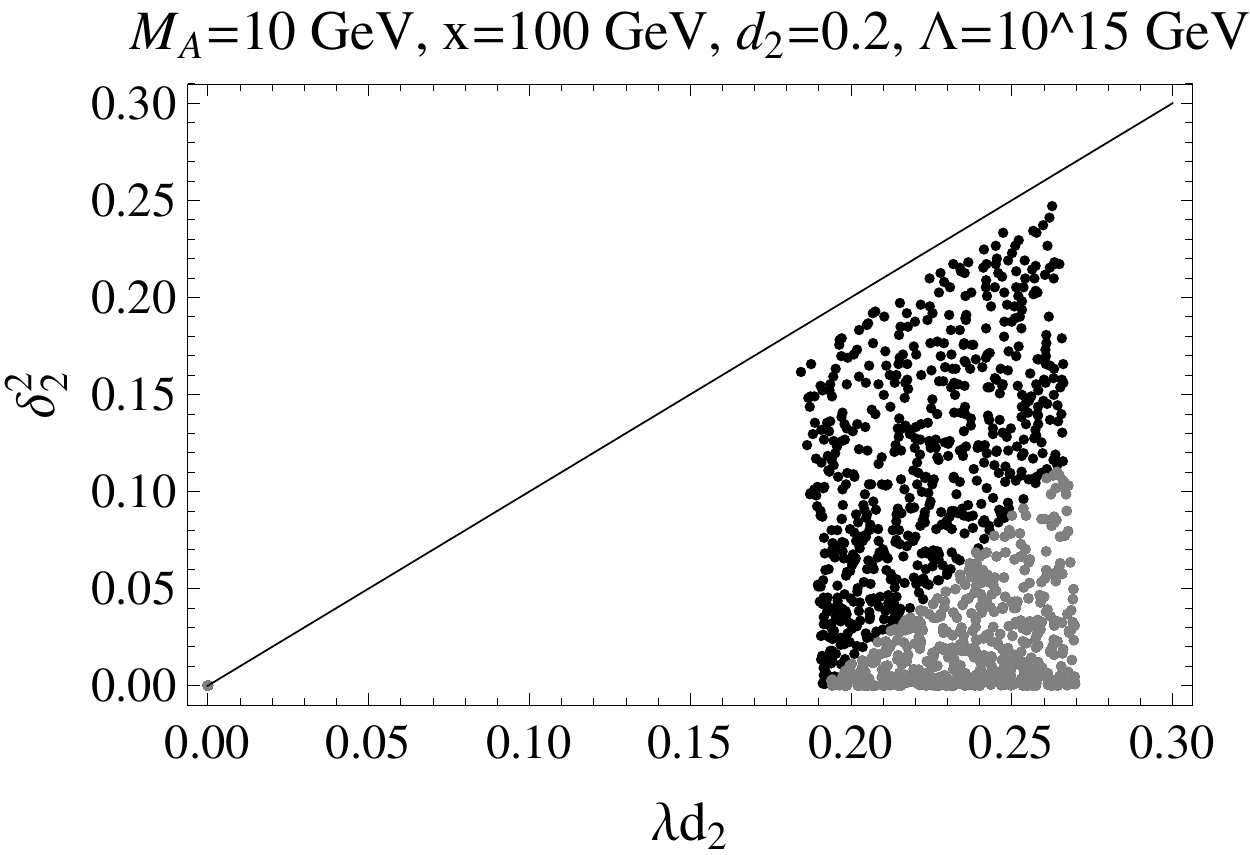}
	\includegraphics[width=.45\textwidth]{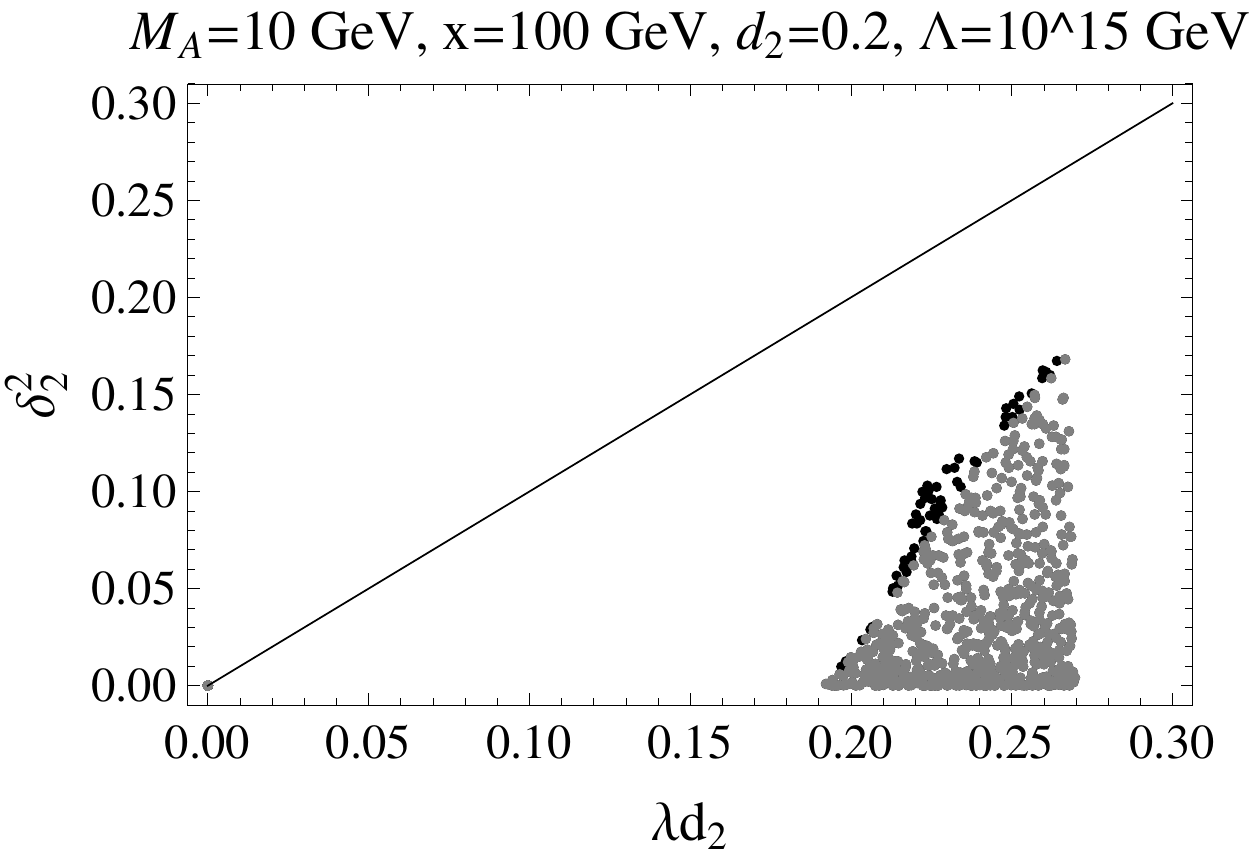}
\caption{Plots of $\delta_2^2\of{M_Z}$ vs. $\lambda\of{M_Z}d_2\of{M_Z}$.  For all plots, $M_{A}=10$~GeV, $x=100$~GeV, and $d_2\of{M_Z}=0.2$.  The tree level vacuum stability requirement, $\del^2<\lambda d_2$, is indicated with the solid line.  Gray points satisfy the constraints on the running couplings, \eqnref{eq:rg_running_stability}, while black points satisfy the effective potential vacuum stability requirement, \eqnref{eq:traditional_vs_req}, in the Landau gauge.  In the left column we take $\del>0$ while in the right column $\del<0$.  The cutoff scale $\Lambda$ is 1~TeV (top row), 1000~TeV (middle row), or $10^{15}$~GeV (bottom row).}
\label{fig:del2_vs_lambdad2}
\end{figure*}

We will use as our example the parameter set $M_{A}=10$~GeV, $x=100$~GeV, and $d_2\of{M_Z}=0.2$.  The restrictions on \themodel~parameters from the vacuum stability and RG analysis vary greatly with the choice of the cutoff scale $\Lambda$ (described in \secref{sec:vacuum_stability}).  This is demonstrated in the plots of $\del^2$ vs. $\lambda d_2$ in \figref{fig:del2_vs_lambdad2}.  In the left column, $\del>0$, and on the right $\del<0$.  The latter choice may accommodate a first order EWPT, as indicated by the work of \cite{Profumo:2007wc}.

In each plot, the solid line indicates the tree level vacuum stability requirement $\del^2<\lambda d_2$.  The plotted points correspond to values of the parameters that satisfy the RG running coupling constraints in \eqnref{eq:rg_running_stability} (gray points) or the traditional Landau gauge one-loop effective potential vacuum stability requirement in \eqnref{eq:traditional_vs_req} (black points).  Values of the cutoff scale $\Lambda$ are taken to be 1~TeV (top row), 1000~TeV (middle row), or $10^{15}~\mathrm{GeV}\simeq M_{\mathrm{GUT}}$ (bottom row).

\Figref{fig:del2_vs_lambdad2} evinces all of the generic features of the vacuum stability analysis discussed in \secref{sec:vacuum_stability}.  Even with the most generous cutoff scale, $\Lambda=1$~TeV, the allowed values of $\del$ for a given $\lambda$ and $d_2$ do not extend up to the tree level bound because the RG evolution of the couplings breaks the condition of \eqnref{eq:rg_running_stability} that $\del\of{\mu}^2<\lambda\of{\mu}d_2\of{\mu}$ at some $\mu<\Lambda$.  When $\del<0$ (right column), this leads to a runaway direction in the Landau gauge effective potential: hence, the effective potential limits (black points) closely match the RG coupling limits (gray points).  However, if $\del>0$ (left column), the potential still appears stable (in the Landau gauge) even if the condition on the RG evolution of $\del$ is not satisfied and so the effective potential bound closely matches the tree level requirement.

Furthermore, as the cutoff scale increases, small values of $\lambda$ are forbidden because of the appearance of deep minima along the $h$-axis of the potential (or, alternatively, $\lambda\of{\mu}<0$), as in the SM.  Large values of $\lambda$ are also forbidden because RG evolution results in non-perturbative values for the quartic couplings in violation of eq. \ref{eq:coupling_pert_req}.

\begin{figure*}
\includegraphics[width=0.45\textwidth]{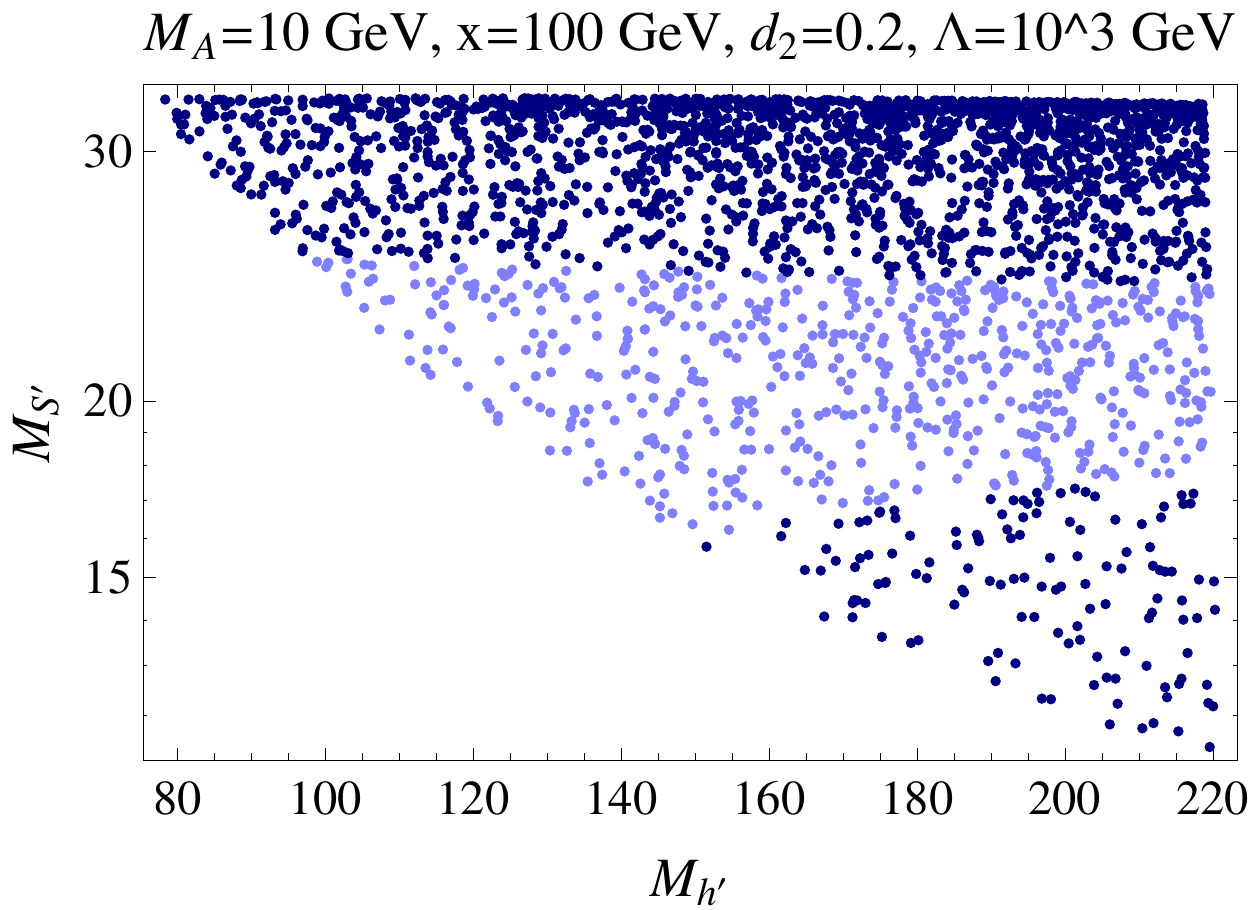}
\includegraphics[width=0.45\textwidth]{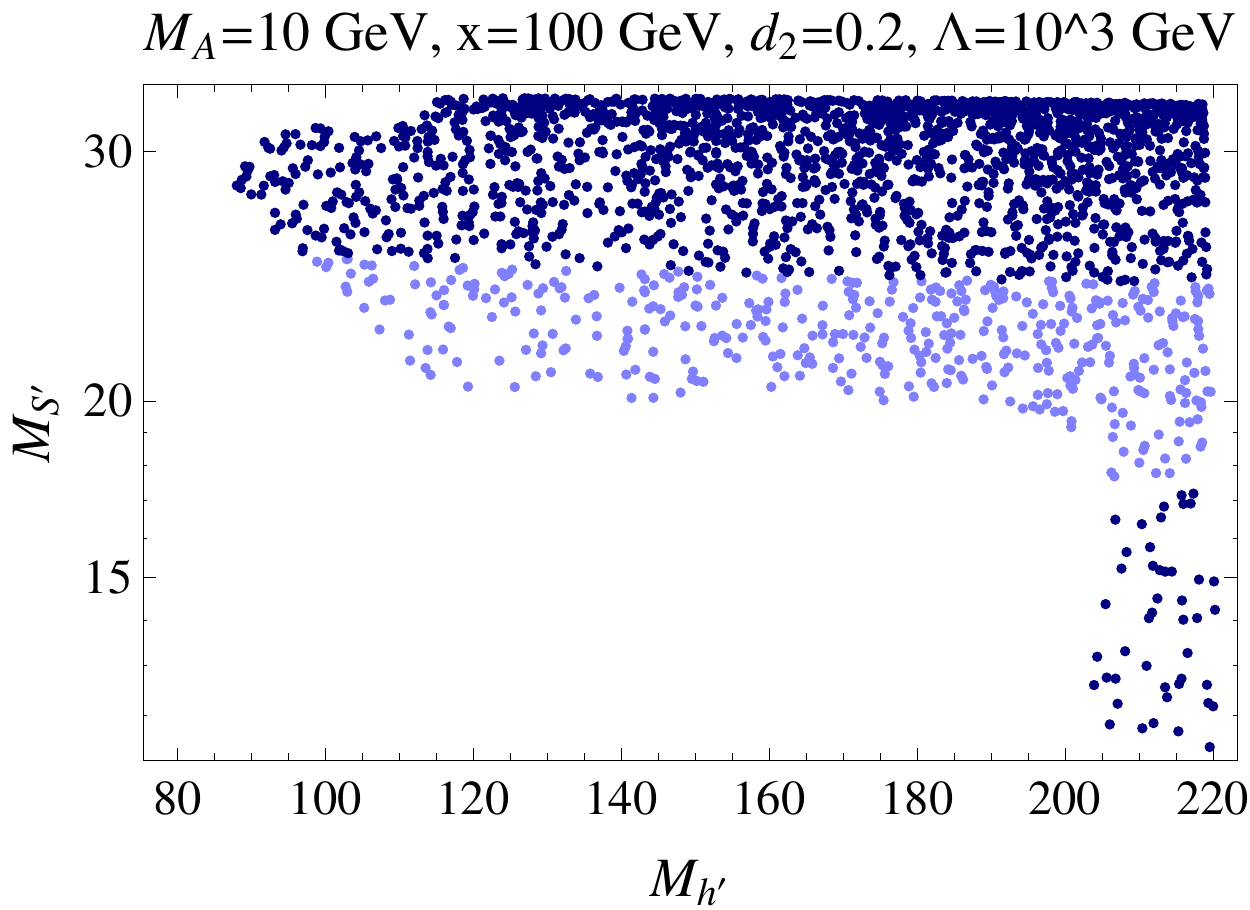}\\
\includegraphics[width=0.45\textwidth]{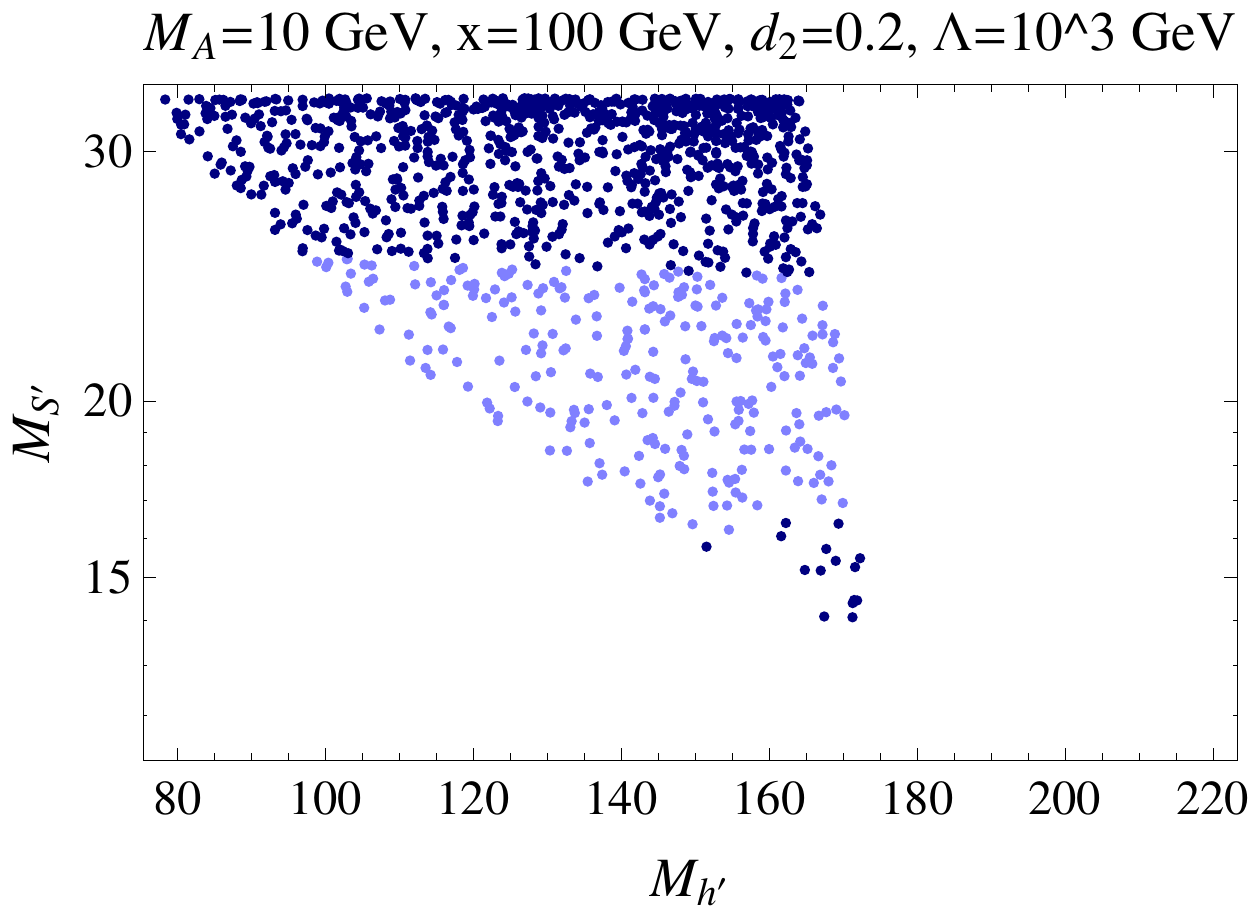}
\includegraphics[width=0.45\textwidth]{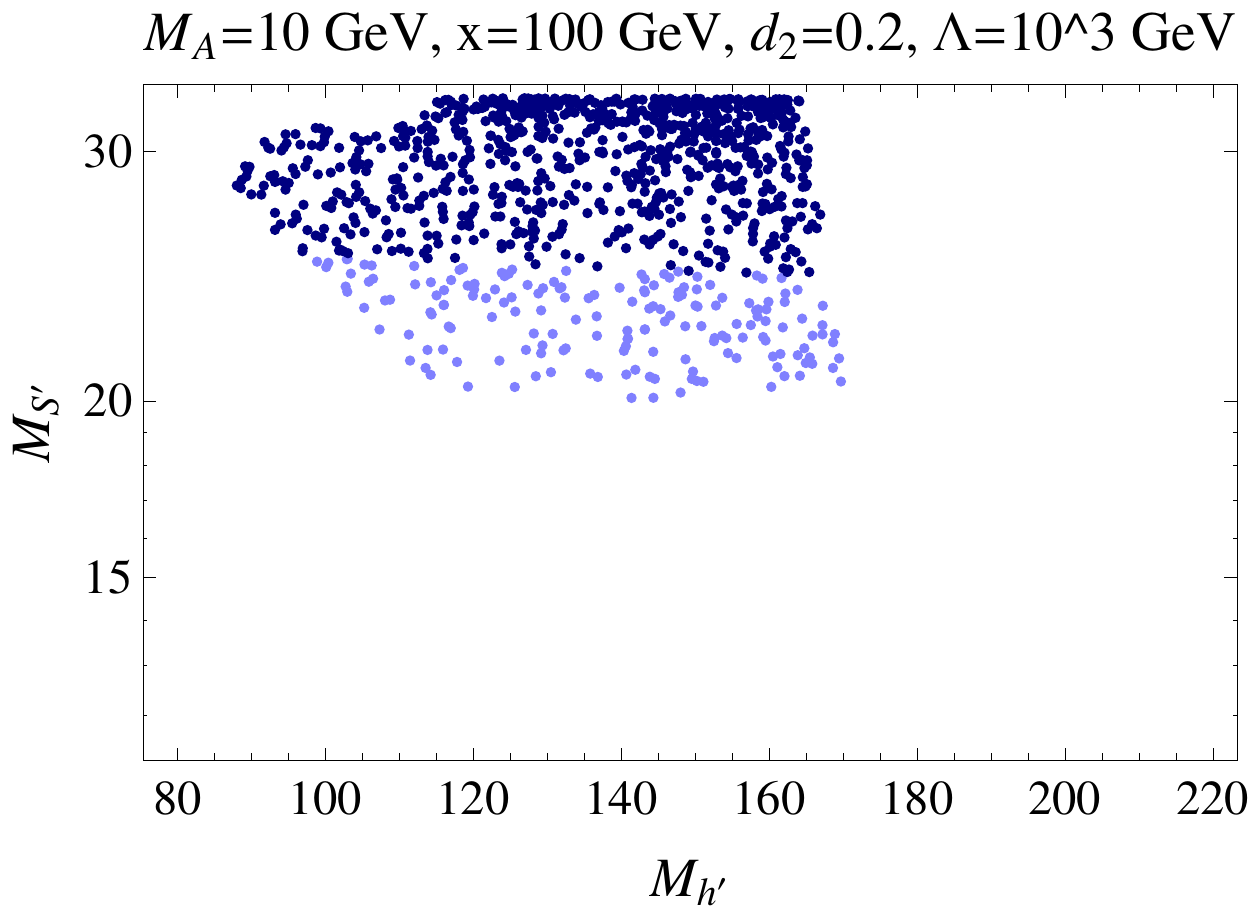}
\caption{Results of the scan for $M_{A}=10$~GeV, $x=100$~GeV, $d_2\of{M_Z}=0.2$, and $\Lambda=1$~TeV with $\del<0$ 
shown in the $M_{S'}$ vs. $M_{h'}$ plane.  Dark colored points oversaturate the relic density, while light colored points (under)saturate.  The top left plot imposes only the RG coupling limits, \eqnref{eq:rg_running_stability}.  RG coupling limits plus either the LEP constraints (top right) or the EWPO constraints (bottom left), and finally all three (bottom right), are also shown.}
\label{fig:ms_vs_mh_lepewpo}
\end{figure*}
 
In the discussion that follows, we consider the more conservative but gauge-independent bounds on the parameter space from the RG evolution of the quartic couplings in \eqnref{eq:rg_running_stability}, rather than the gauge-dependent limits from the effective potential stability requirement in \eqnref{eq:traditional_vs_req}.  The plots for $\del>0$ have similar features so we do not show them here; instead we focus on the $\del<0$ scenario because of the impact on the electroweak phase transition, discussed in \secref{sec:ewpt}.

The allowed masses of the $h',S'$ eigenstates are shown in \figref{fig:ms_vs_mh_lepewpo} for the same set of parameters in \figref{fig:del2_vs_lambdad2}: $M_{A}=10$~GeV, $x=100$~GeV, $d_2\of{M_Z}=0.2$, and also $\Lambda=1$~TeV and $\del<0$.  \Figref{fig:ms_vs_mh_lepewpo} shows the constraints on the masses from LEP searches, EWPO measurements, and the RG evolution of the quartic couplings in \eqnref{eq:rg_running_stability}.  Darker colored points result in a singlet relic density that is above the $1\sigma$ WMAP bound, {\em i.e.}, $\Omega_A h^2>0.118$.  Lighter colored points correspond to saturation or undersaturation of the relic density, $\Omega_A h^2 \leq 0.118$.  We note that the relic density is (under)saturated when there are $s$-channel resonances in the annihilation cross section (for $2M_A = M_{S'}$) or the 4-point interaction dominates (for $M_A > M_{S'}$).

One important feature of note is that increasing the Higgs-singlet coupling $\del$ decreases the mass of the lighter eigenstate --- the singlet-like $M_{S'}$ here --- according to \eqnref{eq:tree_mass_evals} and increases the mixing angle (see \eqnref{eq:tree_mixing_angle}) when the other parameters ($\lambda, d_2, x$) are fixed.  This effect is responsible for three features observed in \figref{fig:ms_vs_mh_lepewpo}:
\begin{enumerate}
\item The EWPO constraints (imposed in \figref{fig:ms_vs_mh_lepewpo}, bottom left) favor a light scalar with SM-like couplings.  As indicated by the slight slope on the right edge of the allowed region, larger $h'$ masses are allowed as $M_{S'}$ decreases due to the increased mixing that allows $S'$ to offset the heavier $h'$.
\item It is possible for the heavier Higgs-like eigenstate to avoid the 114~GeV bound from LEP (see \figref{fig:ms_vs_mh_lepewpo}, top right) as $M_{S'}$ decreases from a maximum of $M_{S'}> 30$~GeV due to an increased mixing angle and reduced $h'$ coupling strength to SM particles.  
\item Also regarding the LEP constraints, a significant number of points corresponding to $M_{S'}<20~\mathrm{GeV}=2M_A$ are eliminated because the decay $S'\rightarrow AA$ is no longer allowed, resulting in a light scalar $S'$ with SM-like branching fractions in violation of the LEP constraints.  Increasing the $h'$ mass above 200~GeV decreases the mixing angle, so the lighter $M_{S'}$ masses are once again allowed by the LEP constraint despite the SM-like branching fractions of the $S'$.
\end{enumerate}

\begin{figure*}
\includegraphics[width=0.45\textwidth]{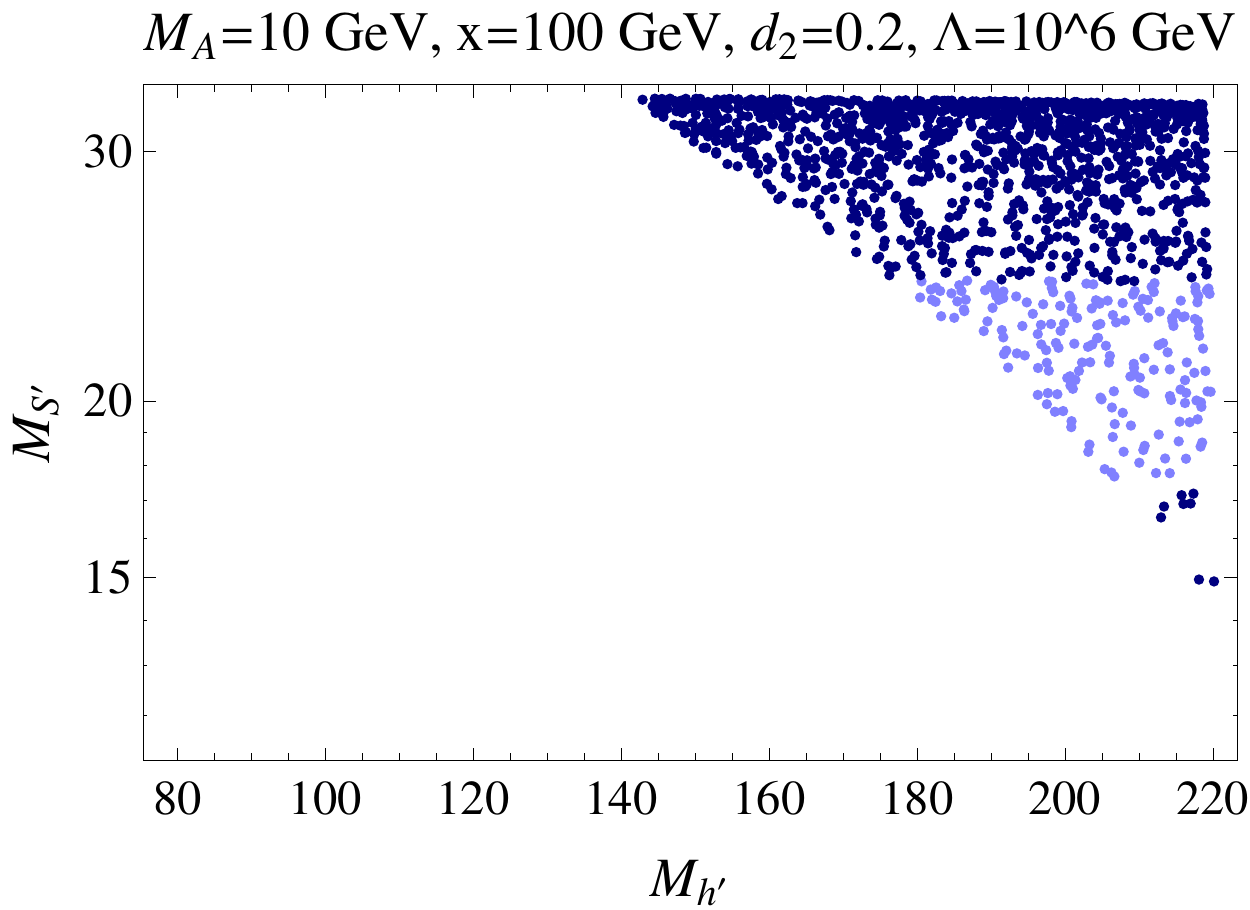}
\includegraphics[width=0.45\textwidth]{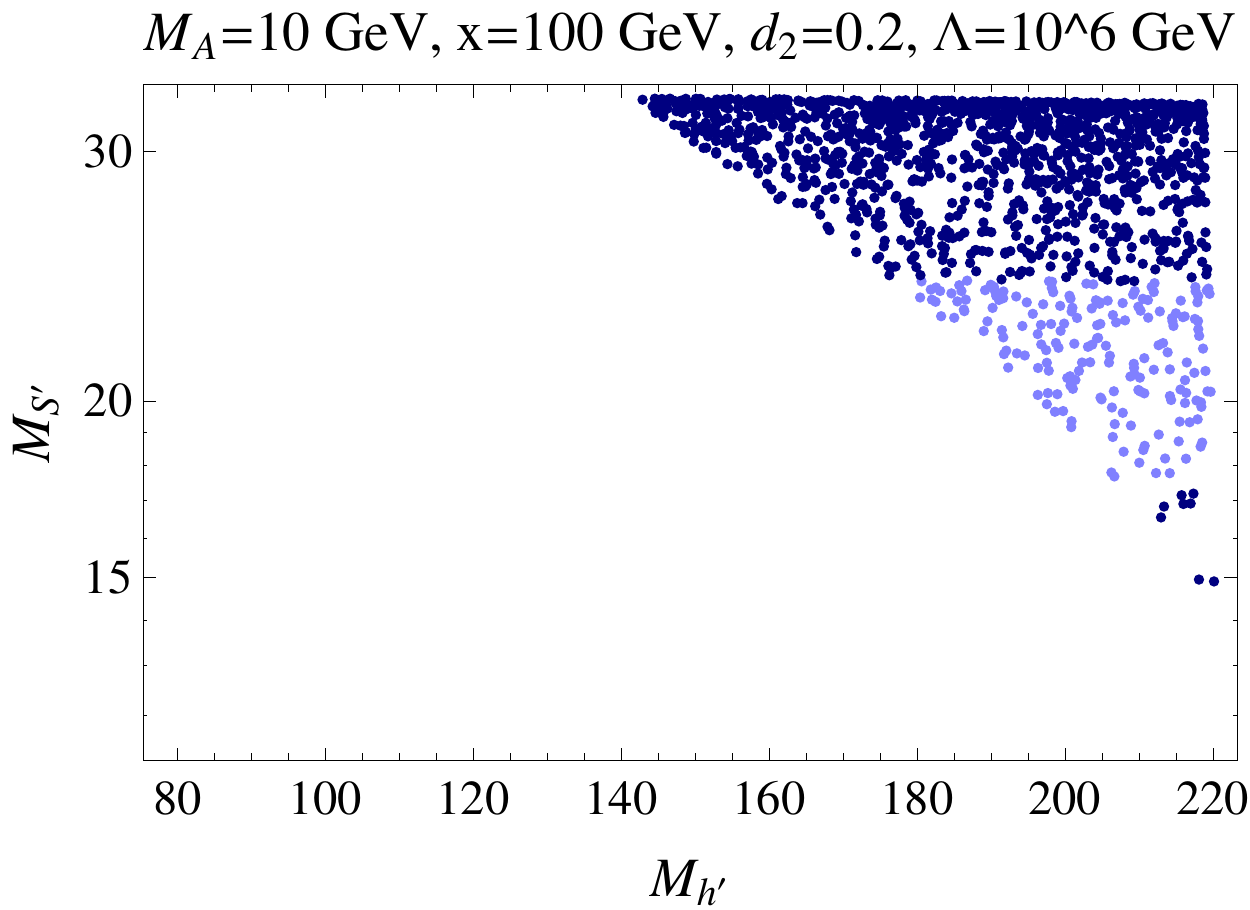}\\
\includegraphics[width=0.45\textwidth]{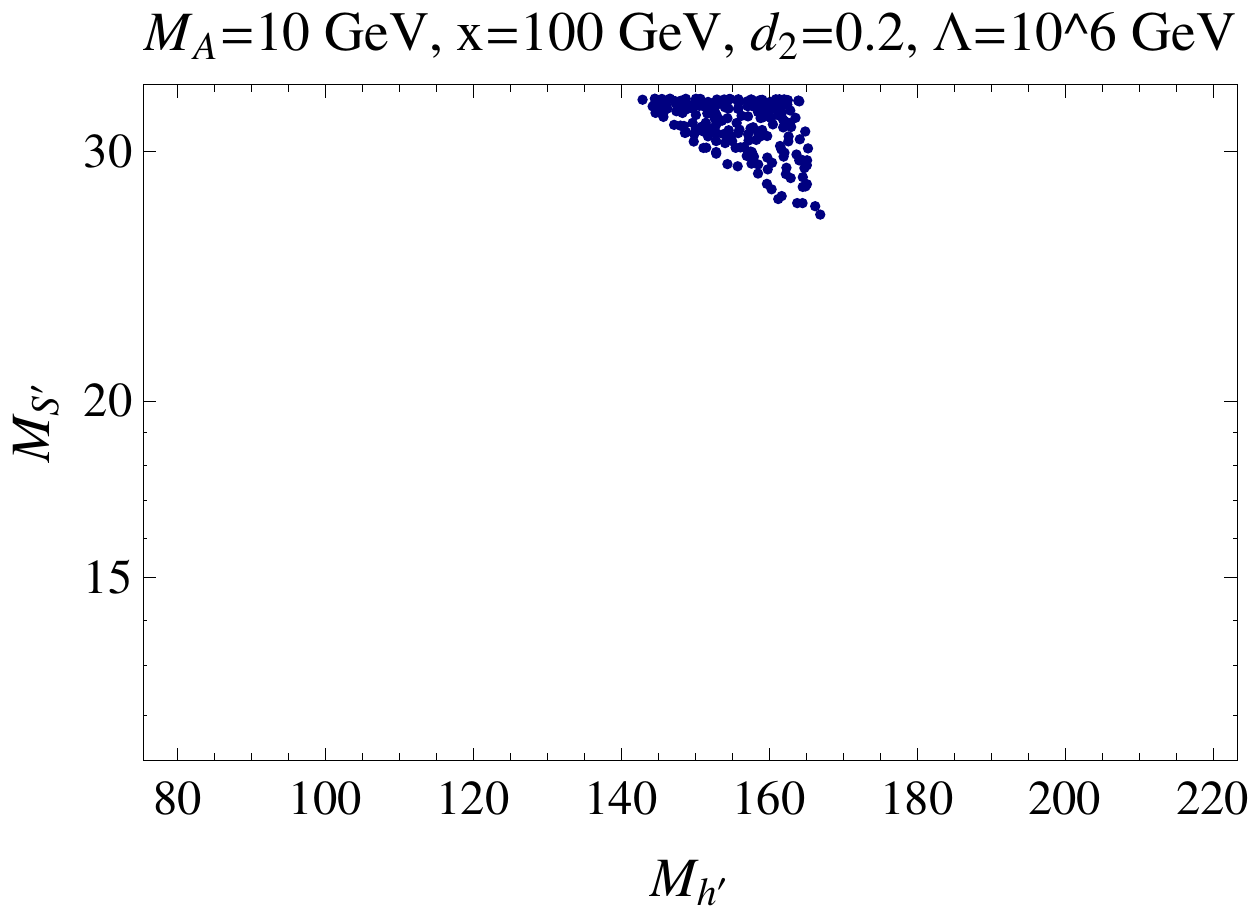}
\includegraphics[width=0.45\textwidth]{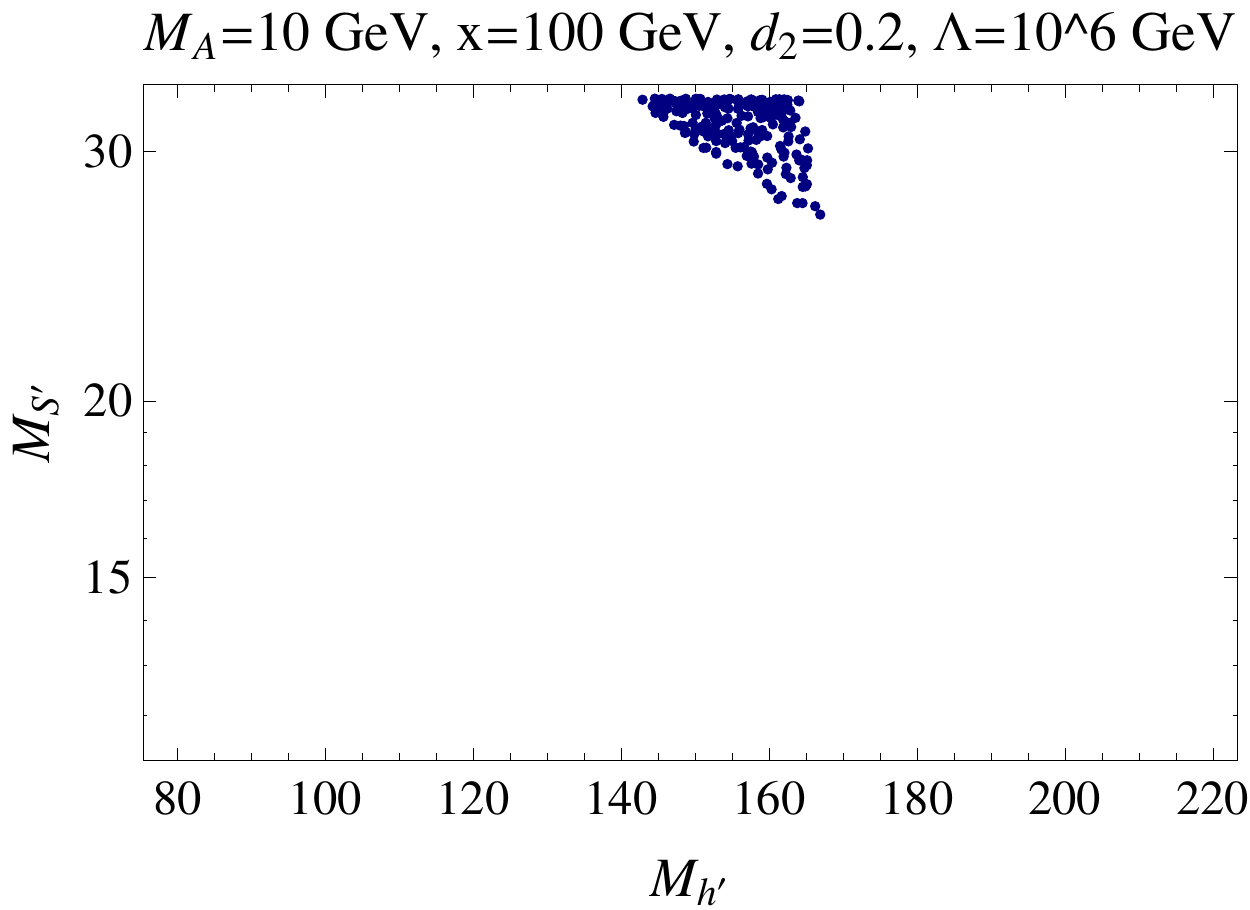}
\caption{Same as \figref{fig:ms_vs_mh_lepewpo} but with $\Lambda=10^6$~GeV.}
\label{fig:ms_vs_mh_lepewpo_6}
\end{figure*}

\begin{figure*}
\includegraphics[width=0.45\textwidth]{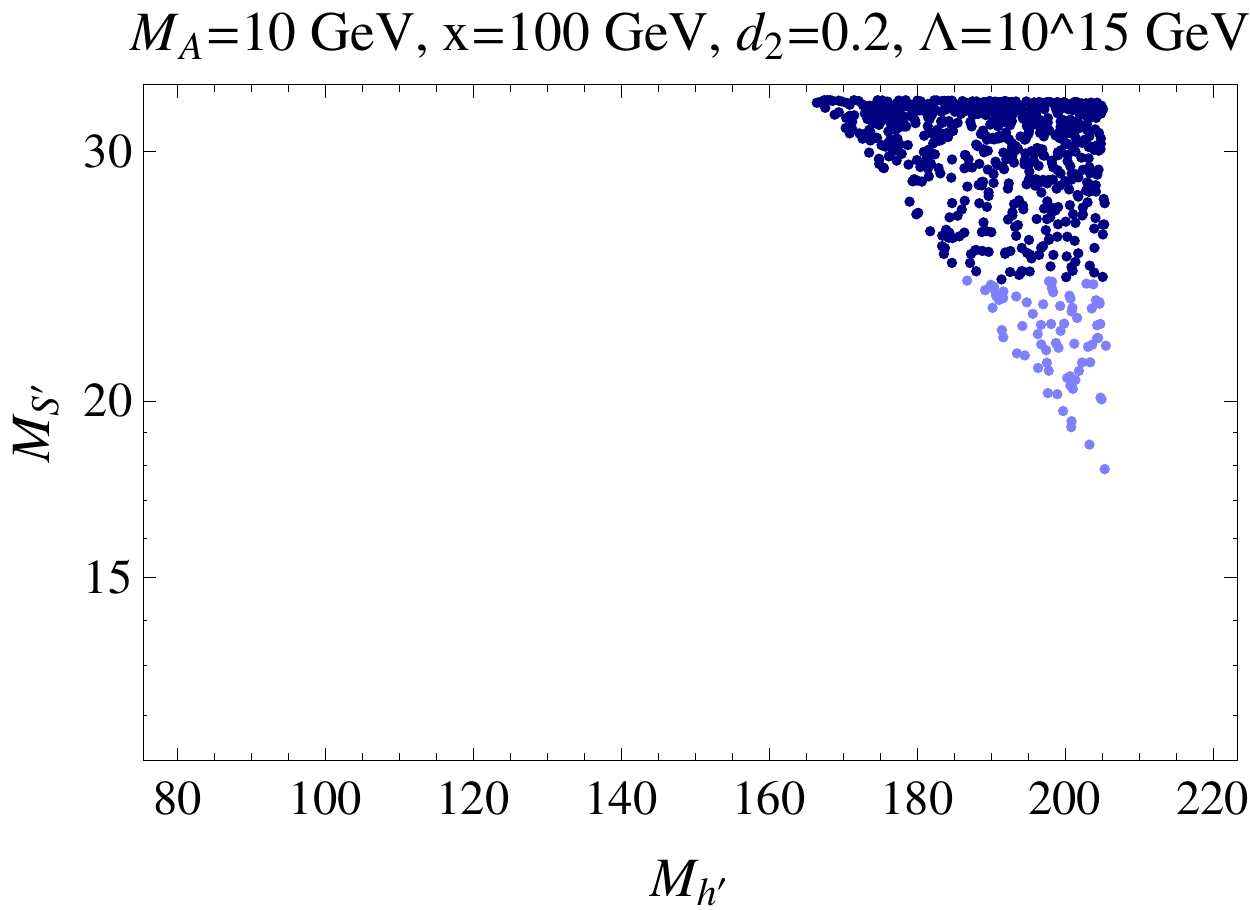}
\includegraphics[width=0.45\textwidth]{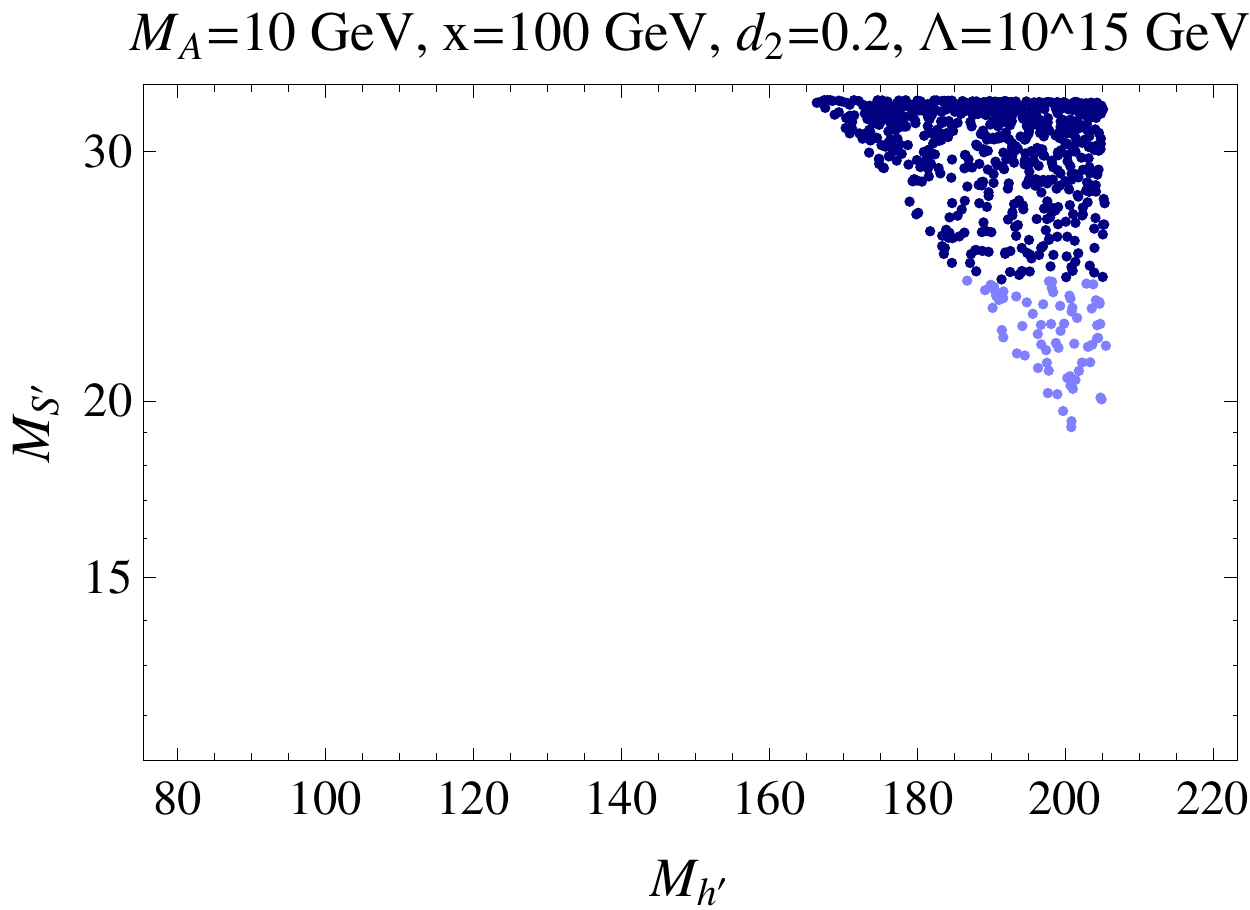}\\
\includegraphics[width=0.45\textwidth]{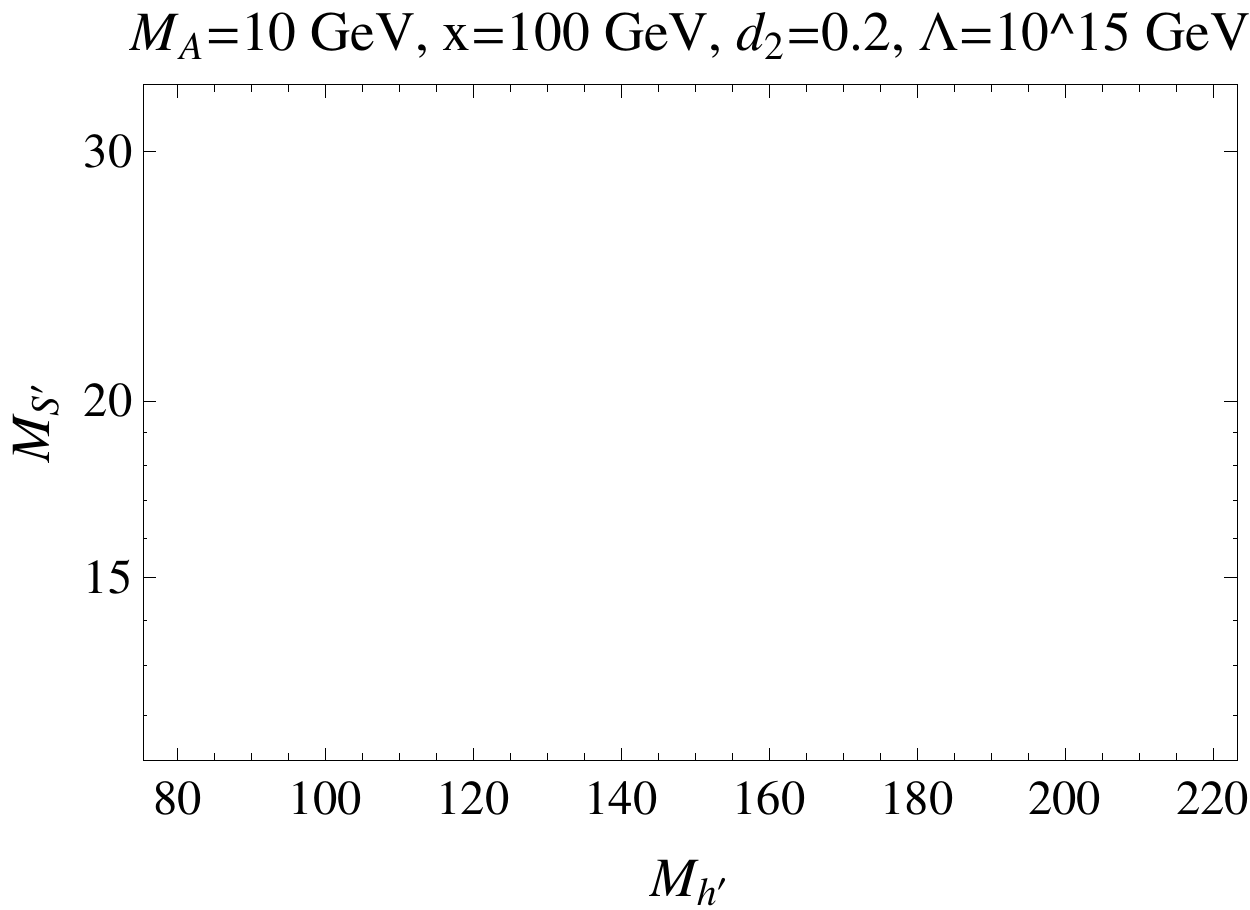}
\includegraphics[width=0.45\textwidth]{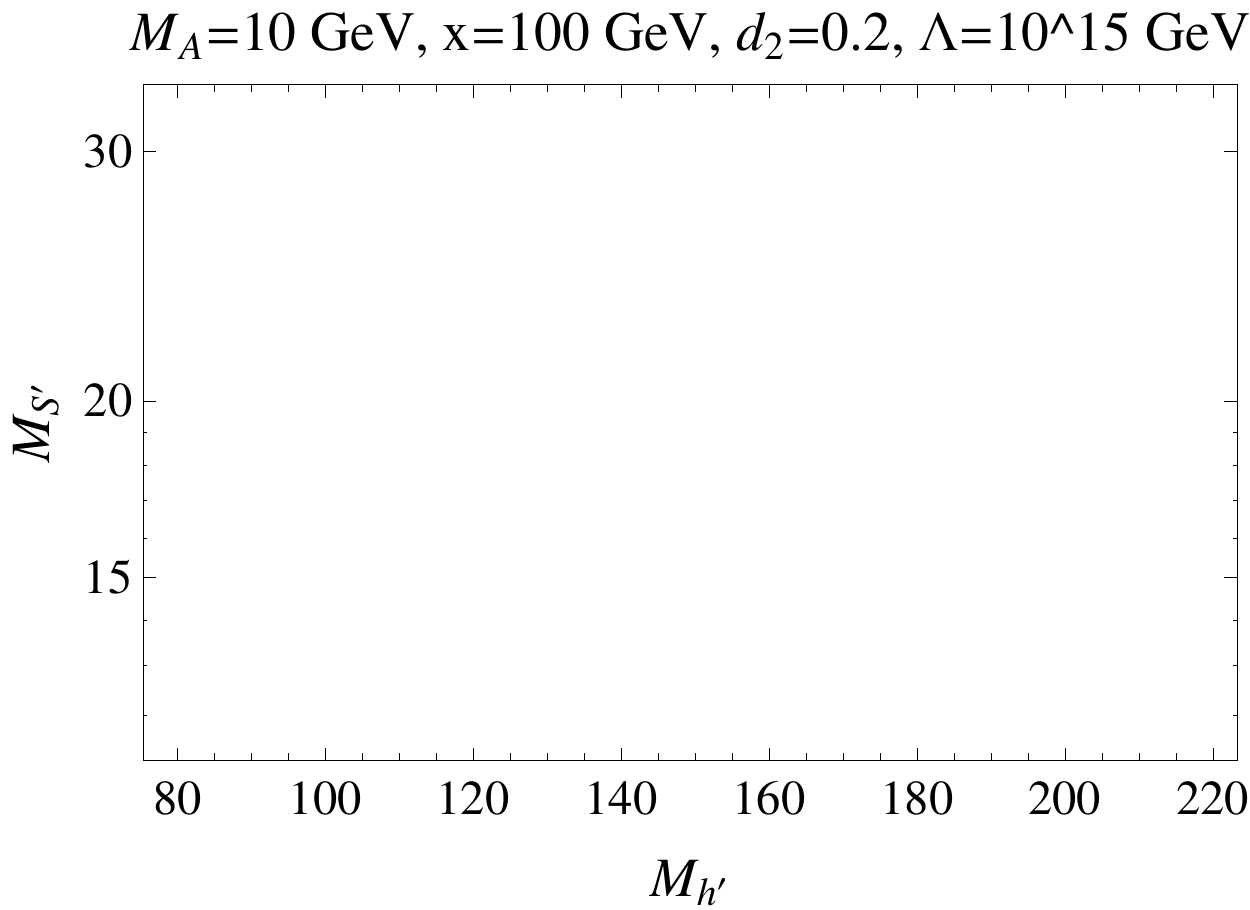}
\caption{Same as \figref{fig:ms_vs_mh_lepewpo} but with $\Lambda=10^{15}$~GeV.}
\label{fig:ms_vs_mh_lepewpo_15}
\end{figure*}

\Figref{fig:ms_vs_mh_lepewpo_6} and \figref{fig:ms_vs_mh_lepewpo_15} are the same as \figref{fig:ms_vs_mh_lepewpo} but with $\Lambda=10^6$~GeV and $10^{15}$~GeV, respectively.  Increasing the cutoff scale forces $\del$ smaller --- as was shown in \figref{fig:del2_vs_lambdad2} --- and results in relatively larger masses for both the scalar eigenstates.  As these plots show, for a 1~TeV cutoff scale the effects of the LEP and EWPO limits are roughly equivalent.  However, as $\Lambda$ increases, the vacuum stability and perturbativity requirements reduce the allowed regions of parameters and masses; of the points that remain at these higher cutoff scales, a smaller number satisfy the EWPO constraint than the LEP bounds.  Thus the EWPO constraint is more significant than the LEP bounds at higher cutoff scales.  Indeed, \figref{fig:ms_vs_mh_lepewpo_15} shows that there are scalar masses that satisfy the RG evolution requirement in \eqnref{eq:rg_running_stability} up to the GUT scale, and the LEP bounds, but not the EWPO constraints.

\begin{figure*}
\includegraphics[width=0.45\textwidth]{ms_vs_mh_scan10c3.pdf}
\includegraphics[width=0.45\textwidth]{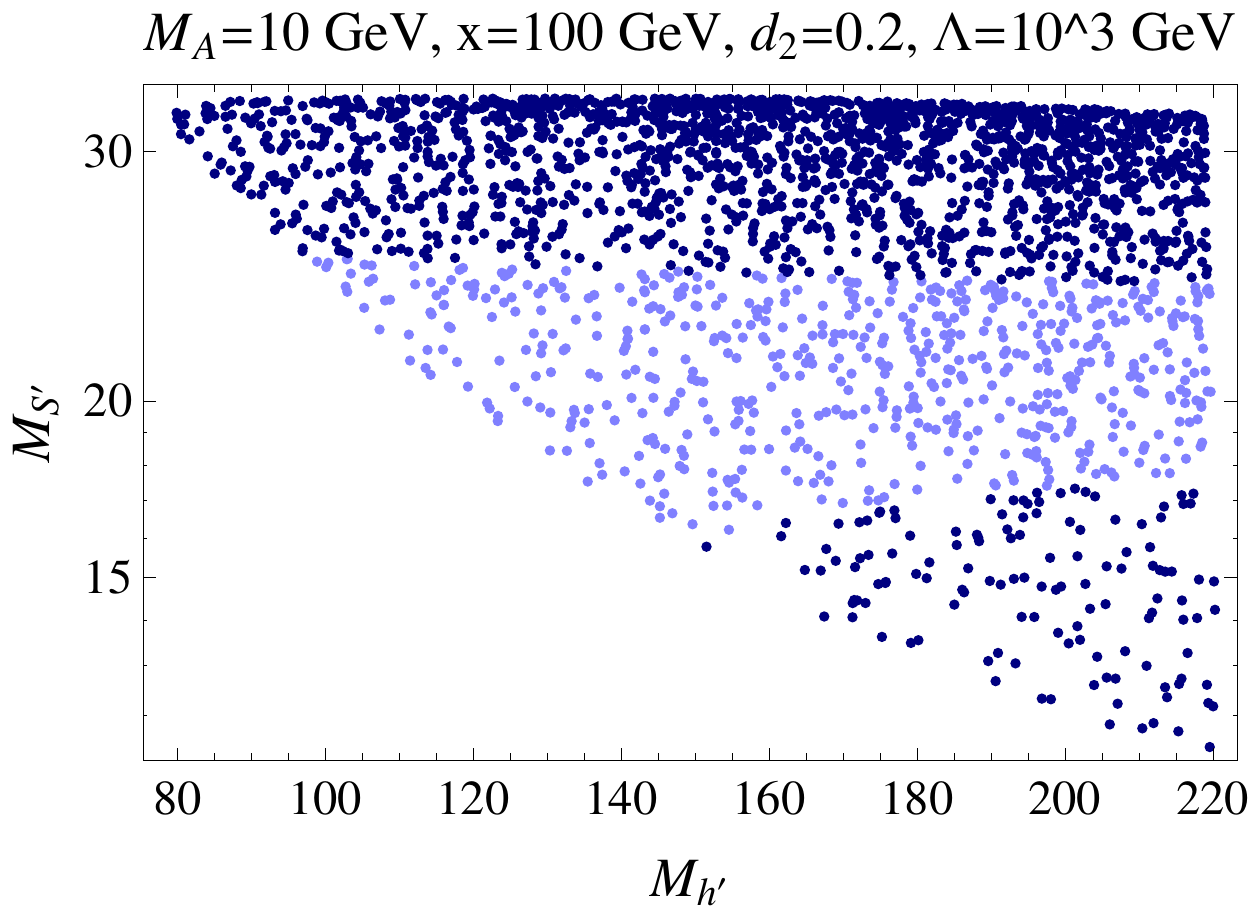}\\
\includegraphics[width=0.45\textwidth]{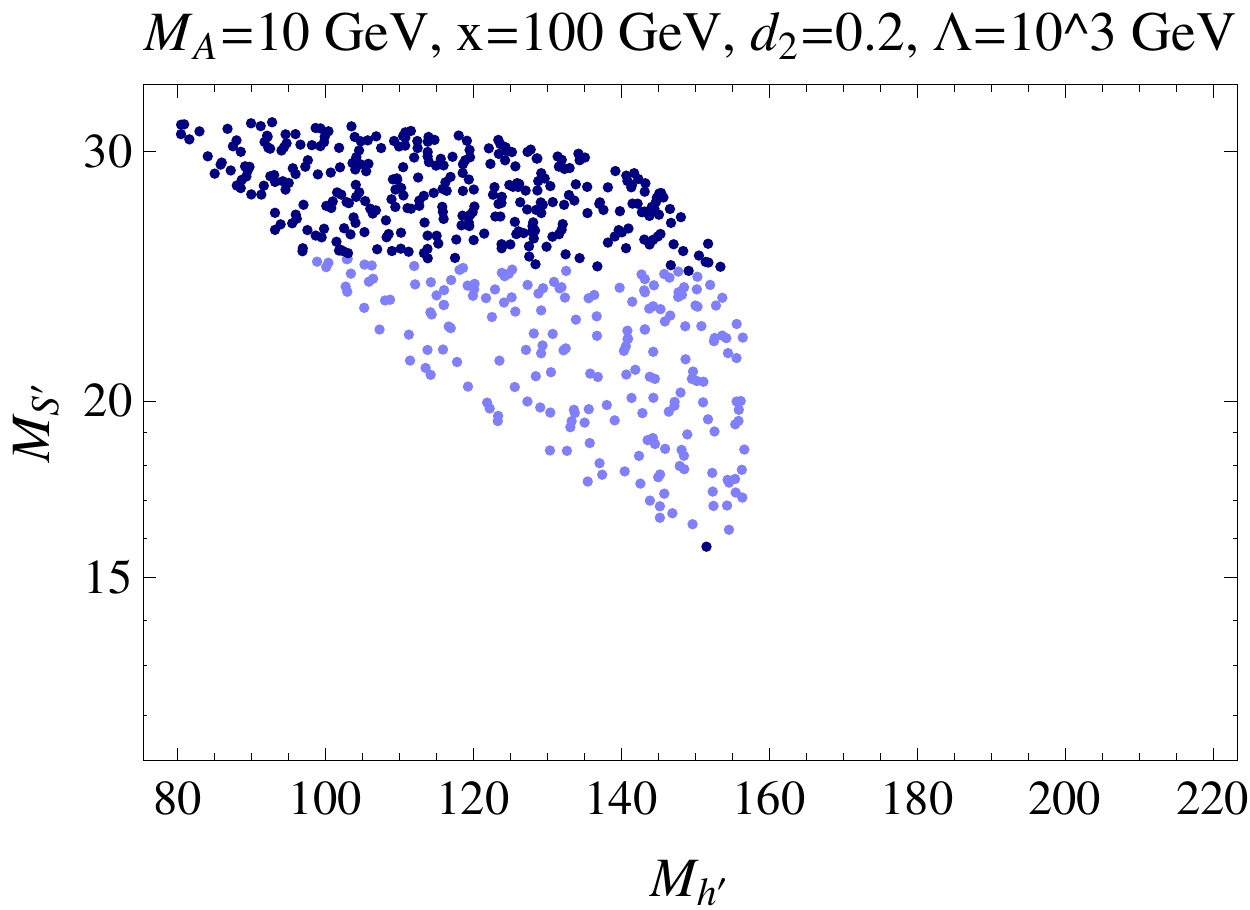}
\includegraphics[width=0.45\textwidth]{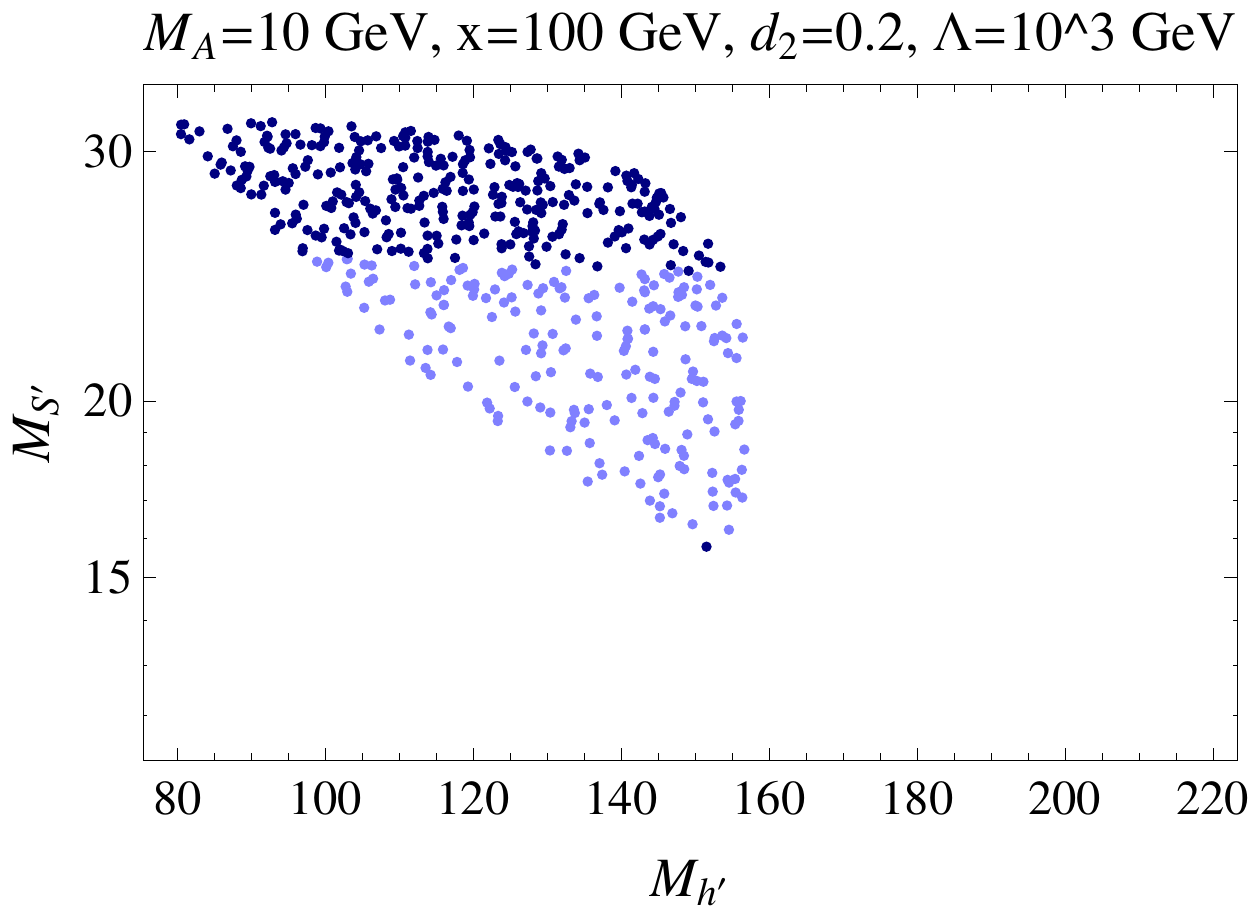}
\caption{Results of the scan for $M_{A}=10$~GeV, $x=100$~GeV, $d_2\of{M_Z}=0.2$, and $\Lambda=1$~TeV with $\del<0$ 
shown in the $M_{S'}$ vs. $M_{h'}$ plane.  Dark colored points oversaturate the relic density, while light colored points (under)saturate.  The top left plot imposes only the RG coupling limits, \eqnref{eq:rg_running_stability}.  RG coupling limits plus either the XENON100 bound (top right) or satisfaction of the requirement for ATLAS invisible Higgs decay searches (bottom left), and finally all three (bottom right), are also shown.}
\label{fig:ms_vs_mh_ddinvis}
\end{figure*}

\begin{figure*}
\includegraphics[width=0.45\textwidth]{ms_vs_mh_scan10c6.pdf}
\includegraphics[width=0.45\textwidth]{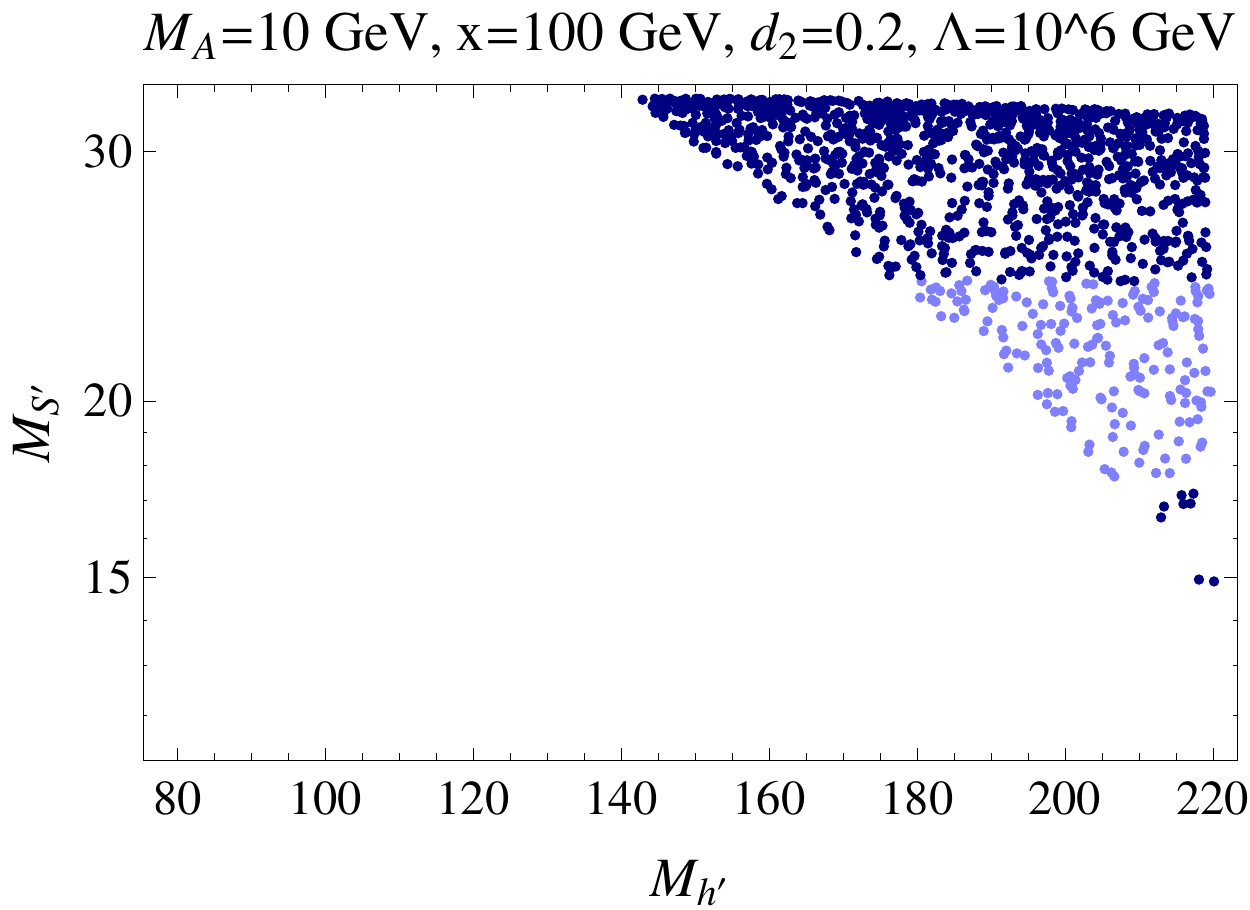}\\
\includegraphics[width=0.45\textwidth]{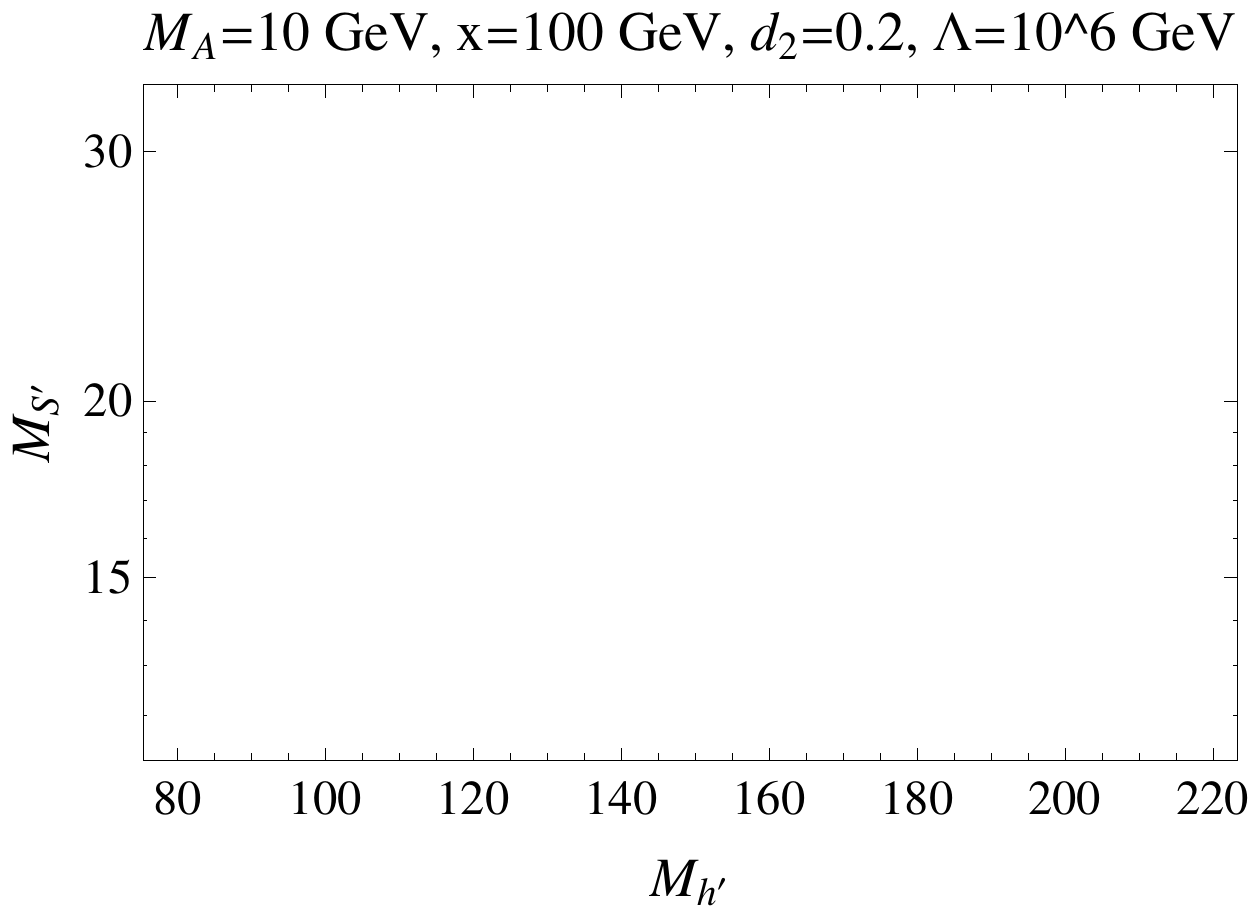}
\includegraphics[width=0.45\textwidth]{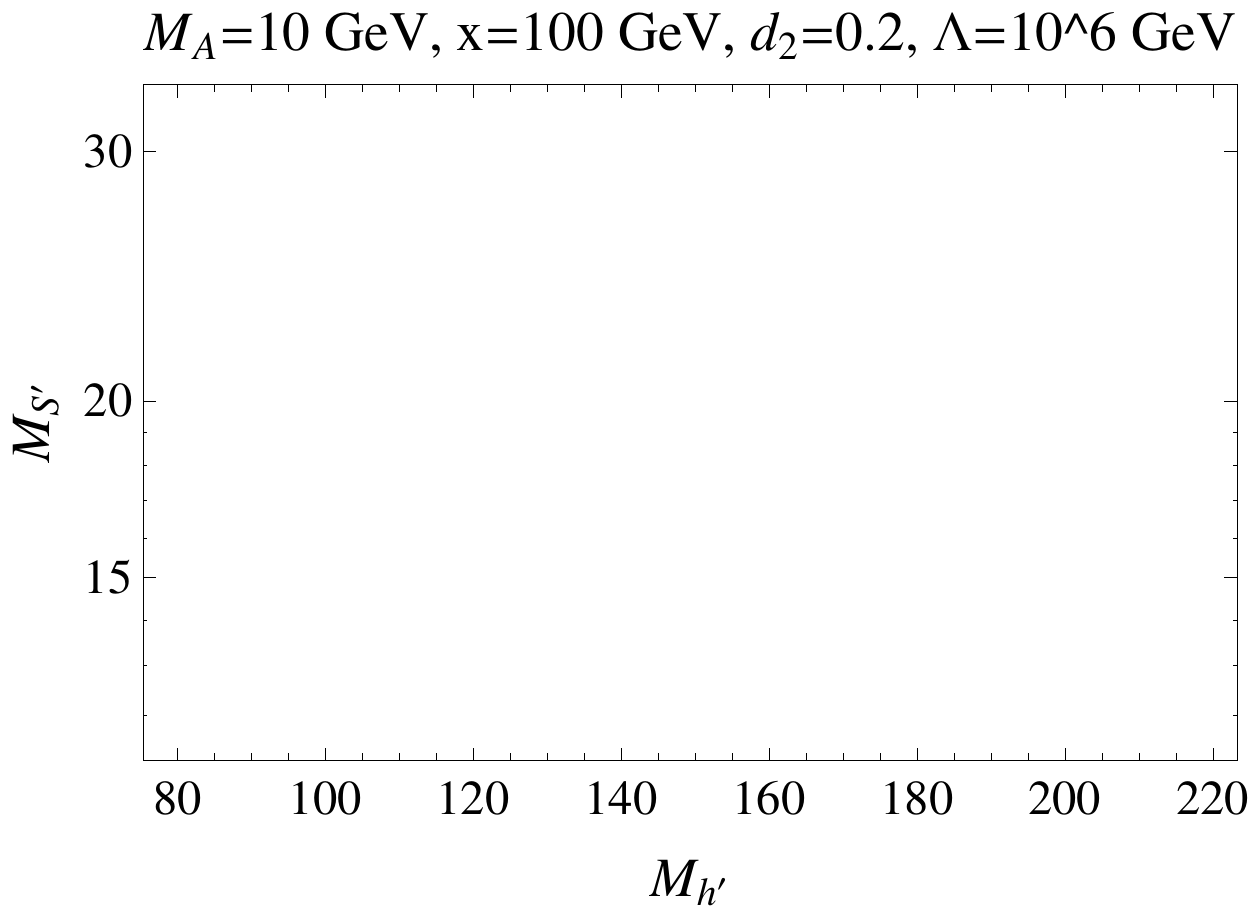}
\caption{Same as \figref{fig:ms_vs_mh_ddinvis} but with $\Lambda=10^6$~GeV.}
\label{fig:ms_vs_mh_6_ddinvis}
\end{figure*}

\begin{figure*}
\includegraphics[width=0.45\textwidth]{ms_vs_mh_scan10c15.pdf}
\includegraphics[width=0.45\textwidth]{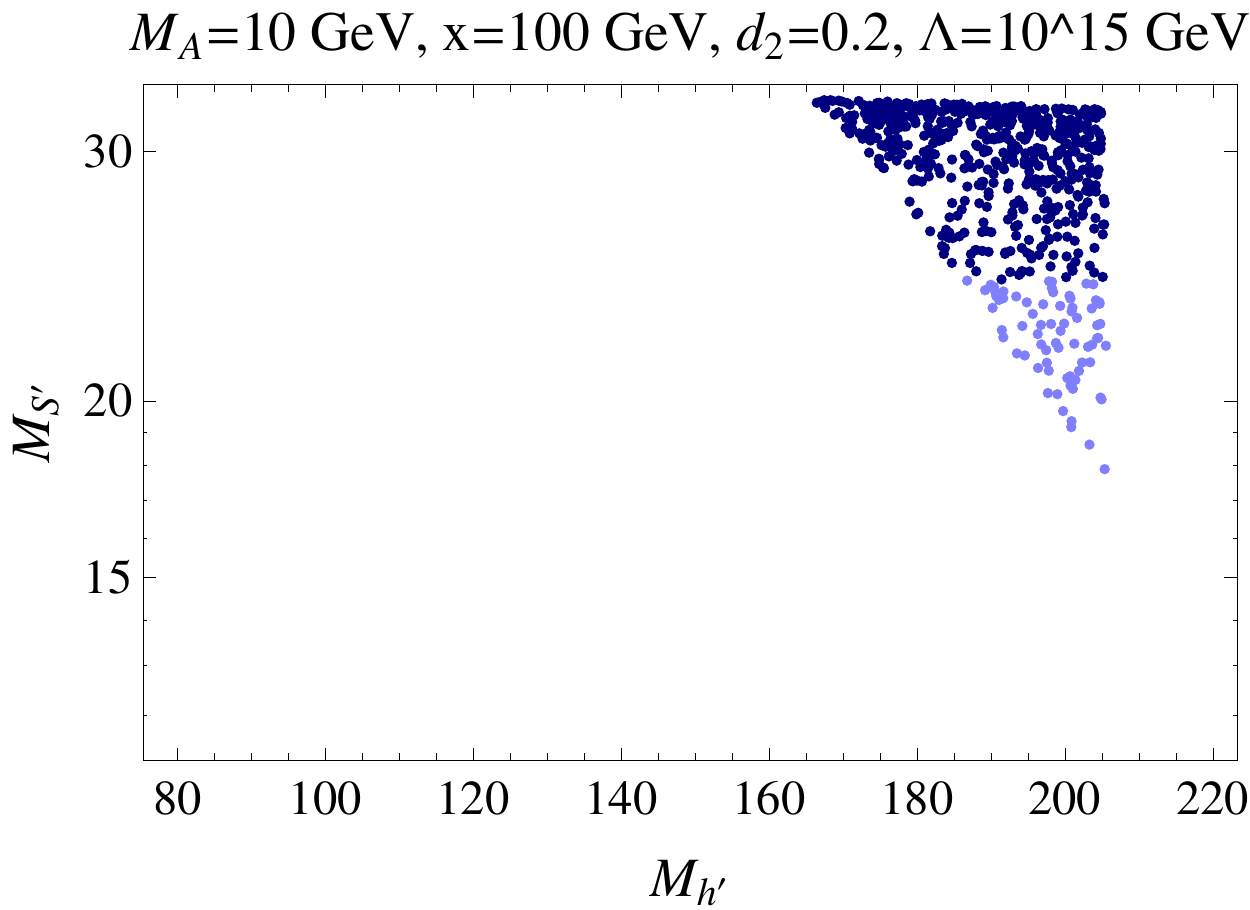}\\
\includegraphics[width=0.45\textwidth]{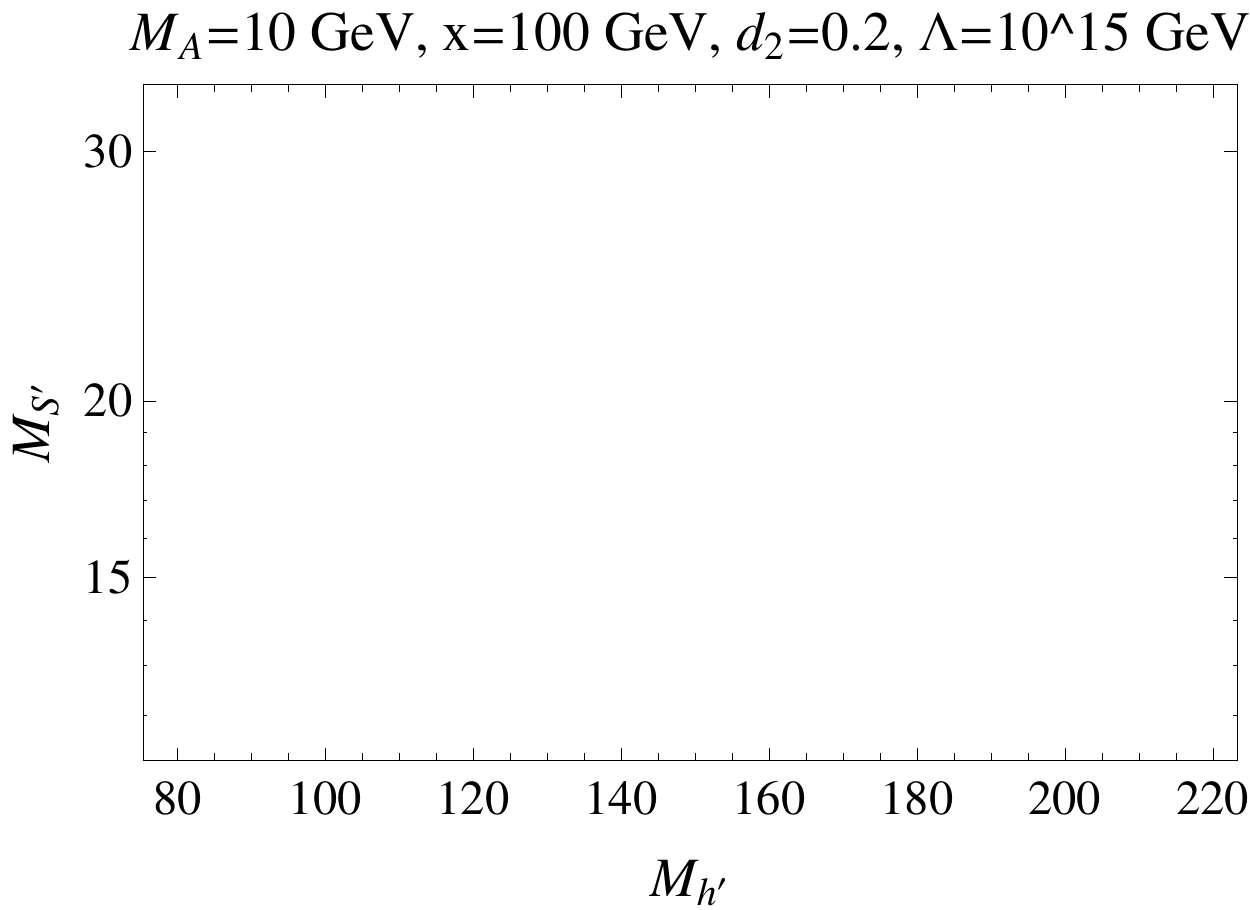}
\includegraphics[width=0.45\textwidth]{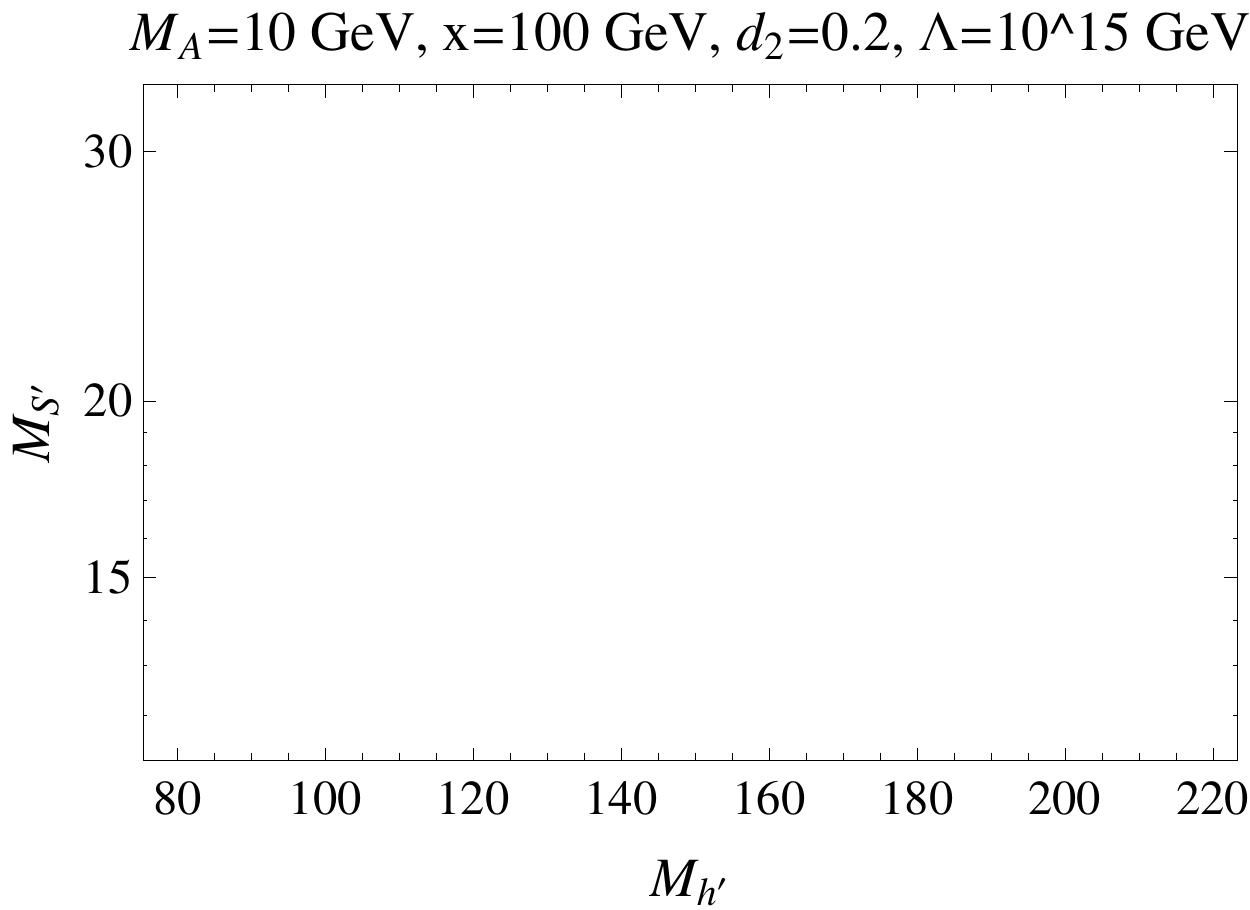}
\caption{Same as \figref{fig:ms_vs_mh_ddinvis} but with $\Lambda=10^{15}$~GeV.}
\label{fig:ms_vs_mh_15_ddinvis}
\end{figure*}

We now show the impact of the dark matter direct detection limits and the condition for invisibly decaying Higgs searches at ATLAS.  \Figref{fig:ms_vs_mh_ddinvis} displays those values of the Higgs-like and singlet-like scalar masses that satisfy \eqnref{eq:rg_running_stability} (top left --- the same as the top left plot in \figref{fig:ms_vs_mh_lepewpo}), plus the XENON100 direct detection cross section bound (top right) or the requirement for ATLAS invisible Higgs decay searches (bottom left), and all three together (bottom right).  \Figref{fig:ms_vs_mh_6_ddinvis} and \figref{fig:ms_vs_mh_15_ddinvis} are similar to \figref{fig:ms_vs_mh_ddinvis} but with $\Lambda=10^6$ and $10^{15}$~GeV respectively.

The choice of parameters here is such that nearly all the points satisfy the XENON100 bound on the direct detection cross section.  Consequently, very few points satisfy the CRESST-II or DAMA regions (and a 10~GeV dark matter particle is incompatible with the result from CoGeNT presented in \cite{Aalseth:2011wp_cogent2011}).  Therefore, here and in what follows in later sections, we impose the XENON100 bound as a more conservative upper bound on the direct detection cross section.  

More restrictive is the requirement for sensitivity to an invisibly decaying Higgs at ATLAS: this condition prefers a lighter Higgs-like eigenstate for which the total decay rate is smaller and hence the $h'\rightarrow AA$ decay has a larger branching fraction.  Though the light $S'$ eigenstate has a large branching fraction to dark matter, the mixing angle is too small to give a $\xi^2$ greater than the requisite 60\%.  The RG evolution bounds and the ATLAS sensitivity become mutually exclusive at higher $\Lambda$.

\subsection{All Results for Light and Heavy Dark Matter}

\begin{figure*}
\includegraphics[width=0.29\textwidth]{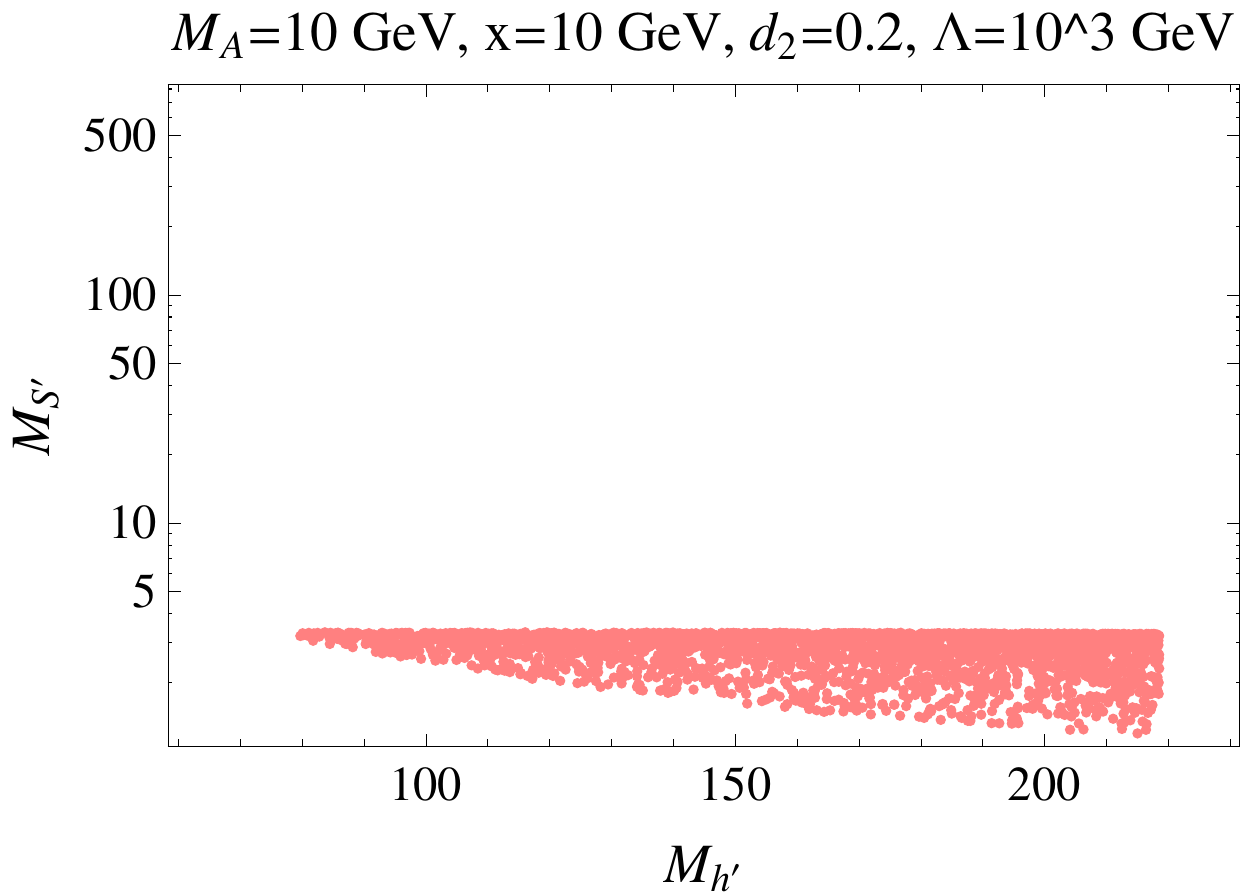}
\includegraphics[width=0.29\textwidth]{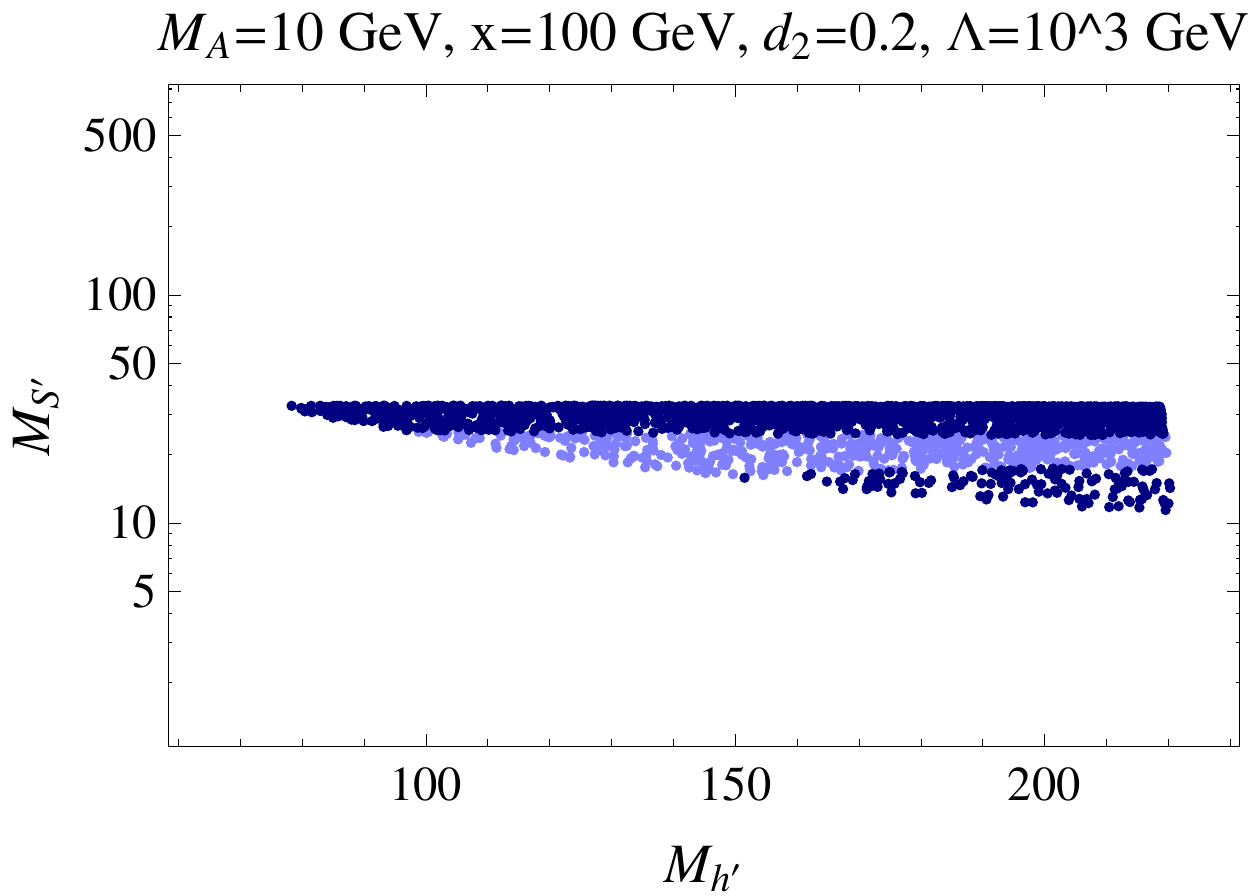}
\includegraphics[width=0.29\textwidth]{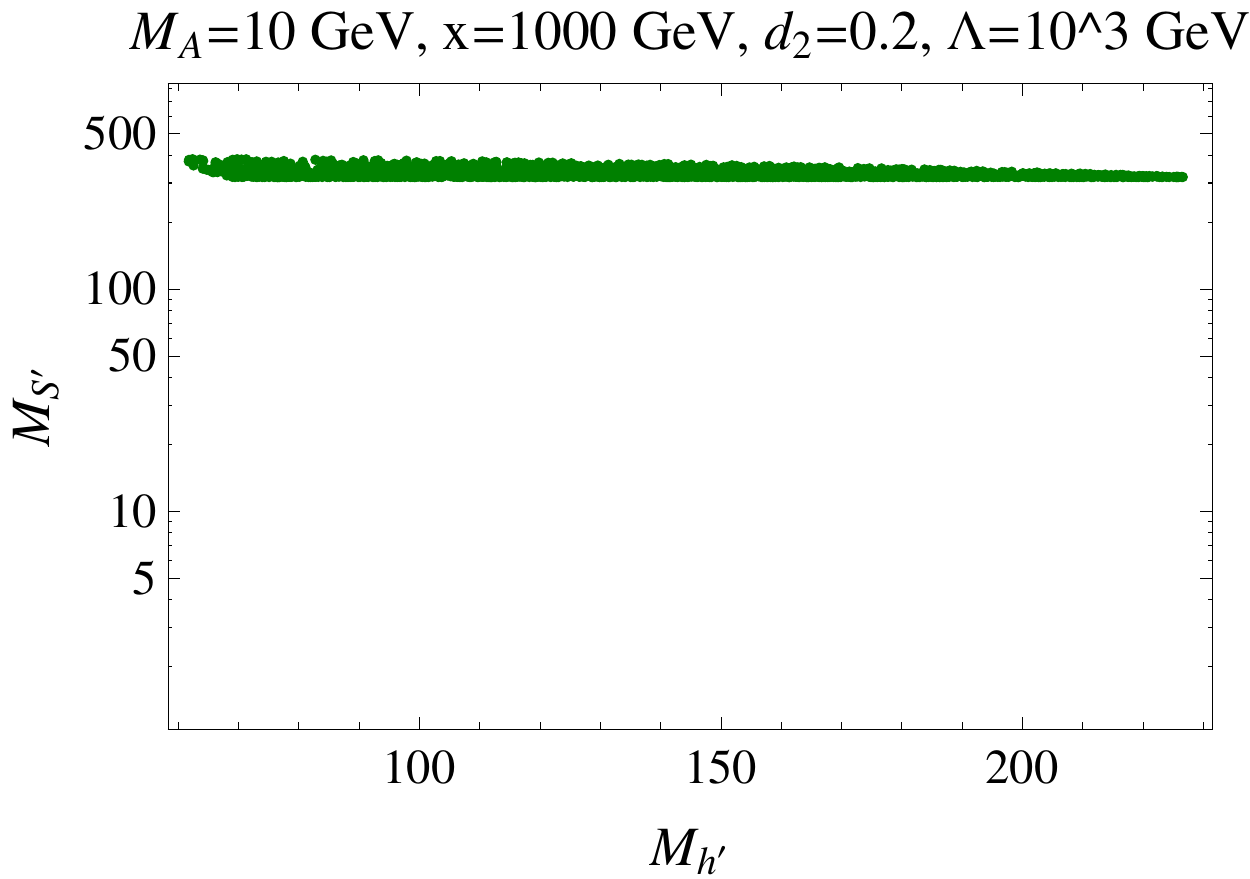}\\
\includegraphics[width=0.29\textwidth]{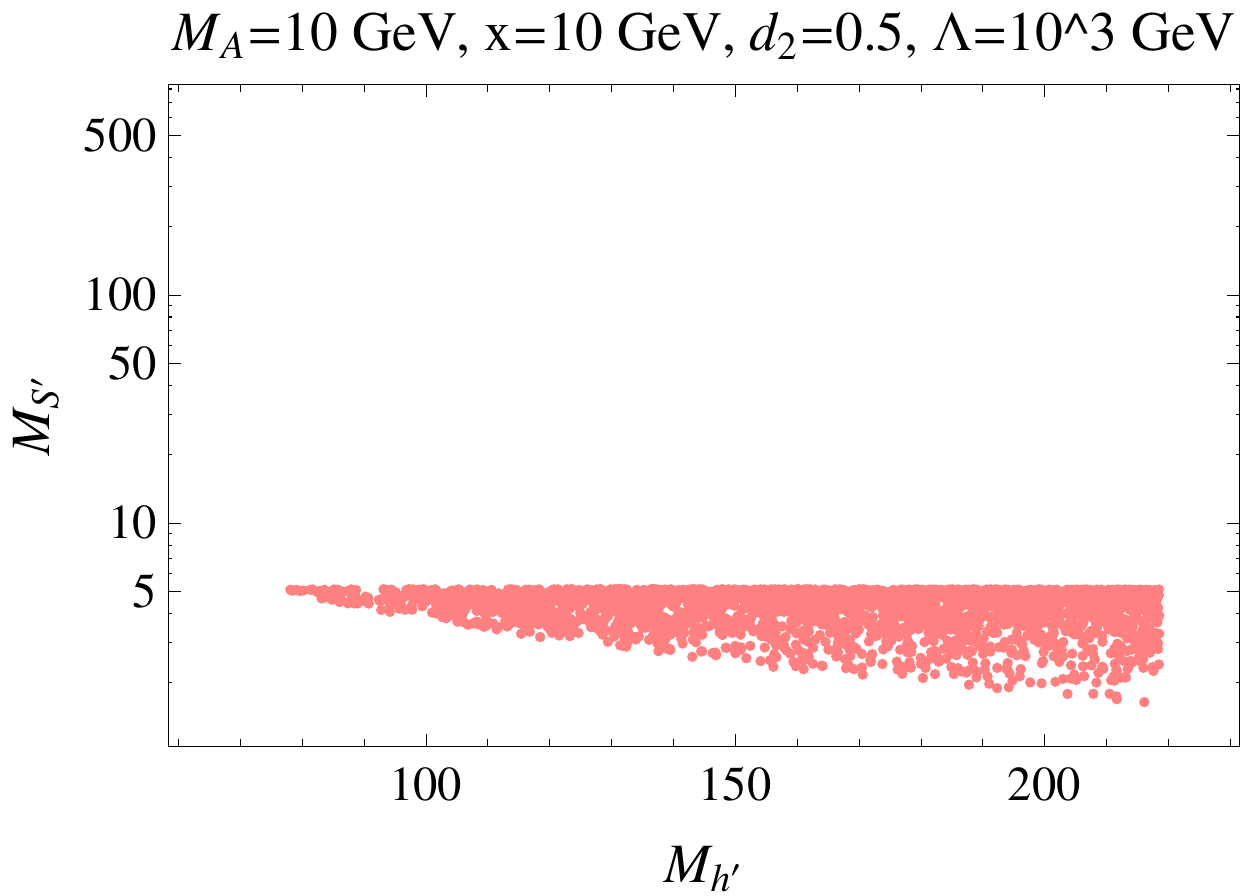}
\includegraphics[width=0.29\textwidth]{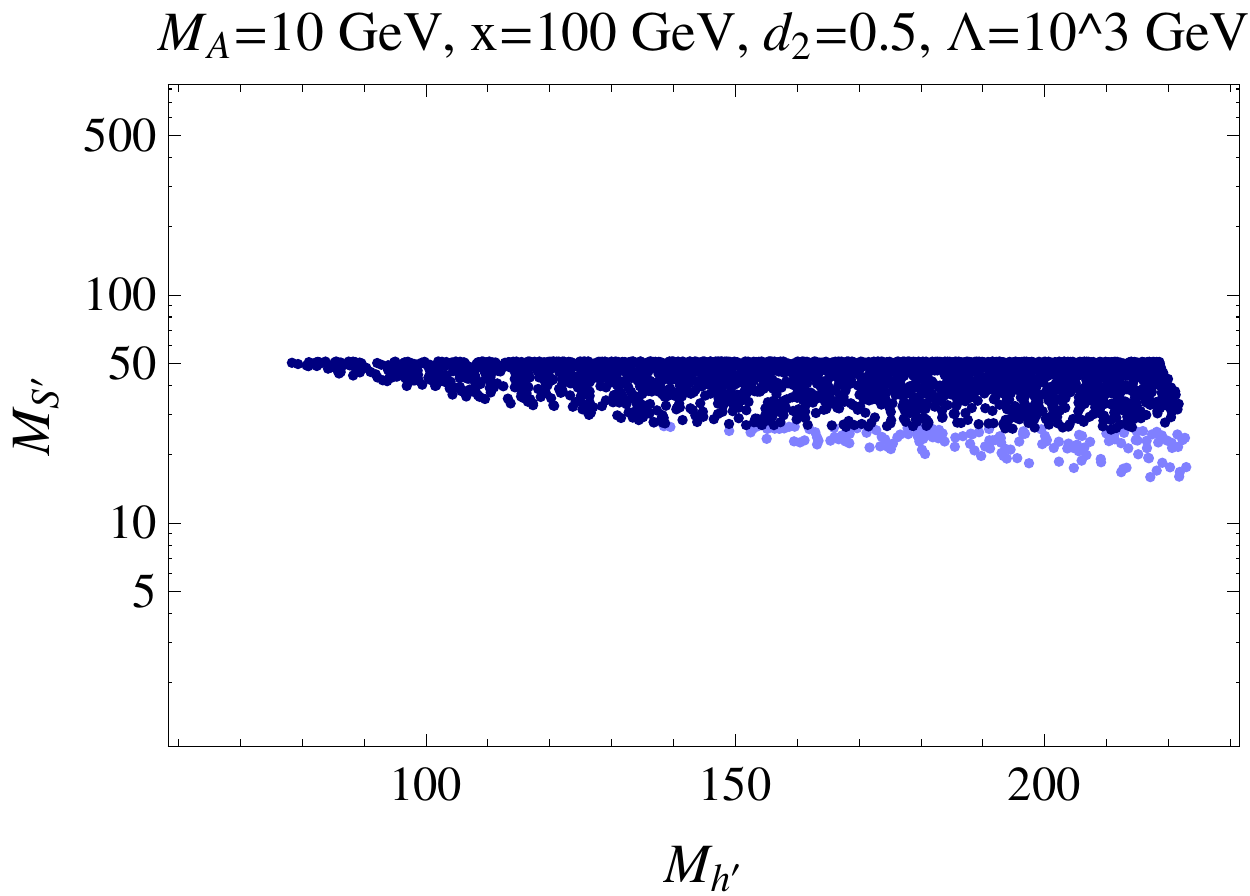}
\includegraphics[width=0.29\textwidth]{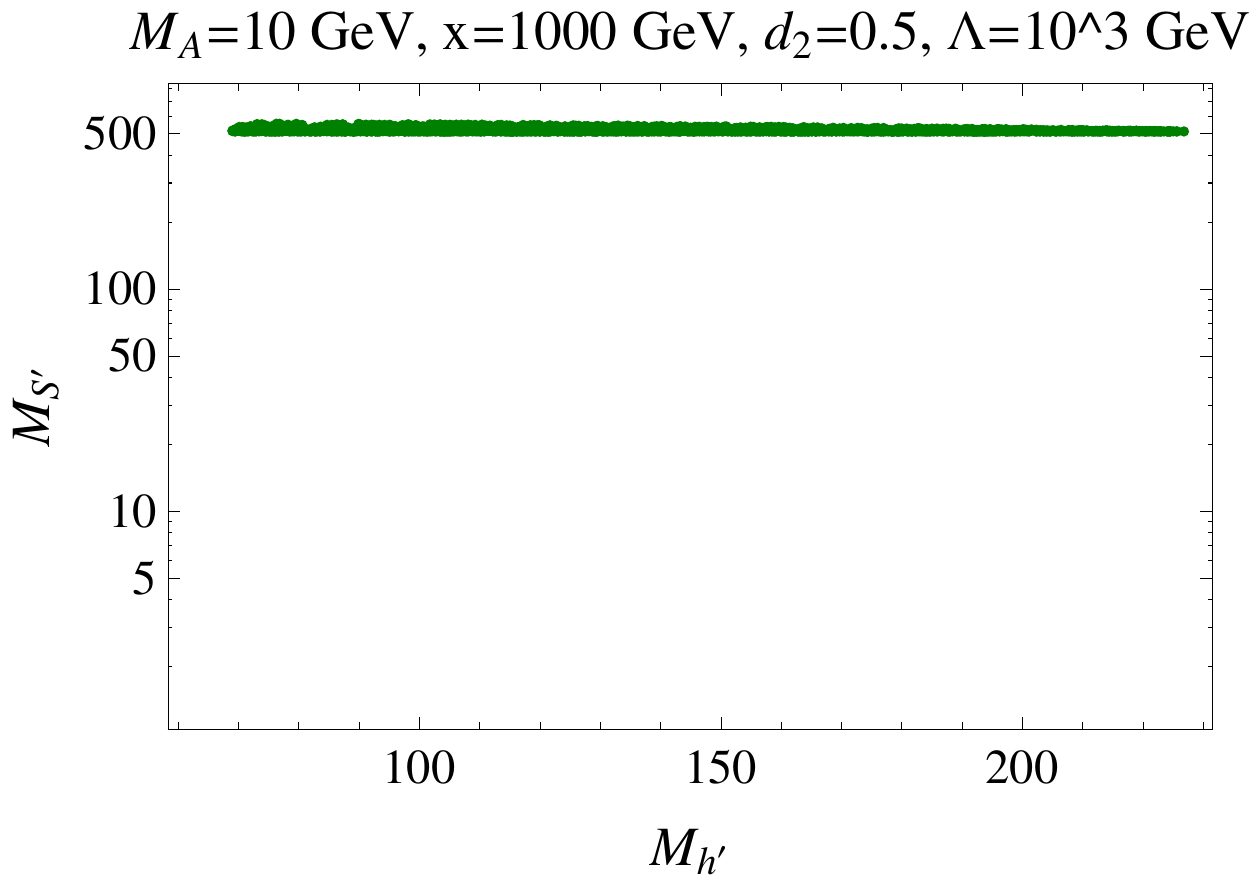}\\
\includegraphics[width=0.29\textwidth]{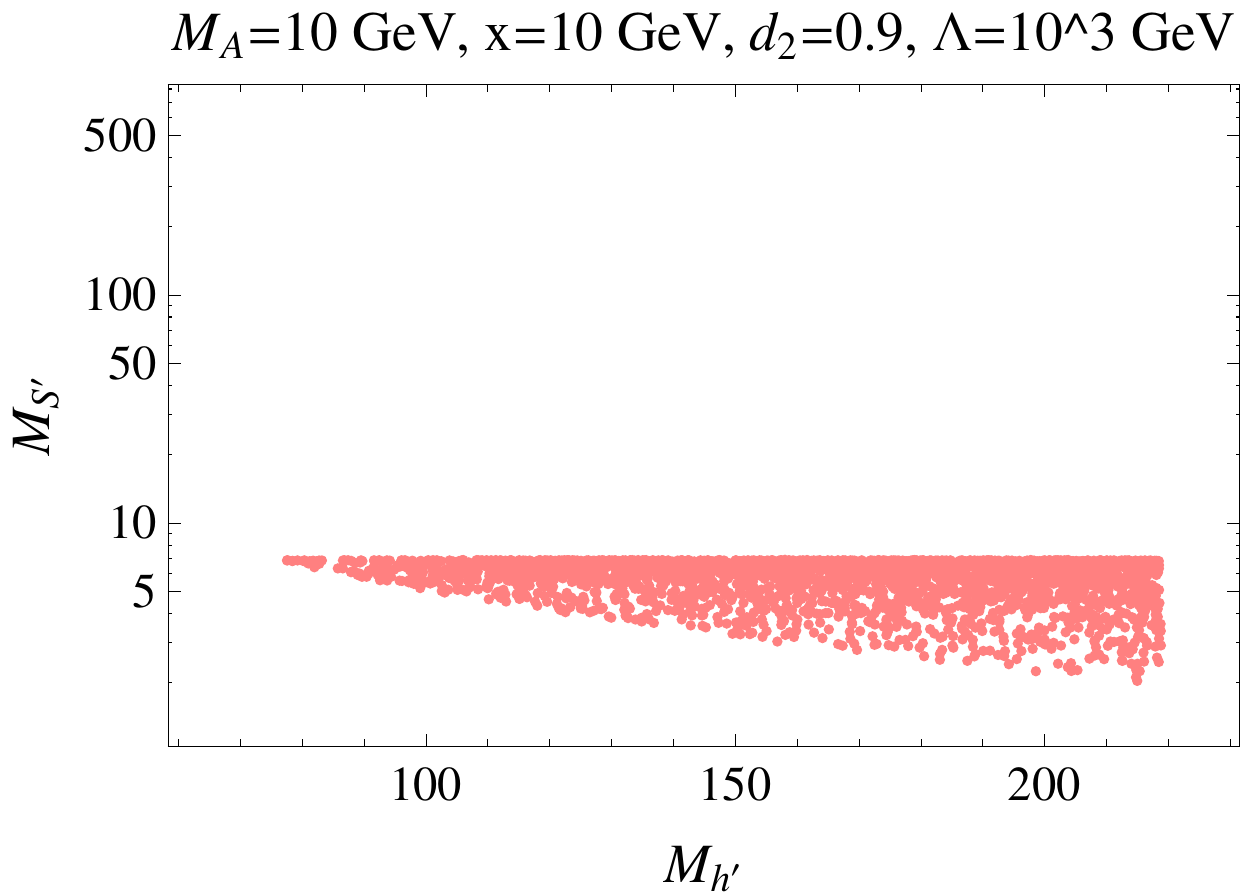}
\includegraphics[width=0.29\textwidth]{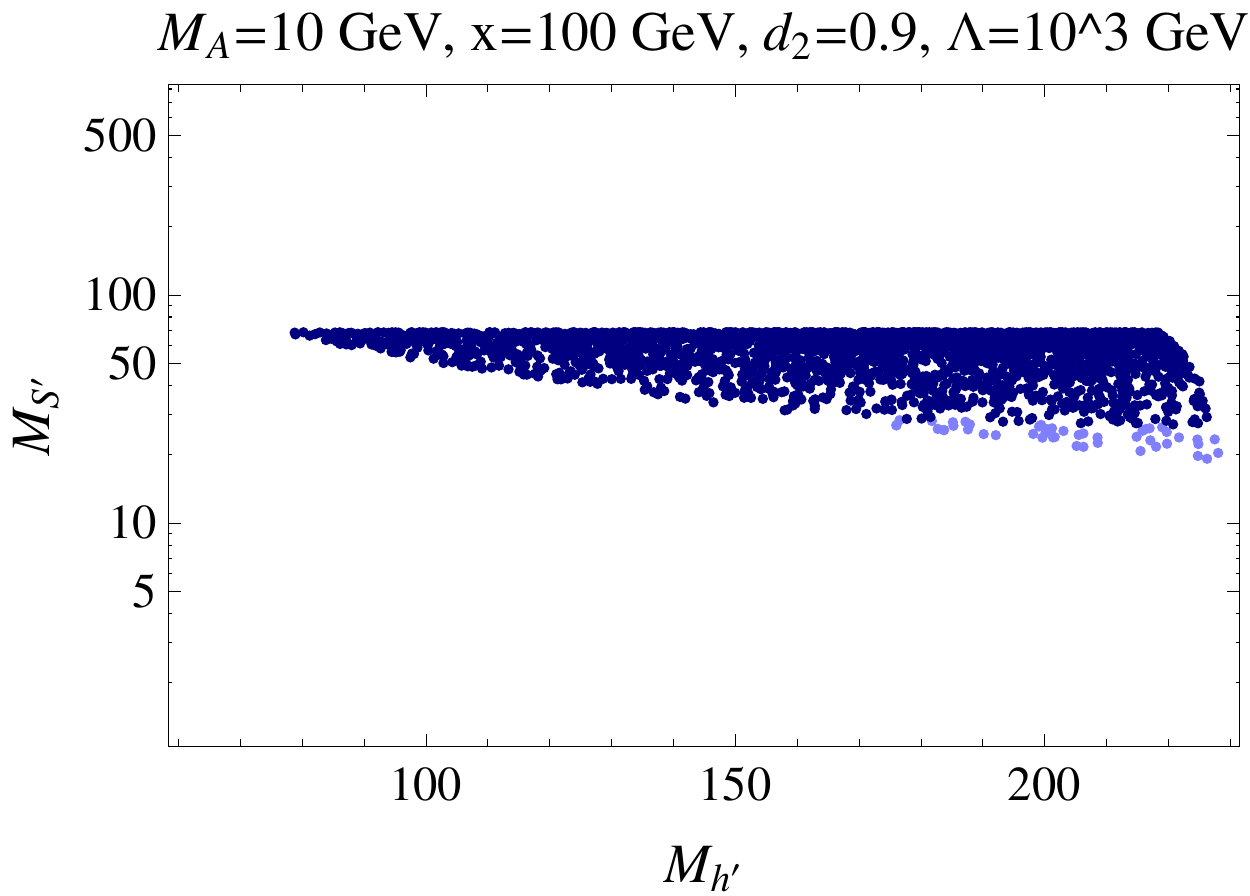}
\includegraphics[width=0.29\textwidth]{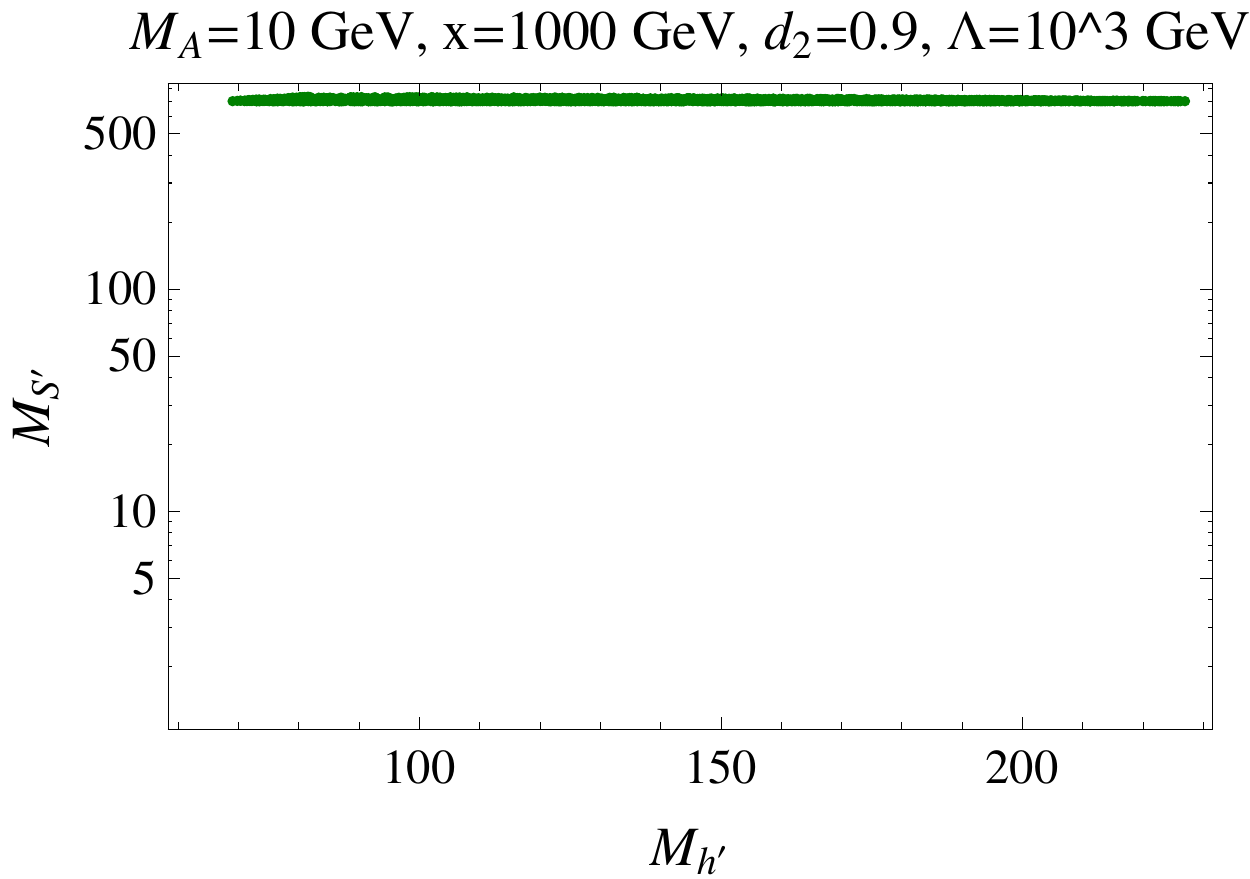}
\caption{Here are shown the effects on the mass eigenvalues and the relic density of changing $d_2$ and $x$ (with $\del <0$).  Only the RG running constraints ($\Lambda=1$~TeV) have been imposed.  Left column (red): $x=10$~GeV; middle column (blue): $x=100$~GeV; right column (green): $x=1000$~GeV.  Top row: $d_2=0.2$; middle row: $d_2=0.5$; bottom row: $d_2=0.9$.  Dark colored points oversaturate the relic density, while light colored points (under)saturate.}
\label{fig:ms_vs_mh_all}
\end{figure*}

We now consider all values of the couplings, singlet vev, and dark matter mass (light being 10~GeV, heavy being 100~GeV) listed in \tabref{table:scans}.  \Figref{fig:ms_vs_mh_all} shows the effect of varying the model parameters on the scalar mass eigenstates when $M_A=10$~GeV; only the RG evolution constraints of \eqnref{eq:rg_running_stability} have been imposed with a 1~TeV cutoff.  The top center plot in \figref{fig:ms_vs_mh_all} is identical to the top left plots in \figref{fig:ms_vs_mh_lepewpo} through \figref{fig:ms_vs_mh_15_ddinvis}.  

Varying the singlet vev $x$ between 10 and 1000~GeV clearly has a greater effect on the singlet-like eigenstate mass than varying the singlet quartic self-coupling $d_2$ between 0.2 and 0.9, as is expected from \eqnref{eq:tree_mass_matrix}.  Furthermore, the $S'$ eigenstate, when it is the lighter state (as in the left and middle columns), has a maximum allowed mass: the smaller eigenvalue of the mass matrix has, for fixed $d_2,x$, a maximum value of $d_2x^2/2$ (plus loop corrections) even as $\lambda$ increases.  The Higgs-like eigenstate has a maximum value because we have limited our scan of $\lambda$.  Finally, the general trend for the relic density is oversaturation when $M_{h'},M_{S'} > 2M_A$ --- due to an off-resonance $s$-channel scalar exchange in the dark matter annihilation cross section --- and undersaturation when $2M_{S'}\simeq M_A$ or $M_{S'} \lesssim M_A$.

\begin{figure*}
\includegraphics[width=0.29\textwidth]{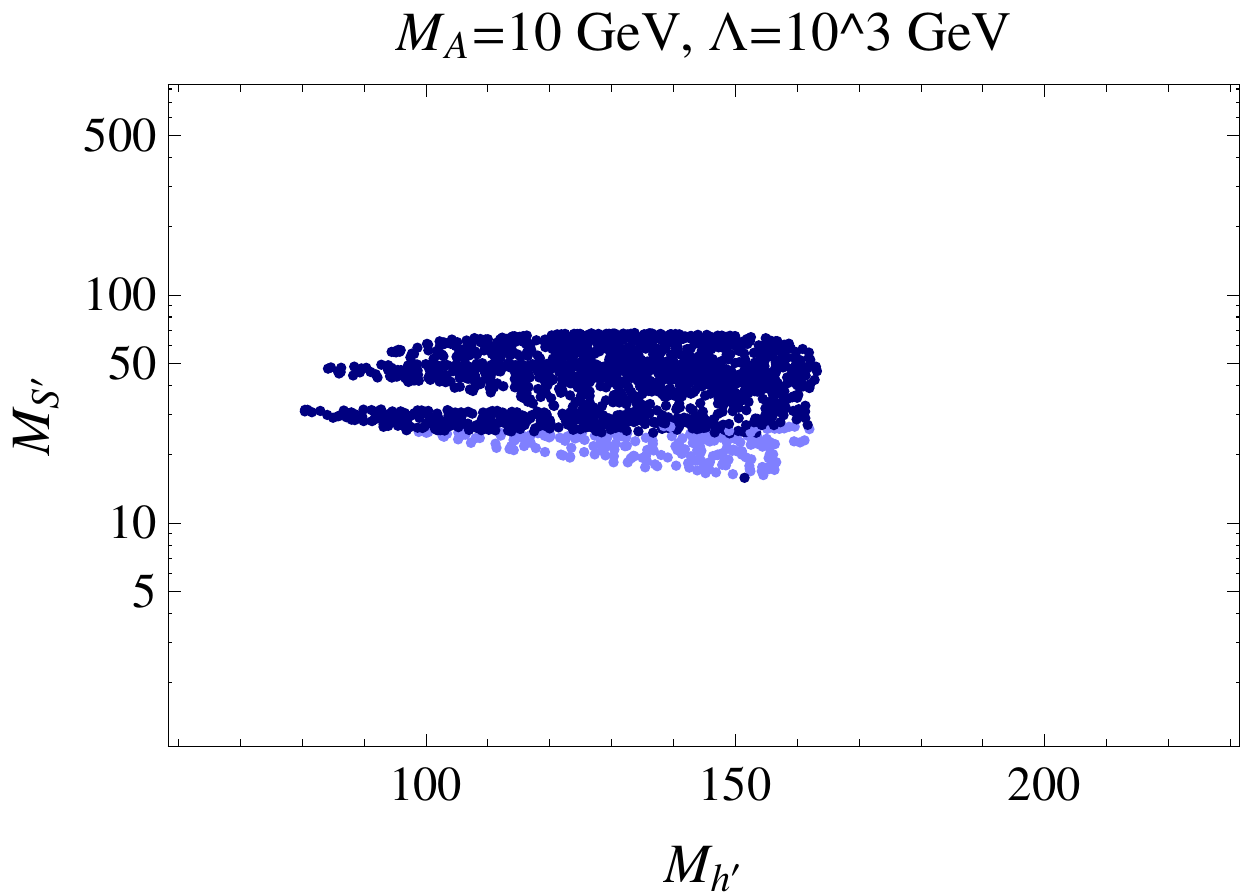}
\includegraphics[width=0.29\textwidth]{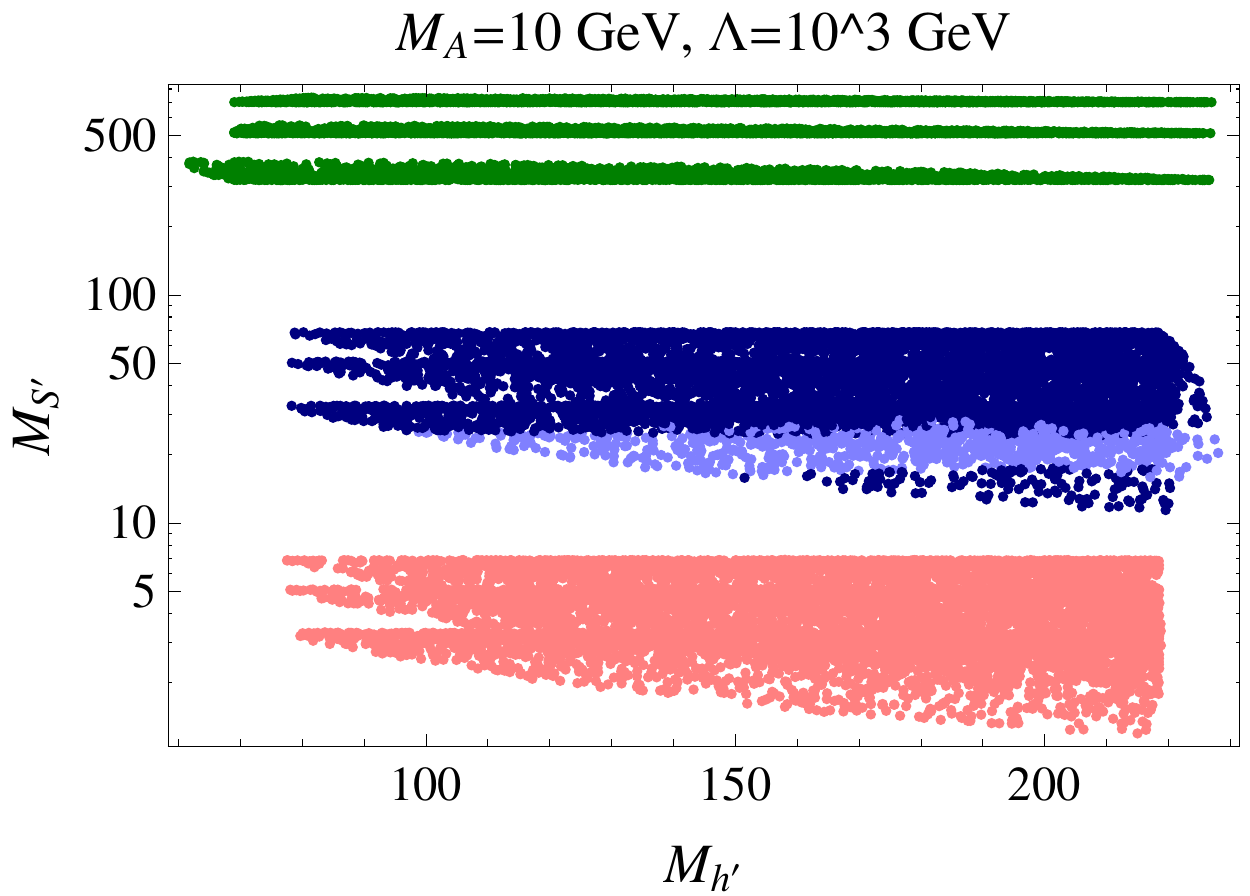}
\includegraphics[width=0.29\textwidth]{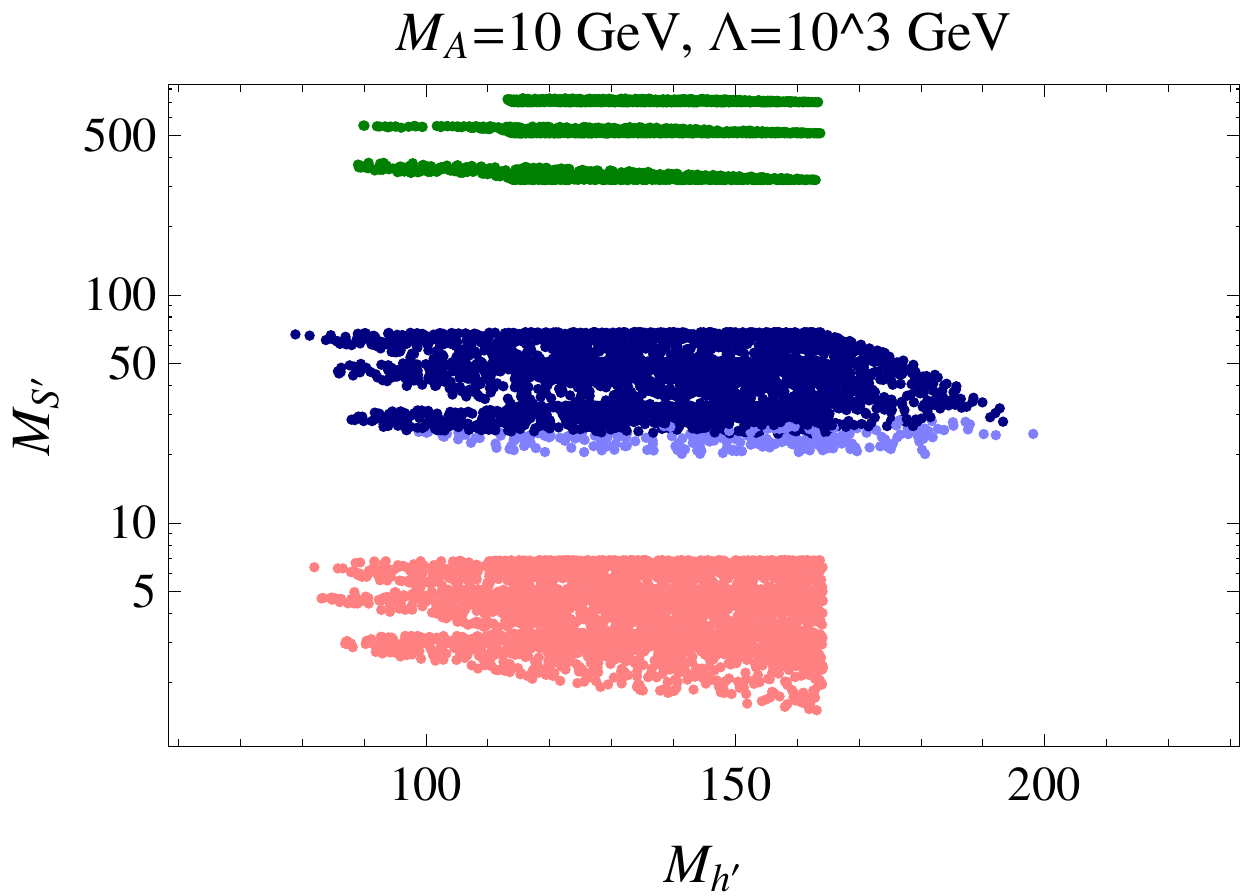}\\
\includegraphics[width=0.29\textwidth]{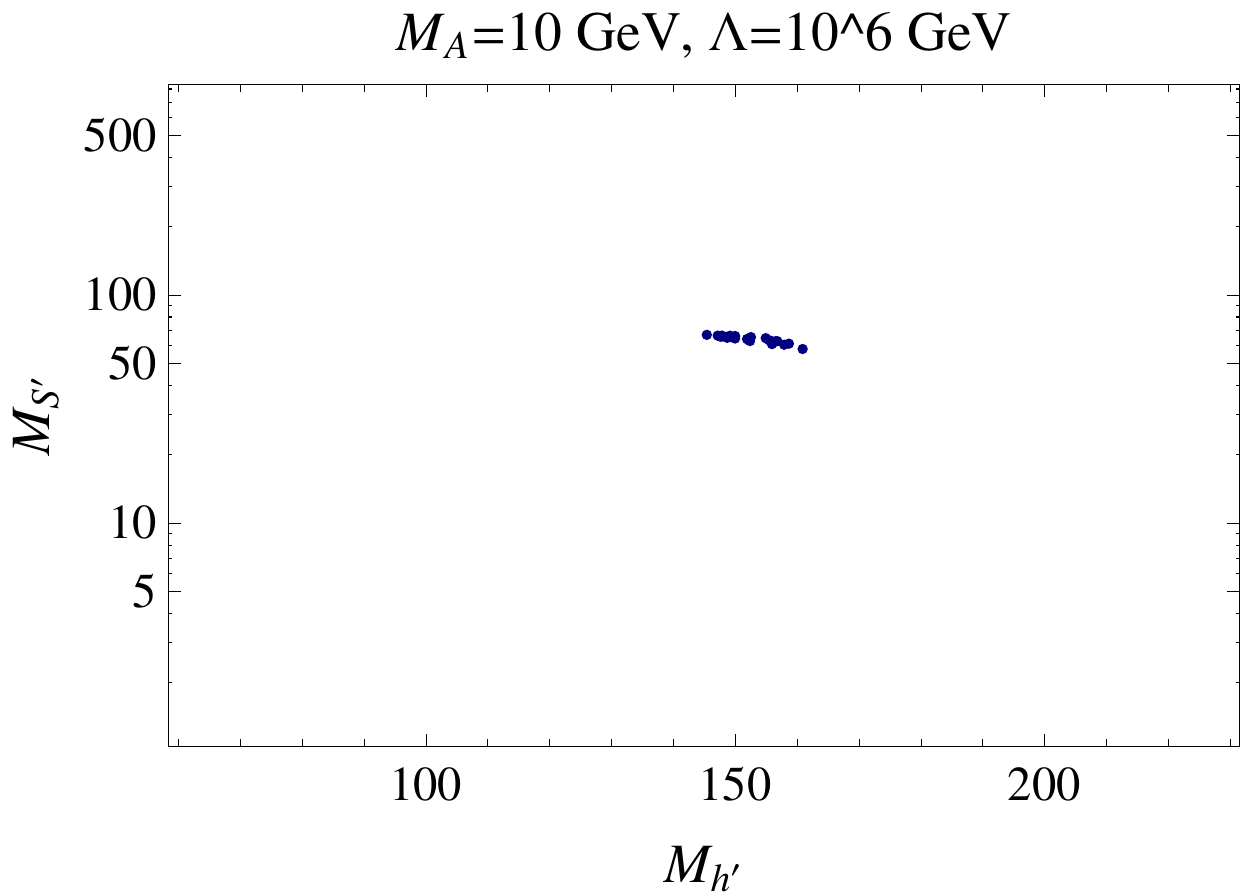}
\includegraphics[width=0.29\textwidth]{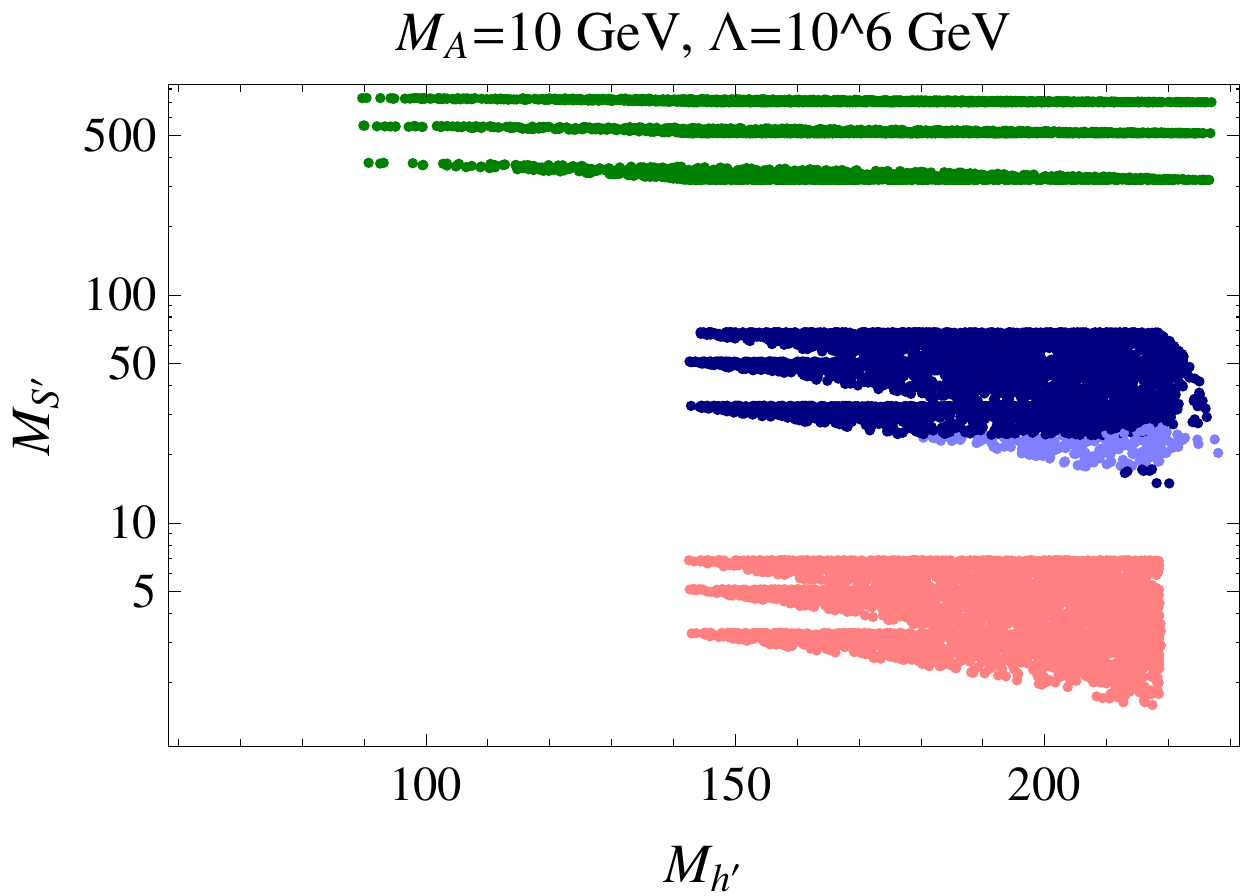}
\includegraphics[width=0.29\textwidth]{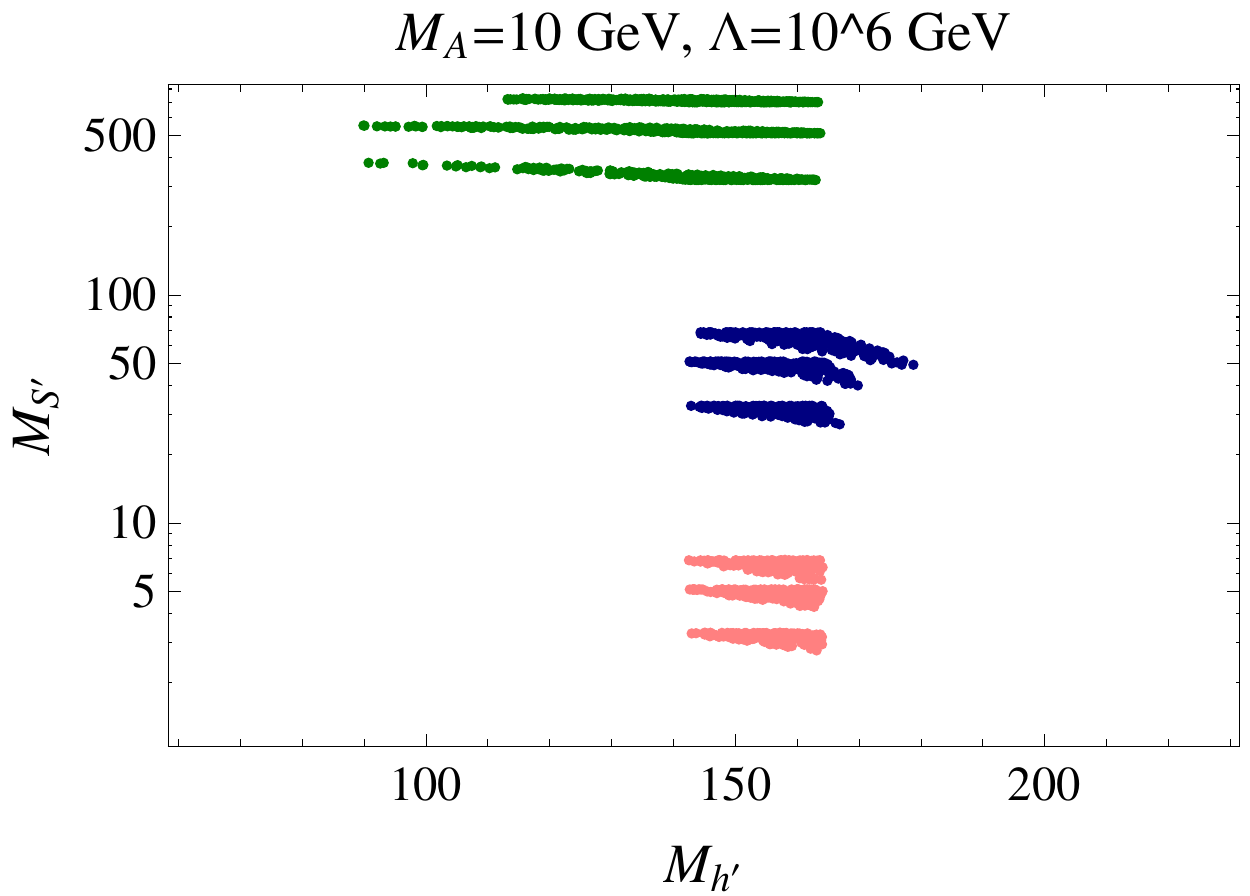}\\
\includegraphics[width=0.29\textwidth]{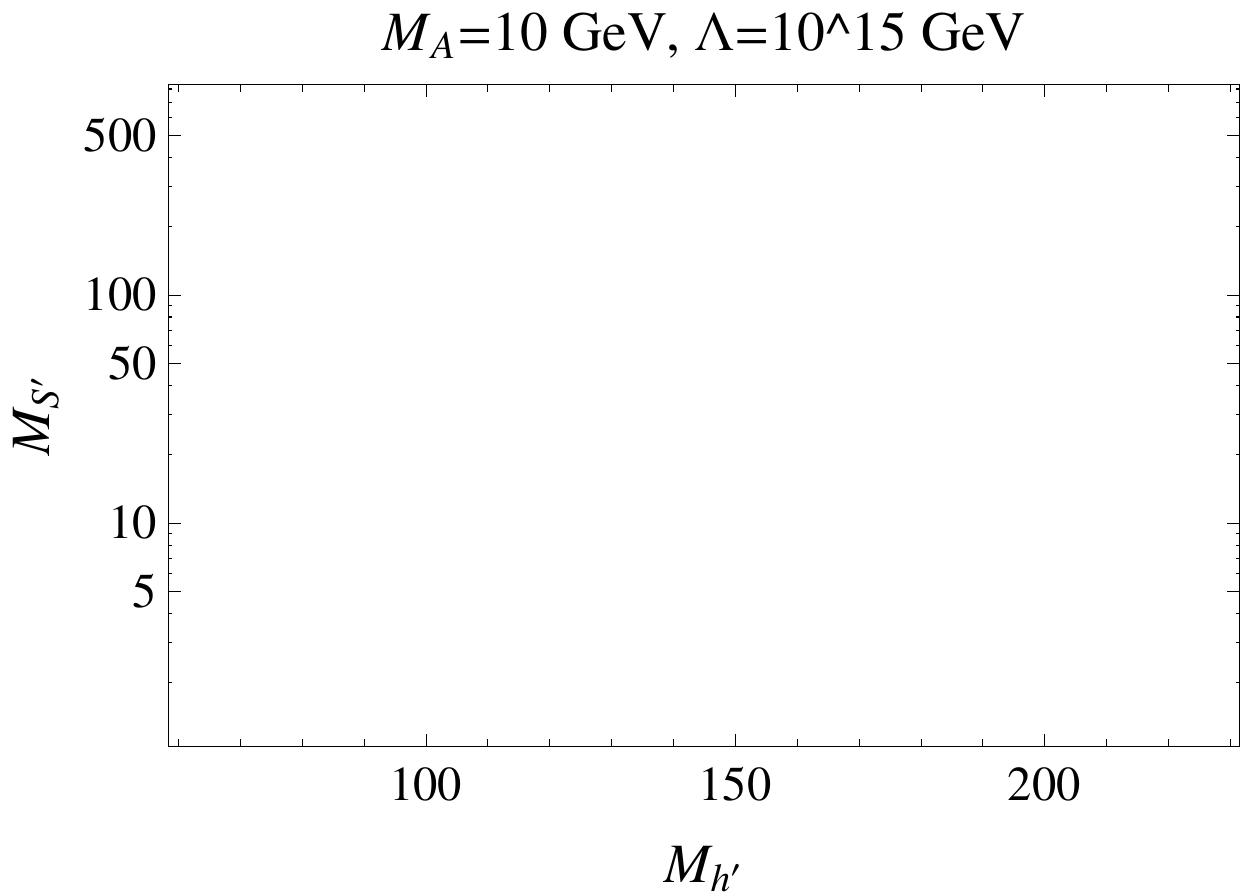}
\includegraphics[width=0.29\textwidth]{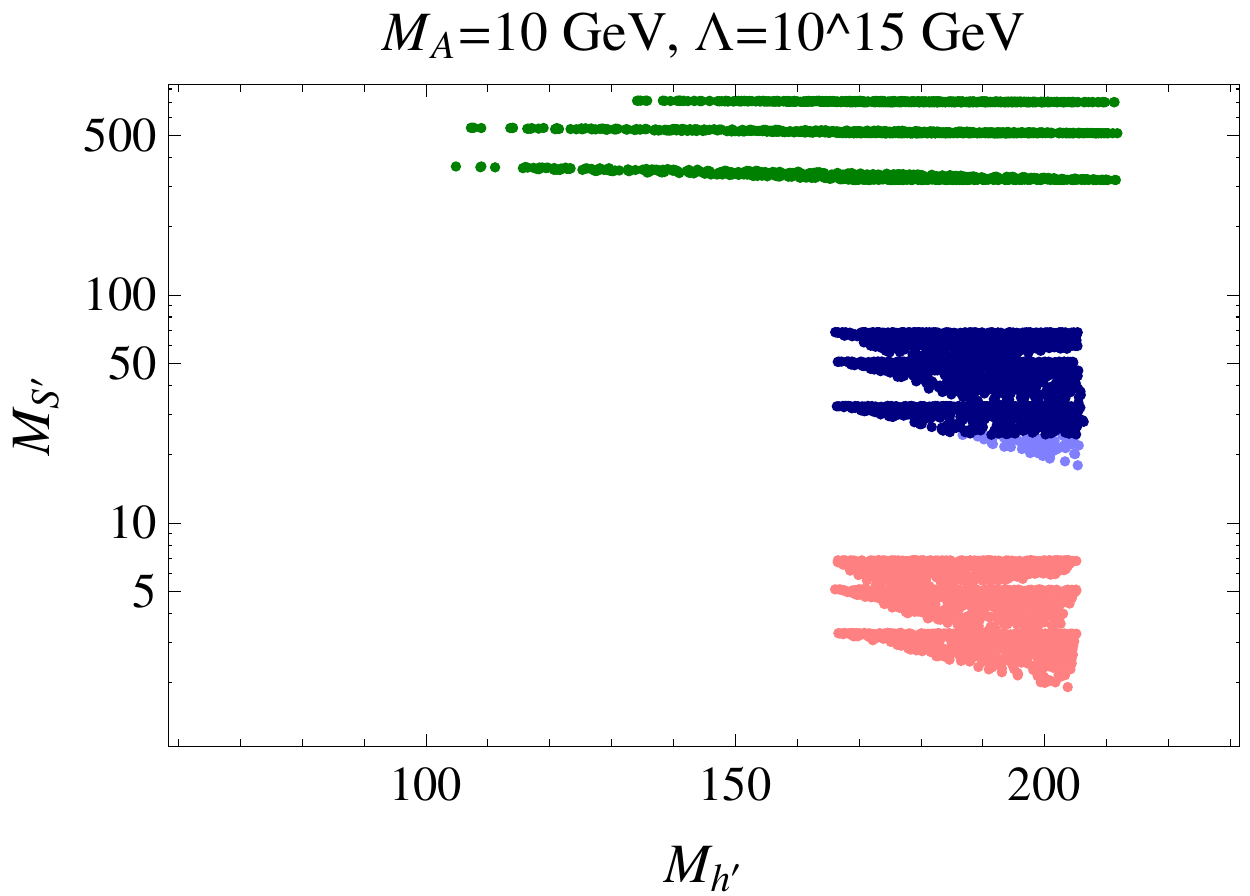}
\includegraphics[width=0.29\textwidth]{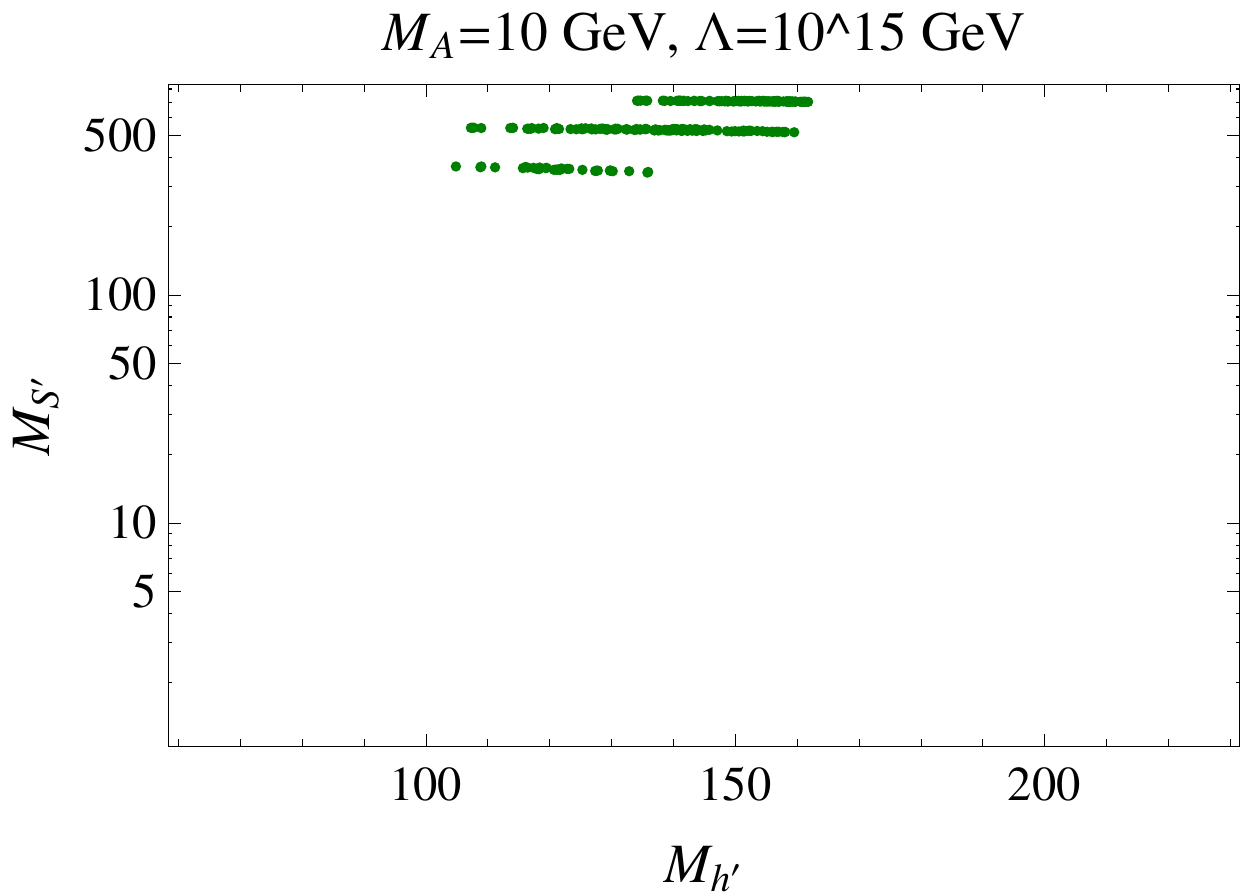}
\caption{Here are shown the allowed mass eigenvalues ($M_A=10$~GeV) after imposing the RG evolution requirement in \eqnref{eq:rg_running_stability} (central column) plus the LEP and EWPO constraints (right column) or, alternately, the direct detection bound from XENON100 and the invisible decay requirements (left column).  The effect of increasing the cutoff scale is also seen: $\Lambda=1$~TeV (top row), $10^6$~GeV (middle row), and $10^{15}$~GeV (bottom row).  Note that the top central plot corresponds to overlaying all plots in \figref{fig:ms_vs_mh_all}.  The three colors indicate the value of $x$ as in \figref{fig:ms_vs_mh_all}.  The appearance of discrete bands for a given $x$ is the result of the three different choices for $d_2$.}
\label{fig:ms_vs_mh_all_lepewpo_ddinvis}
\end{figure*}

\begin{figure*}
\includegraphics[width=0.29\textwidth]{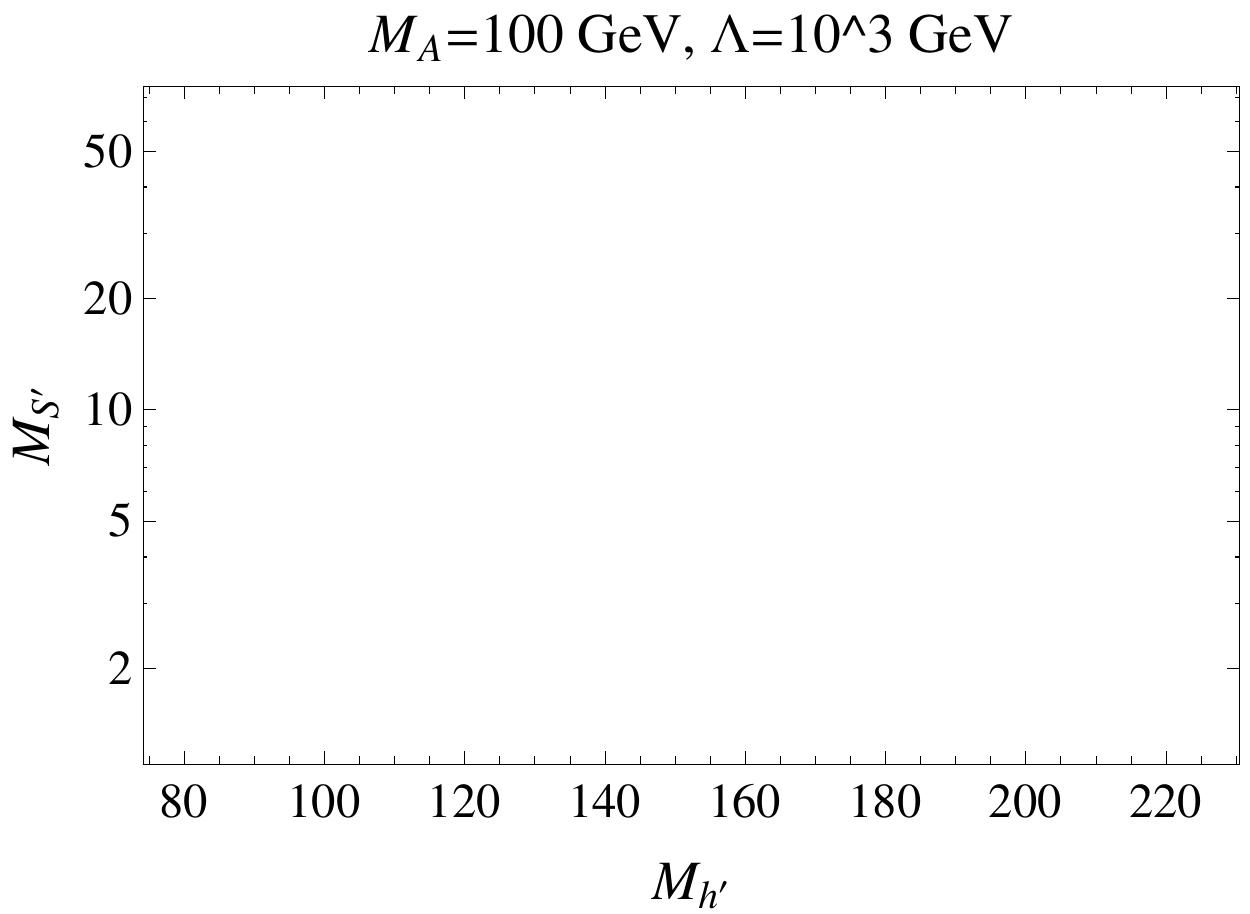}
\includegraphics[width=0.29\textwidth]{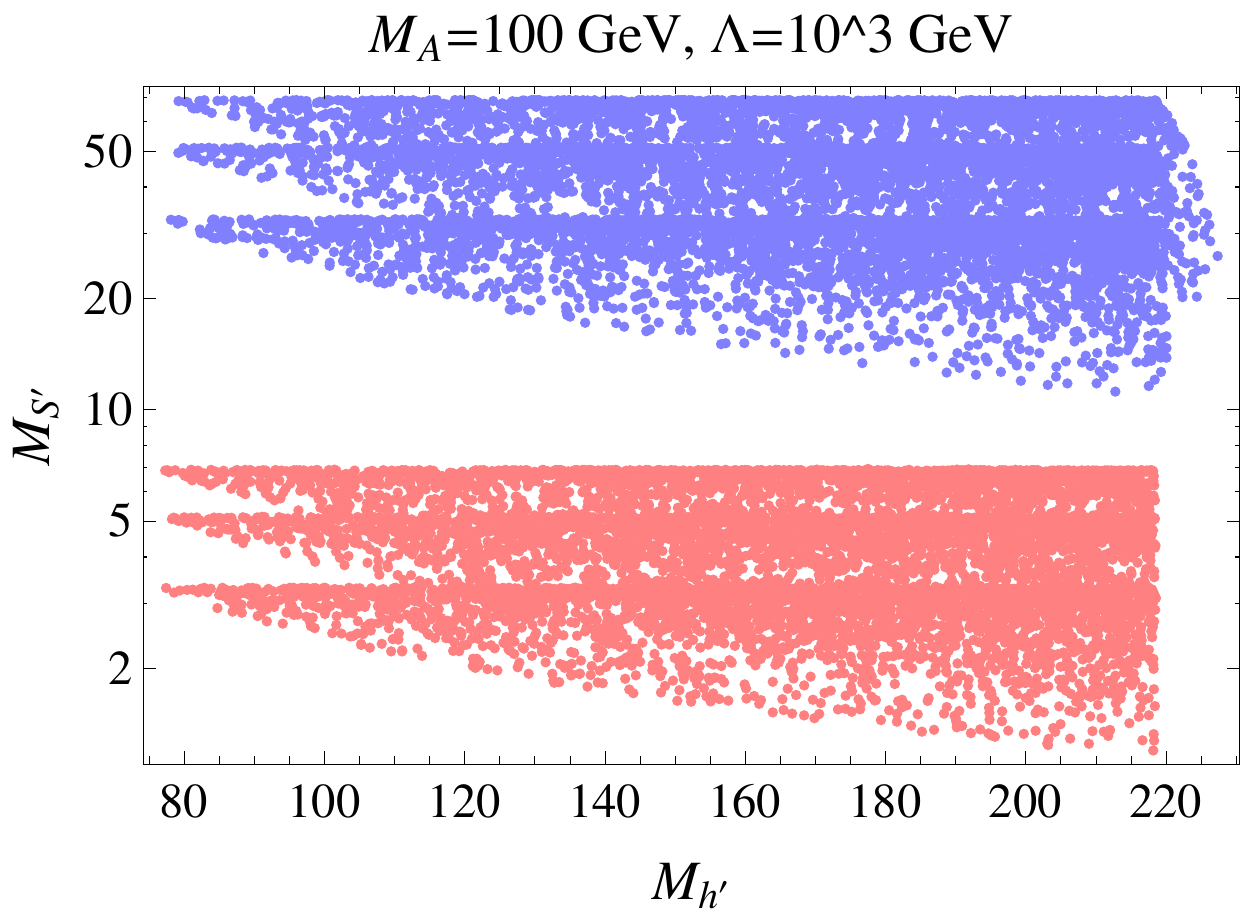}
\includegraphics[width=0.29\textwidth]{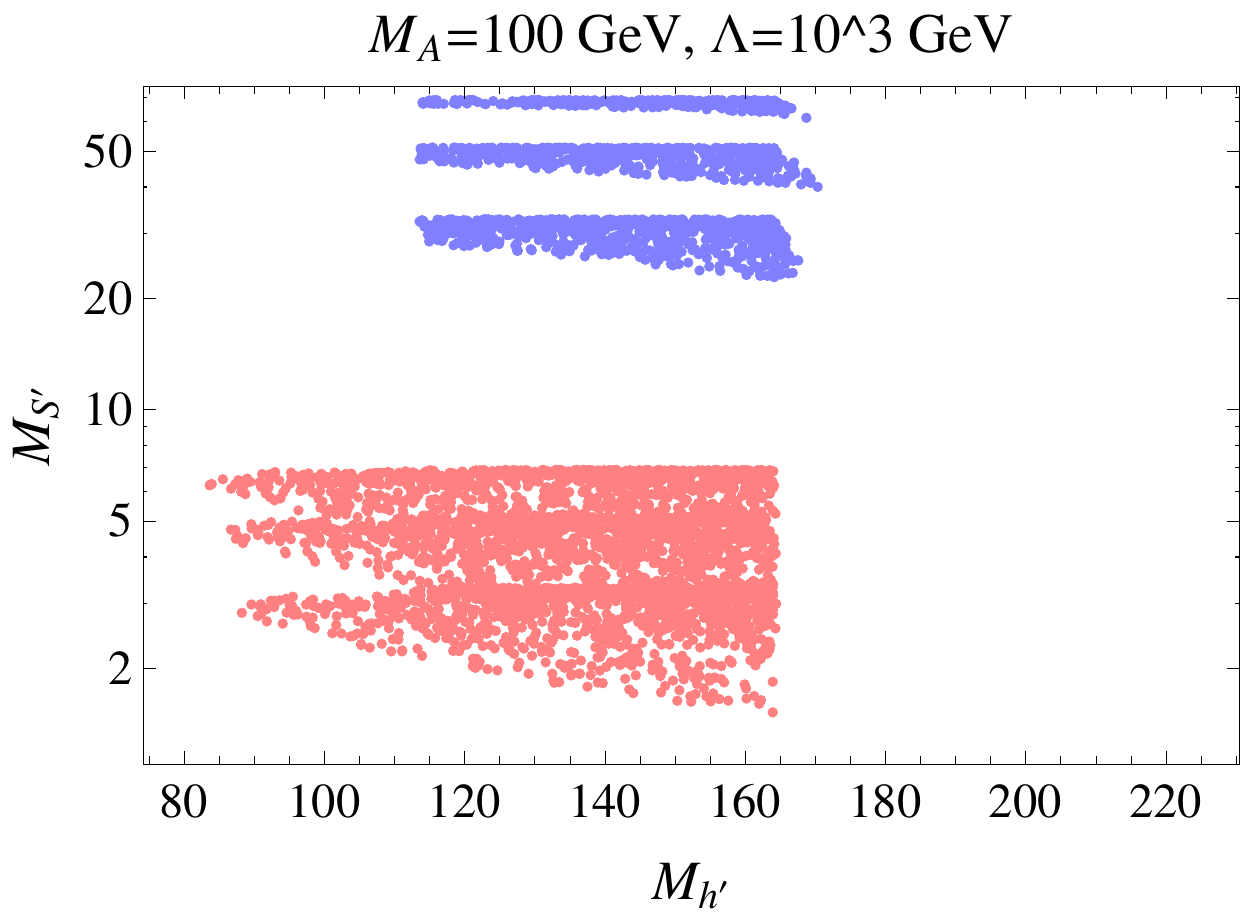}\\
\includegraphics[width=0.29\textwidth]{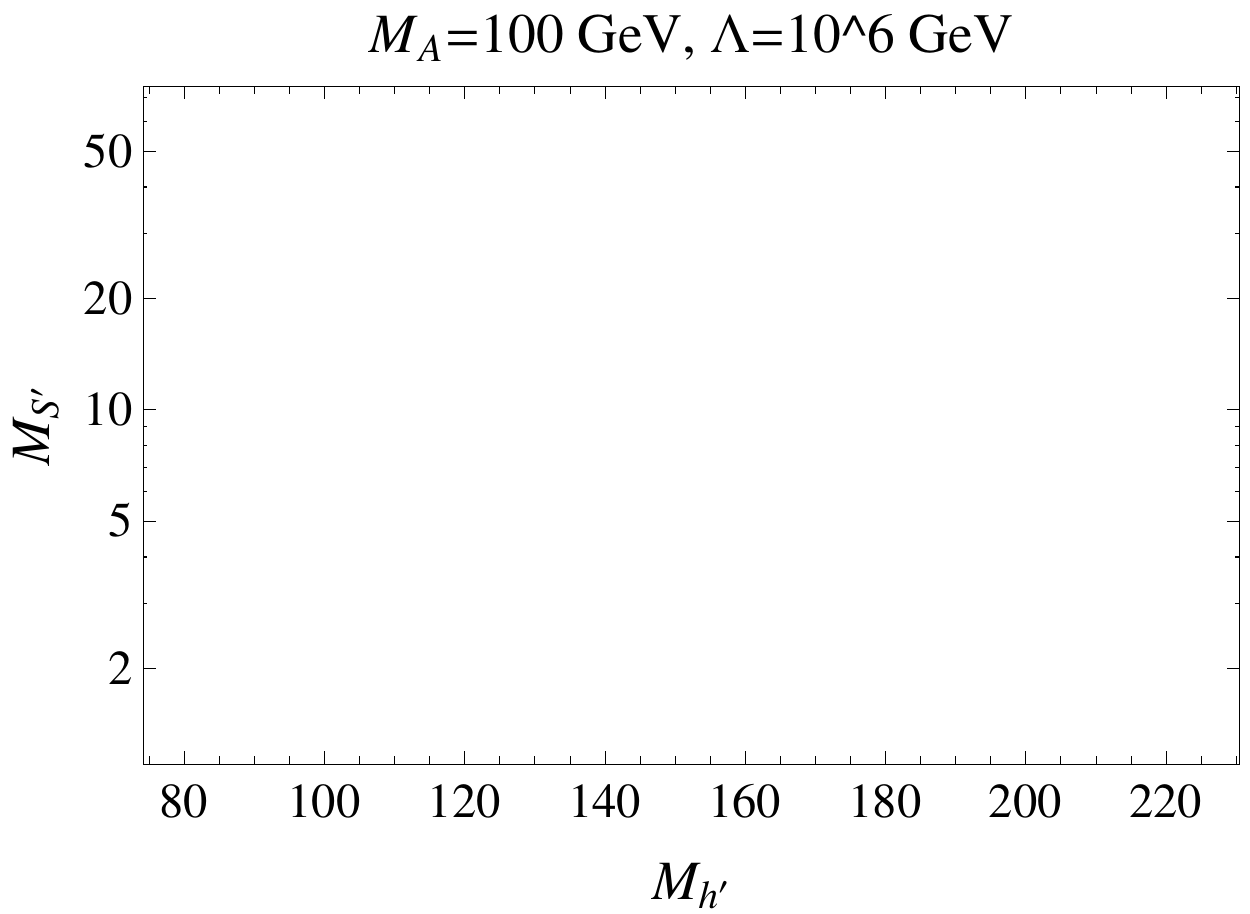}
\includegraphics[width=0.29\textwidth]{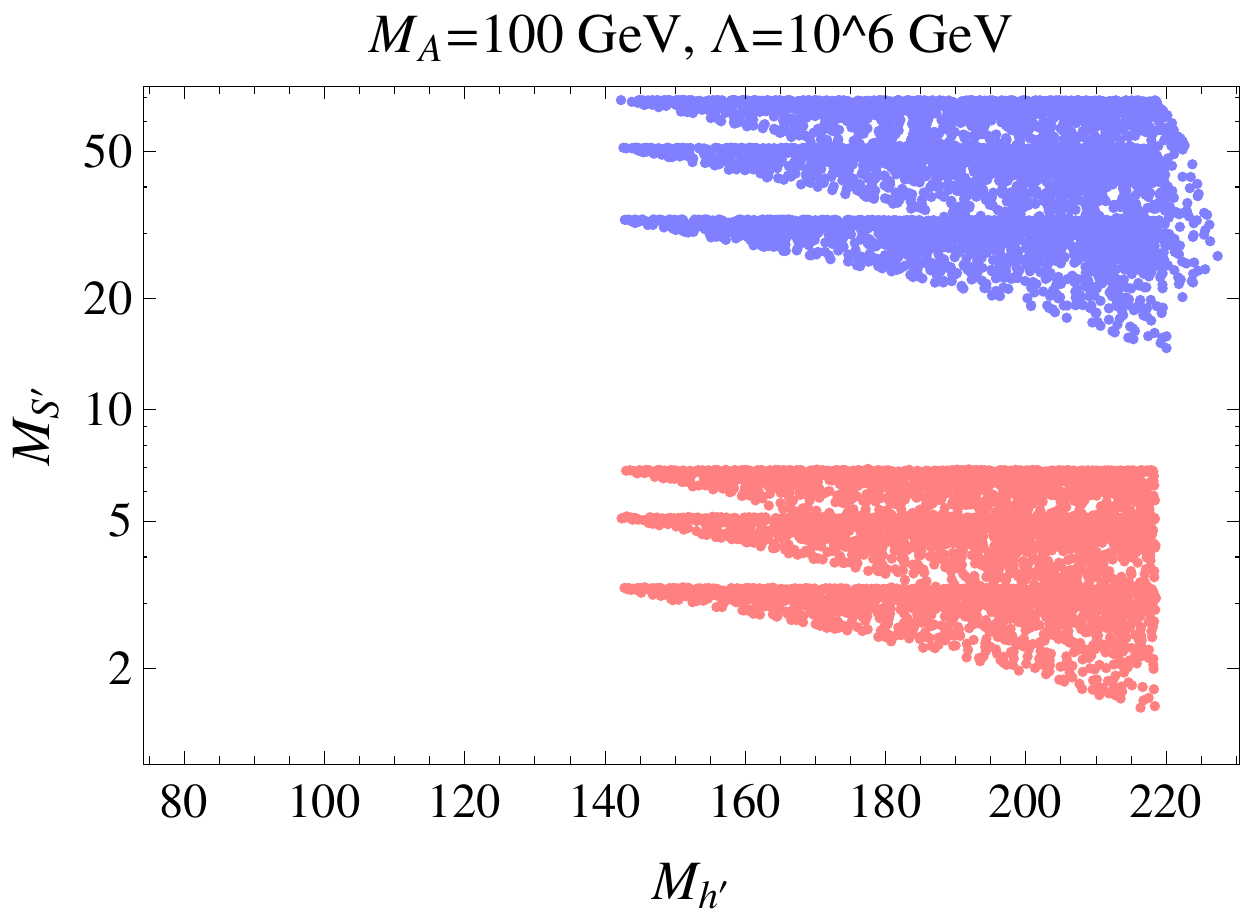}
\includegraphics[width=0.29\textwidth]{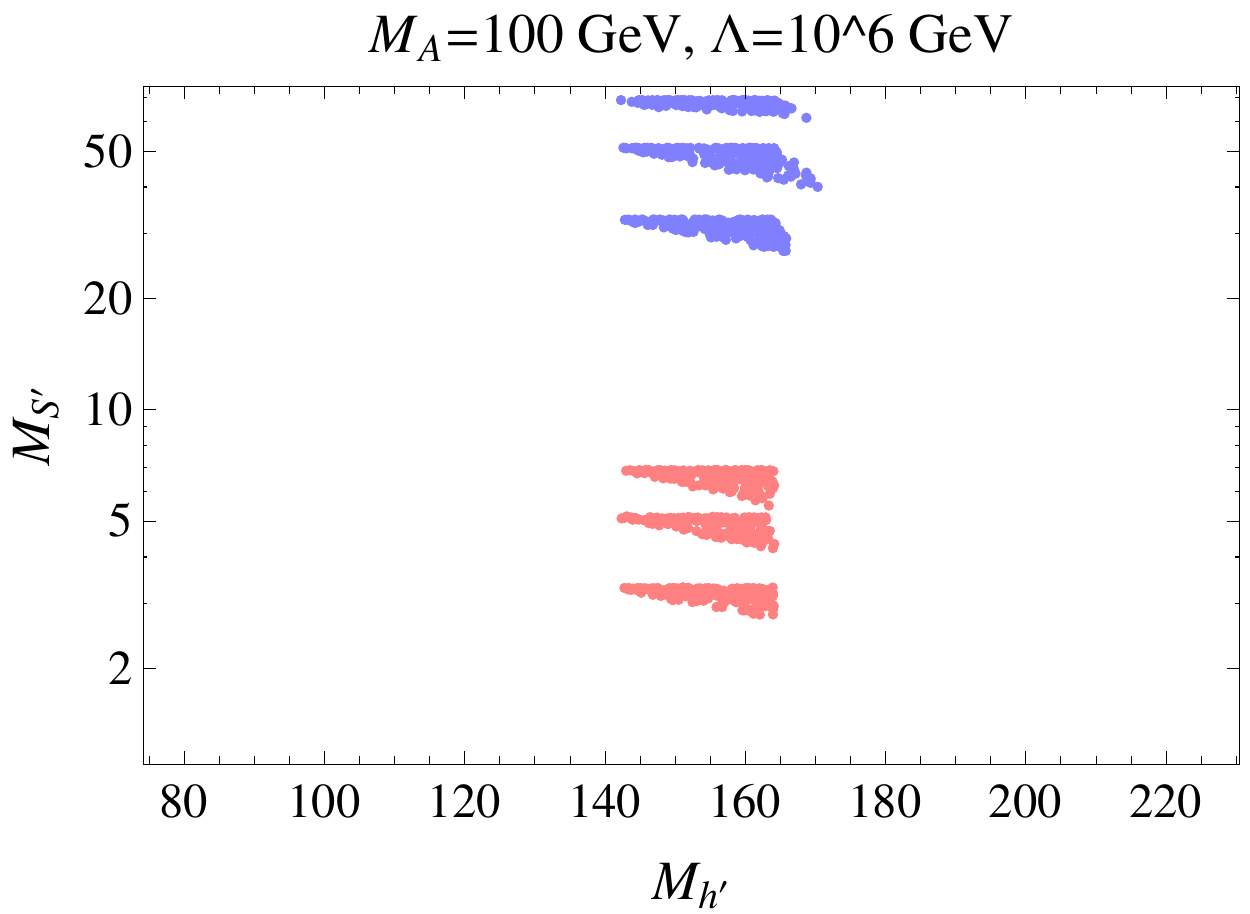}\\
\includegraphics[width=0.29\textwidth]{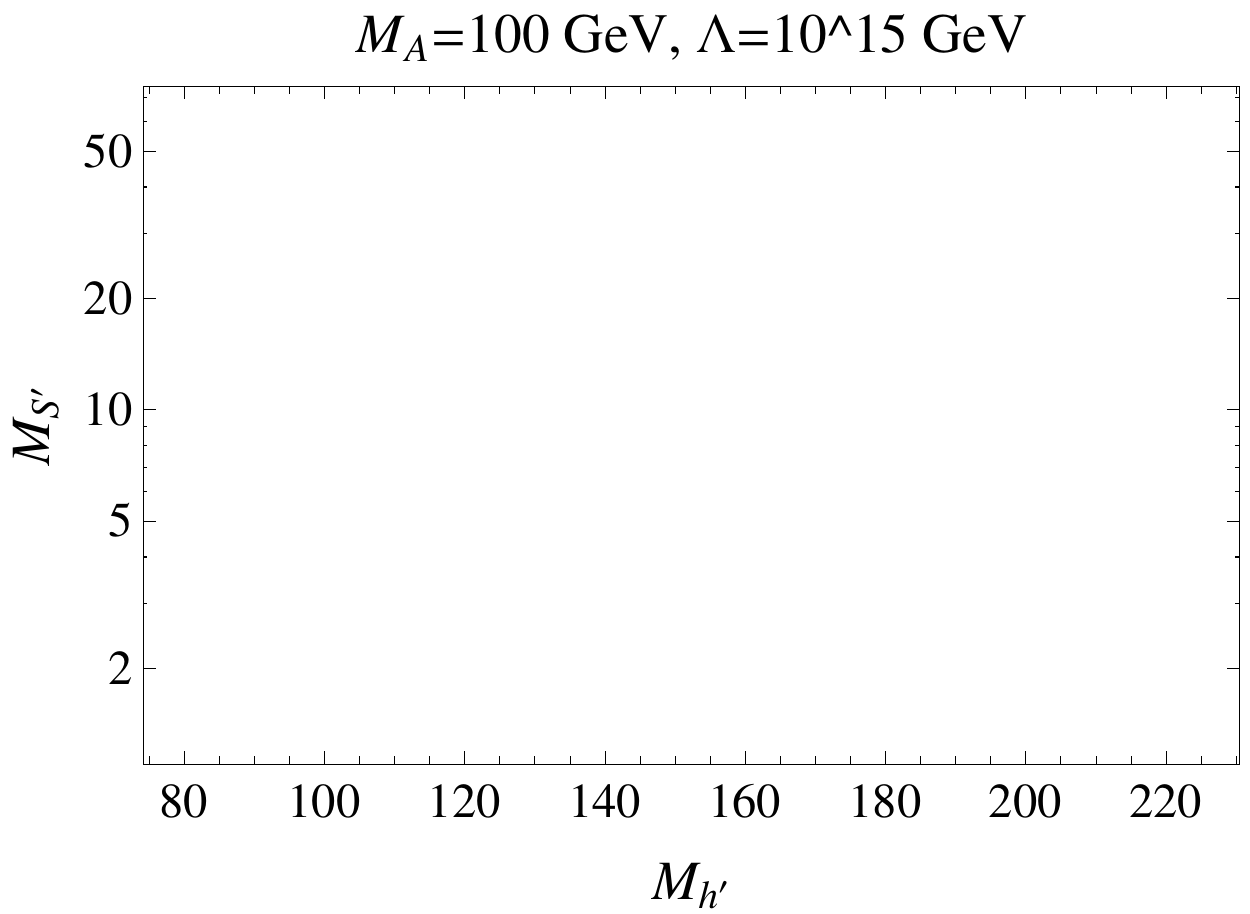}
\includegraphics[width=0.29\textwidth]{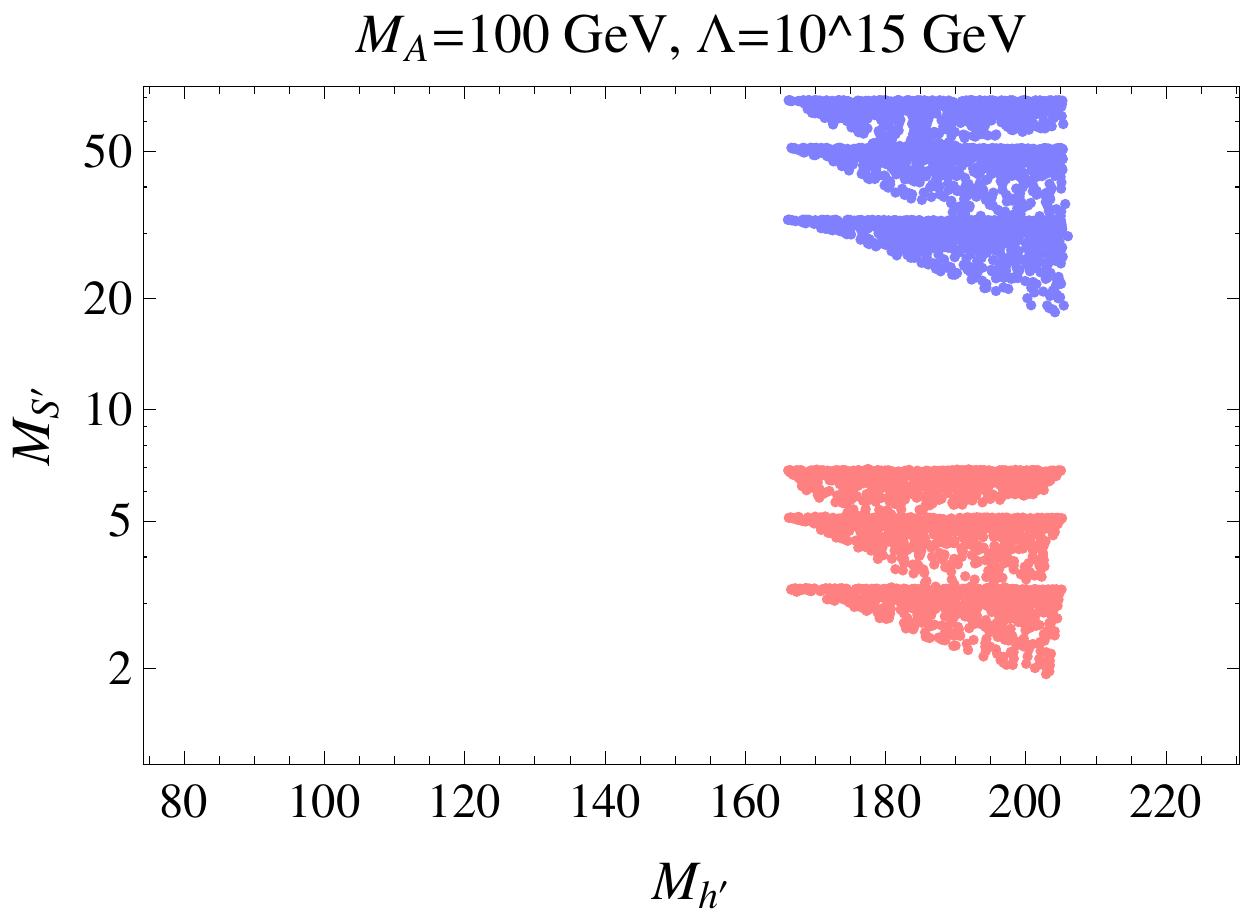}
\includegraphics[width=0.29\textwidth]{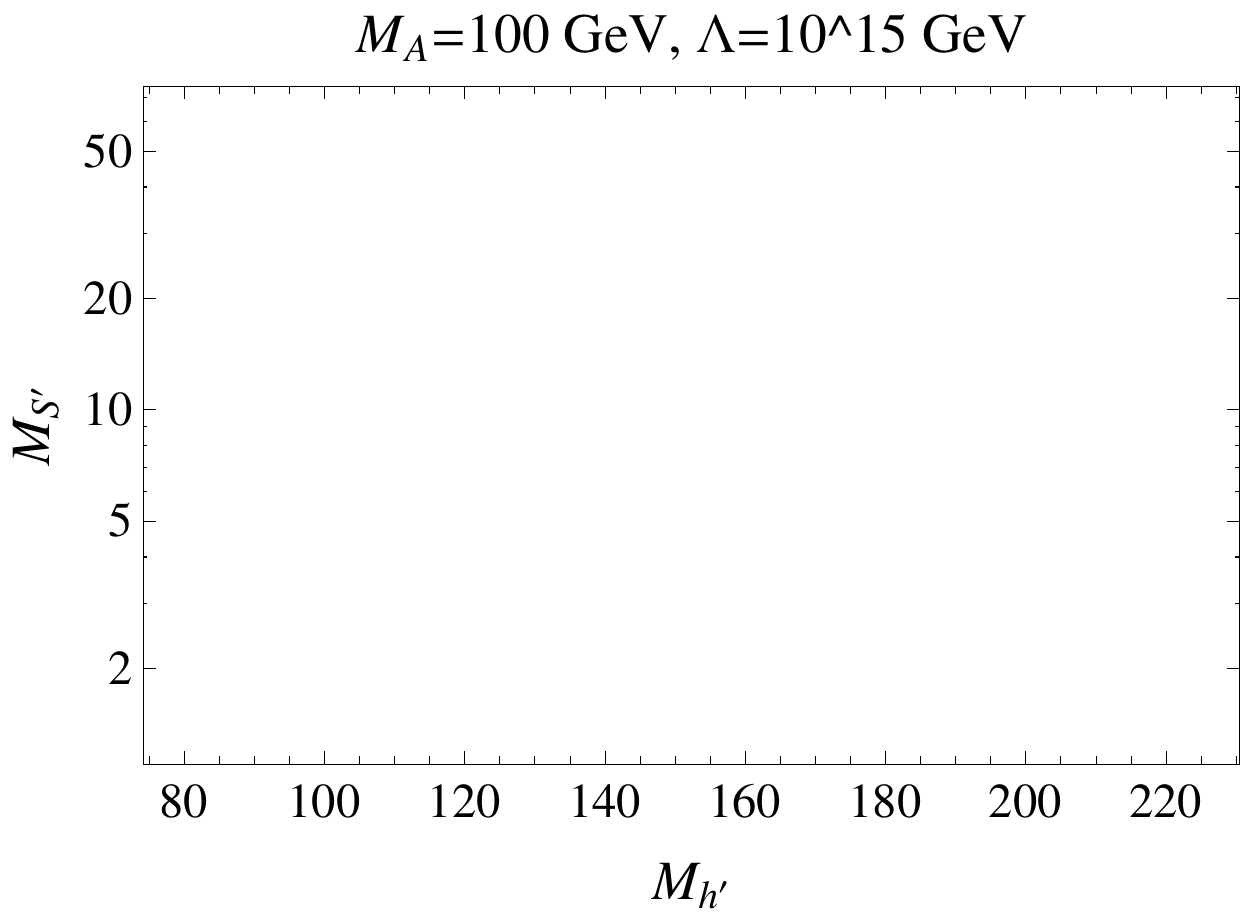}
\caption{The same as \figref{fig:ms_vs_mh_all_lepewpo_ddinvis} but with a dark matter mass of 100~GeV.}
\label{fig:ms_vs_mh_all100_lepewpo_ddinvis}
\end{figure*}

\Figref{fig:ms_vs_mh_all_lepewpo_ddinvis} and \figref{fig:ms_vs_mh_all100_lepewpo_ddinvis} show the inclusion of LEP and EWPO constraints (right columns), direct detection and invisible search constraints (left columns), and higher cutoff scales of the effective theory for $M_A=10$~GeV and 100~GeV respectively.  Most of the discussion in \secref{sec:light_dark_matter} generalizes to the other choices of the parameters.  We summarize the main results from these two figures as follows:
\begin{itemize}
\item Avoiding oversaturation of the relic density requires at least one scalar eigenstate to be lighter than the dark matter except in the vicinity of a resonance in the annihilation cross section.  Thus, if any of the dark matter direct detection experiments --- in particular, DAMA/LIBRA, CoGeNT, or CRESST-II --- unambiguously detects lighter ($M_A\simeq 10$~GeV) or heavier ($M_A\simeq 100$~GeV) dark matter, it would be natural to consider the possibility of other light scalars weakly coupled to the SM.
\item The RG evolution and vacuum stability analysis requires heavier Higgs- and singlet-like eigenstates, {\em i.e.}, less mixing, at larger cutoff scales (central columns).  This is primarily to avoid the runaway direction in the potential corresponding to $\del<0$ and $\del\of{\mu}^2>\lambda\of{\mu}d_2\of{\mu}$.
\item \themodel~requires the existence of additional new physics below the GUT scale, regardless of whether the dark matter is lighter or heavier.  If the $S'$ eigenstate is lighter than the $h'$ state, the RG evolution constraints and the LEP and EWPO limits (right columns) become mutually exclusive at high cutoff scales.  If the $S'$ eigenstate is heavier than the $h'$, it is possible to satisfy LEP and EWPO constraints with a lighter Higgs, but having a heavier $S'$ results in oversaturation of the relic density.
\item The regions of parameter space most favorable for the invisible decay channel in the ATLAS detector correspond to light (10~GeV) dark matter, a light singlet-like eigenstate, and a low (1~TeV) new physics scale.
\end{itemize}

\subsection{Discussion}

Putting together the trends from all of the plots in \figref{fig:ms_vs_mh_lepewpo} through \figref{fig:ms_vs_mh_all100_lepewpo_ddinvis}, we conclude that in order for \themodel~to be natural, the singlet vev cannot be very large (1000~GeV) and the scale of new physics has to be at most $10^{12}$~GeV.  Moreover, if ATLAS observes the invisible decay mode of the Higgs, the new physics scale has to be in fact much smaller, at most 10~TeV, and the singlet self-coupling $d_2$ cannot be too large. To illustrate, for $x = 10$~GeV, the maximum value of $\Lambda$ for points that satisfy all constraints (LEP, EWPO, relic density, and direct detection) is $10^{12}$~GeV.  However, as \figref{fig:ms_vs_mh_all_lepewpo_ddinvis} shows, none of these points satisfy the requirements for the ATLAS Higgs search via invisible decays. For $x = 100$~GeV, the maximum $\Lambda$ for points that satisfy all constraints is about 100~TeV for $d_2 = 0.2$, 10~TeV for $d_2 = 0.5$, and slightly greater than 1~TeV for $d_2 = 0.9$.  When we consider those points that also satisfy the requirements for the ATLAS Higgs search via invisible decays, the maximum $\Lambda$ is 10~TeV for $d_2 = 0.2$, approximately 5~TeV for $d_2 = 0.5$, and no points for $d_2 = 0.9$. For $x = 1000$~GeV, there are no points that satisfy all constraints at any cutoff scale.

It is interesting, then, to consider the additional impact of specifying the mass of the Higgs-like scalar to the range suggested by the recent ATLAS and CMS results\footnote{We note that the significance of the reported excess will vary with the singlet-doublet mixing angle that affects the production cross section and the value of the dark matter mass that could allow additional decay channels to open.}. If the mass of the Higgs-like scalar is 125~GeV and ATLAS does not detect the invisible decay mode, then the region near $x = 10$~GeV is favored (the singlet-like state is very light) and the maximum $\Lambda$ is $\sim 100$~TeV.\footnote{In these scenarios for which ATLAS is not sensitive to the invisible decay mode, the Higgs branching fraction to dark matter may nonetheless be large and the branching fractions to visible final states reduced.  This may be in conflict with the results from ATLAS and CMS, as pointed out in \cite{He:2011gc} for the $Z_2x$SM.}  On the other hand, if the Higgs-like scalar  mass is 125~GeV and ATLAS does detect the invisible decay mode, this seems to be compatible only with a singlet vev near $x = 100$~GeV and $d_2 = 0.2$. In this case, the singlet is again very light, with mass of order 20-25~GeV, and \themodel~requires new physics at 5~TeV, well within reach of the LHC.

\section{Conclusions}\label{sec:concl}

Two of the significant outstanding questions in particle physics are the nature of dark matter and  the energy scale associated with  physics beyond the Standard Model.  Both of these topics are relevant for general simple scalar extensions of the SM including the model we have studied, a complex scalar singlet extension.  In \themodel, through spontaneous and soft breaking of a global $U(1)$ symmetry, we obtain a massive stable dark matter candidate and two scalars which mix at the minimum of the potential, one Higgs-like and the other singlet-like.  Rather than a traditional gauge-dependent vacuum stability analysis of the one-loop effective potential, we have chosen to place constraints on the parameters of \themodel~using a gauge-independent analysis of the renormalization group evolution of the quartic couplings, motivated by requiring the potential to be bounded below for the vacuum to be absolutely stable (metastable vacua may also be allowed, but we have not considered this possibility in our analysis).  Our analysis shows that constraints on the RG running of the couplings gives results quite similar to the traditional vacuum stability analysis of the Landau gauge effective potential when the Higgs-dark matter coupling, $\del$, is negative as can be favorable to an EWPT; for $\del>0$, the RG running constraints may be more conservative than is strictly necessary for vacuum stability.  

We have also considered constraints from relic density measurements, the electroweak phase transition, LEP, EWPO data, and dark matter direct detection experiments.  Additionally, we have considered the sensitivity of the ATLAS experiment to a scalar which decays invisibly to dark matter.  We have found that if the scale of new physics (the effective theory cutoff of the RG evolution of \themodel~parameters) is a TeV,  then it is possible to satisfy all phenomenological constraints with a Higgs-like scalar mass in the region allowed by recent results from ATLAS and CMS and a light singlet-like scalar with a mass roughly twice the dark matter mass or less.  Under these conditions, the mixing between the Higgs and the complex scalar singlet is very small, so the singlet-like scalar couples very weakly to SM particles.  If the dark matter is light (10~GeV), the ATLAS detector may have sufficient sensitivity to a Higgs that decays to dark matter with a large branching fraction.

If new physics does not appear at the TeV scale but instead arises at higher scales ($10^6$~GeV or the GUT scale, $10^{15}$~GeV), \themodel~is severely restricted by the combined vacuum stability and phenomenological bounds.  In particular, limits from EWPO data are most in conflict with vacuum stability constraints (when the singlet-like eigenstate is lighter than the Higgs-like state) since the former generally favors a lighter Higgs while the latter requires a heavier Higgs at higher cutoff scales.  

With the continued operation of the LHC, a definitive statement on the existence of a SM-like Higgs and TeV scale new physics is expected.  Determining whether or not the Higgs exists and whether or not it has SM production cross sections and branching fractions will shed light on the scalar sector of fundamental particle physics.  Conclusive and consistent results from dark matter direct detection experiments will also provide information crucial for determining whether or not dark matter is a scalar particle and how dark matter couples to the SM.  Until such results become available, our vacuum stability and phenomenology analysis has shown that further study of simple scalar extensions of the SM --- in particular \themodel~with a light singlet-like scalar --- is worthwhile.

\section*{Acknowledgements}
MG thanks M. McCaskey for useful discussion and technical assistance, and also A. Long, P. McGuirk, and H. Patel for useful discussion. This work was supported in part in by the U.S. Department of Energy Contract DE-FG02-08ER41531 and the Wisconsin Alumni Research Foundation. 

\appendix 
\section{One-Loop Potential \& RGEs}\label{app:rges}

The one-loop potential is given by
\begin{equation}
V_1\of{h,S,A}=\frac{1}{64\pi^2}\sum_i n_i\text{Tr}\lc M_i^4\of{\log\frac{M_i^2}{\mu^2}-c_i}\rc \ \ .
\end{equation}
The sum $i$ runs over the scalar, fermion, and vector boson contributions.  The field-dependent mass matrices are given in \eqnsref{eq:scalar_msq}{eq:vector_msq}.  The number of degrees of freedom associated with each contribution, $n_i$, are given in \eqnref{eq:dof}, and the numerical factors are given in \eqnref{eq:c_num_factors}.  In the scalar sector, we include the contributions from the would-be Goldstone bosons.  We use the notation $g$ and $g'$ for the $SU(2)_L$ and $U(1)_Y$ gauge couplings, respectively, and $y_t$ is the top quark Yukawa coupling.  Due to the smallness of the other fermion Yukawa couplings, we exclude them from the one-loop effective potential.
\begin{widetext}
\begin{align}
&M_{\text{scalar}}^2 = 
	\text{Diag}\of{
		M_\phi^2\of{h,S,A},
		\lb\begin{array}{ccc}
			m_{hh}^2\of{h,S,A} & m_{hS}^2\of{h,S,A} & m_{hA}^2\of{h,S,A}\\
			m_{hS}^2\of{h,S,A} & m_{SS}^2\of{h,S,A} & m_{SA}^2\of{h,S,A}\\
			m_{hA}^2\of{h,S,A} & m_{SA}^2\of{h,S,A} & m_{AA}^2\of{h,S,A}\\
		\end{array}\rb 
	}\label{eq:scalar_msq}
\end{align}
where
\begin{align}
&M_\phi^2\of{h,S,A}=\of{\frac{m^2}{2}+\frac{\delta_2}{4}\of{S^2+A^2}+\frac{\lambda}{4}h^2}\mathbb{I}_{3\times 3}\\
&m_{hh}^2\of{h,S,A}=\frac{m^2}{2}+\frac{\delta_2}{2}\of{S^2+A^2}+\frac{3\lambda}{4}h^2\\
&m_{hS}^2\of{h,S,A}=\frac{\delta_2}{2}hS\\
&m_{hA}^2\of{h,S,A}=\frac{\delta_2}{2}hA\\
&m_{SS}^2\of{h,S,A}=\frac{1}{2}\of{b_2-b_1}+\frac{\delta_2}{4}h^2 + \frac{d_2}{4}\of{3S^2+A^2}\\
&m_{SA}^2\of{h,S,A}=\frac{d_2}{2}SA\\
&m_{AA}^2\of{h,S,A}=\frac{1}{2}\of{b_2+b_1} + \frac{\delta_2}{4}h^2 + \frac{d_2}{4}\of{S^2+3A^2}
\end{align}

\begin{align}
&M_{\text{fermion}}^2 = \frac{1}{2}y_t^2h^2\label{eq:fermion_msq}\\
&M_{\text{vector}}^2 = \text{Diag}\of{\frac{1}{4}g^2h^2,\frac{1}{4}g^2h^2,\frac{1}{4}\of{g^2+g'^{2}}h^2}\label{eq:vector_msq}\\
&n_{\text{scalar}}=1,\ n_{\text{fermion}}=-2,\ n_{\text{vector}}=3\label{eq:dof}\\
&c_{\text{scalar}}=\frac{3}{2},\ c_{\text{fermion}}=\frac{3}{2},\ c_{\text{vector}}=\frac{5}{6}\label{eq:c_num_factors}
\end{align}

The $\beta$ and $\gamma$ functions are defined as
\begin{equation}\label{eq:beta_fns}
\beta_X\equiv\mu\frac{dX}{d\mu},\ \gamma_Y\equiv \frac{\mu}{2}\frac{d\log Z_Y}{d\mu}
\end{equation}
for some coupling or mass parameter $X$ and some field $Y$ with wave function renormalization $Z_Y$.  The $\beta$ and $\gamma$ functions for \themodel~are shown in \eqnsref{eq:beta_lambda}{eq:beta_omega}.  Since there are no interactions between scalars involving derivatives, at one-loop order there is no contribution to $\gamma_h,\ \gamma_S$, or $\gamma_A$ from scalar loops.  Thus $\gamma_S=\gamma_A=0$ and $\gamma_h$ is unchanged from the SM result.

\begin{align}
\beta_{\lambda} &=\ \loopfactor\of{6\lambda^2 + \delta_2^2 - 36y_t^4 + \frac{9}{4}{g'}^4 + \frac{9}{2}g^2{g'}^2 + \frac{27}{4}g^4} + 4\lambda\gamma_h \label{eq:beta_lambda}\\
\beta_{\delta_2} &=\ \loopfactor\of{2d_2\delta_2 + 2\delta_2^2 + 3\delta_2\lambda} + 2\delta_2\of{\gamma_h + \gamma_S}\label{eq:beta_delta2}\\
\beta_{d_2} &=\ \loopfactor\of{5d_2^2+2\delta_2^2}+4d_2\gamma_S\\
\beta_{m^2} &=\ \loopfactor\of{b_2\delta_2 + 3m^2\lambda} + 2m^2\gamma_h\\
\beta_{b_1} &=\ \loopfactor\of{b_1d_2} + 2b_1\gamma_S\\
\beta_{b_2} &=\ \loopfactor\of{2b_2d_2 + 2m^2\delta_2} + 2b_2\gamma_S\\
\beta_{a_1} &=\ a_1\gamma_S\\
\gamma_h &=\ \loopfactor\of{3y_t^2 - \frac{9}{4}g^2 - \frac{3}{4}{g'}^2}\\
\gamma_S &=\ \gamma_A = 0\\
\end{align}

Also included is the running of the vacuum energy,
\begin{equation}\label{eq:beta_omega}
\mu\frac{d\Omega}{d\mu} = \loopfactor\of{\frac{b_1^2}{4} + \frac{b_2^2}{4} + \frac{m^4}{2}}
\end{equation}

\section{Oblique Parameters}\label{app:STU}

Here we present the oblique parameters $S, T$, and $U$, in terms of the gauge boson propagator corrections.  We use the usual shorthand $c_w=\cos\theta_w = M_W/M_Z, s_w=\sin\theta_w, t_w=\tan\theta_w$.
\begin{equation}\label{eq:def_oblique_params}
\begin{aligned}
&\alpha T = && -\frac{g^2}{\of{4\pi}^2}\frac{1}{M_W^2}\of{c_W^2\lb \Pi_{ZZ}\of{0} + 2t_w\Pi_{Z\gamma}\of{0}\rb - \Pi_{WW}\of{0}}\\
&\alpha S = && -\frac{g^2}{\of{4\pi}^2}\frac{4s_w^2c_w^2}{M_Z^2}\ \text{Re}\of{\Pi_{ZZ}\of{0}-\Pi_{ZZ}\of{M_Z^2} + \frac{c_w^2-s_w^2}{c_ws_w}\lb\Pi_{Z\gamma}\of{M_Z^2}-\Pi_{Z\gamma}\of{0}\rb + \Pi_{\gamma\gamma}\of{M_Z^2}}\\
&\alpha U = && -\frac{g^2}{\of{4\pi}^2}\left(\frac{4s_w^2}{M_W^2}\lb \Pi_{WW}\of{0}-\Pi_{WW}\of{M_W^2}\rb + \frac{4s_w^2c_w^2}{M_Z^2}\lb\Pi_{ZZ}\of{M_Z^2}-\Pi_{ZZ}\of{0}\rb\right.\\
& &&\qquad \left. + \frac{2s_wc_w}{M_Z^2}\lb\Pi_{Z\gamma}\of{M_Z^2}-\Pi_{Z\gamma}\of{0}\rb + \frac{s_w^2}{M_Z^2}\Pi_{\gamma\gamma}\of{M_Z^2}\right)
\end{aligned}
\end{equation}

When calculating the differences $\Delta O$ between \themodel~and the SM values of $S, T$, and $U$, all terms except those dependent upon the scalar masses cancel.  The relevant terms are given in \eqnref{eq:oblique_params_masses}.
\begin{equation}\label{eq:oblique_params_masses}
\begin{aligned}
&T = && -\frac{1}{16\pi}\frac{3}{s_w^2}\lb\frac{1}{c_w^2}\frac{M_h^2}{M_h^2-M_Z^2}\ln\of{\frac{M_h^2}{M_Z^2}} - \frac{M_h^2}{M_h^2-M_W^2}\ln\of{\frac{M_h^2}{M_W^2}}\rb + \cdots \\
&S = && -\frac{1}{4\pi}\lb \frac{3M_h^2}{M_h^2-M_Z^2}\ln\of{\frac{M_h^2}{M_Z^2}}+\frac{M_h^2}{2M_Z^2} + 2F_1\of{M_h^2,M_Z^2,M_Z^2} +\frac{2M_h^2}{M_Z^2}F_1\of{M_Z^2,M_h^2,M_Z^2}\right.\\
& &&\qquad \left. -\frac{M_h^2}{M_Z^2}\ln\of{M_h^2} - 4F_0\of{M_Z^2,M_h^2,M_Z^2} - 2\Delta F_{12}\of{M_Z^2,M_h^2,M_Z^2}\rb + \cdots \\
&U = && -\frac{1}{4\pi}\lb \frac{3M_h^2}{M_h^2-M_W^2}\ln\of{\frac{M_h^2}{M_W^2}}+\frac{M_h^2}{2M_W^2} + 2F_1\of{M_h^2,M_W^2,M_W^2} + \frac{2M_h^2}{M_W^2}F_1\of{M_W^2,M_h^2,M_W^2}\right.\\
& &&\qquad \left. - \frac{M_h^2}{M_W^2}\ln\of{M_h^2} - 4F_0\of{M_W^2,M_h^2,M_W^2} - 2\Delta F_{12}\of{M_W^2,M_h^2,M_W^2}\rb - S + \cdots
\end{aligned}
\end{equation}
The functions $F_0, F_1,$ and $F_2$ ($\Delta F_{12}\equiv F_1 - F_2$) are given by \eqnref{eq:oblique_params_F_fns}.
\begin{equation}\label{eq:oblique_params_F_fns}
\begin{aligned}
F_i\of{M_1^2,M_2^2,q^2}=\int_0^1 dz\ z^i\ln\lb\of{1-z} M_1^2+z M_2^2-z\of{1-z}q^2\rb
\end{aligned}
\end{equation}
\end{widetext}

\bibliography{cxsm_vs}
\bibliographystyle{utcaps}

\end{document}